\documentclass[pdftex,twocolumn,epjc3_preprint,runningheads]{svjour3}

\usepackage[colorlinks,citecolor=blue,urlcolor=blue,linkcolor=blue,breaklinks=true]{hyperref}
\usepackage[dvipsnames]{xcolor}

\pdfoutput=1

\usepackage[T1]{fontenc}
\usepackage{lmodern}
\usepackage{calc}
\usepackage{graphicx}
\usepackage{booktabs}
\usepackage{textcomp}
\usepackage{xspace}
\usepackage{relsize}
\usepackage{amssymb}
\usepackage{amsmath}
\usepackage{listings}
\usepackage{microtype}
\usepackage{multirow}
\usepackage{tabularx}
\usepackage{array}
\usepackage{placeins}
\usepackage{cuted}
\usepackage{soul} % only for \st; delete if this causes you problems.
\usepackage{fixltx2e}
\usepackage{slashed}
\usepackage{bm}
\usepackage[numbers,sort&compress]{natbib}
\usepackage[labelfont=bf,font=small]{caption}
\usepackage[skip=-2pt]{subcaption}
\usepackage[clockwise,figuresright]{rotating}
\usepackage{tikz}
\usepackage[normalem]{ulem}
\usepackage[utf8]{inputenc}

\usepackage{etoolbox}
\AfterEndEnvironment{strip}{\leavevmode}

\allowdisplaybreaks

\newcolumntype{L}{>{\raggedright\let\newline\\\arraybackslash\hspace{0pt}}X}
\newcolumntype{R}{>{\raggedleft\let\newline\\\arraybackslash\hspace{0pt}}X}
\newcolumntype{C}{>{\centering\let\newline\\\arraybackslash\hspace{0pt}}X}

\newcommand{\gambitinstitute}[1]{\expandafter\csname #1\endcsname\label{#1}}
\newcommand{\gi}[1]{\gambitinstitute{#1}\and}
\newcommand{\last}[1]{\gambitinstitute{#1}}

\makeatletter

\newcommand{\preprintnumber}[1]{\gdef\@preprintnumber{\begin{flushright}{#1}\end{flushright}}}

\g@addto@macro\bfseries{\boldmath}
\makeatother

\let\underscore\_
\renewcommand{\_}{\discretionary{\underscore}{}{\underscore}}

\makeatletter
\let\orgdescriptionlabel\descriptionlabel
\renewcommand*{\descriptionlabel}[1]{%
  \let\orglabel\label
  \let\label\@gobble
  \phantomsection
  \protected@edef\@currentlabel{#1}%
  \let\label\orglabel
  \orgdescriptionlabel{#1}%
}
\makeatother

\newcommand\postnewlinemarker{\hbox{\ensuremath{\hookrightarrow}}}
\newcommand\cpp[1]{{\lstinline!#1!}}  % Apparently curly braces are only "experimental"

\newcommand\yaml[1]{{\lstset{style=yaml}\lstinline!#1!\lstset{style=cpp}}}

\newcommand\term[1]{{\lstset{style=terminal}\lstinline!#1!\lstset{style=cpp}}}
\newcommand\termalt[1]{{\lstset{style=terminalalt}\lstinline!#1!\lstset{style=cpp}}}
\newcommand\fortran[1]{{\lstset{style=fortran}\lstinline!#1!\lstset{style=cpp}}}
\newcommand\py[1]{{\lstset{style=python}\lstinline!#1!\lstset{style=cpp}}}
\newcommand\customtilde{{\raisebox{0.2ex}{\scalebox{0.6}{\boldmath$\sim$}}}}
\newcommand\mathematica[1]{{\lstset{style=Mathematica}\lstinline!#1!\lstset{style=cpp}}}
\newcommand\guminline[1]{{{\lstset{style=gum}\lstinline!#1!}}}
\newcommand\textinline[1]{{{\lstset{style=text}\lstinline!#1!}}}

\def\be{\begin{equation}}
\def\ee{\end{equation}}
\def\ba{\begin{eqnarray}}
\def\ea{\end{eqnarray}}
\newcommand{\bea}{\begin{eqnarray}}
\newcommand{\eea}{\end{eqnarray}}

\lstnewenvironment{lstlistingyaml}{\lstset{style=yaml}}{\lstset{style=cpp}}
\lstnewenvironment{lstlistingterm}{\lstset{style=terminal}}{\lstset{style=cpp}}
\lstnewenvironment{lstlistingfortran}{\lstset{style=fortran}}{\lstset{style=cpp}}
\lstnewenvironment{lstcpp}{\lstset{style=cpp}}{\lstset{style=cpp}}
\lstnewenvironment{lstcppalt}{\lstset{style=cppalt}}{\lstset{style=cpp}}
\lstnewenvironment{lstcppnum}{\lstset{style=cppnum}}{\lstset{style=cpp}}
\lstnewenvironment{lstyaml}{\lstset{style=yaml}}{\lstset{style=cpp}}
\lstnewenvironment{lstgum}{\lstset{style=gum}}{\lstset{style=cpp}}
\lstnewenvironment{lstterm}{\lstset{style=terminal}}{\lstset{style=cpp}}
\lstnewenvironment{lsttermalt}{\lstset{style=terminalalt}}{\lstset{style=cpp}}
\lstnewenvironment{lsttext}{\lstset{style=text}}{\lstset{style=cpp}}
\lstnewenvironment{lstfortran}{\lstset{style=fortran}}{\lstset{style=cpp}}
\lstnewenvironment{lstpy}{\lstset{style=python}}{\lstset{style=cpp}}
\lstnewenvironment{lstmathematica}{\lstset{style=mathematica}}{\lstset{style=cpp}}

\newcommand{\tmpname}{}
\newcommand{\tmplistingname}{}
\makeatletter
\newif\ifATOlabelname
\lst@Key{labelname}{Listing}{\def\ATOlabelname{#1}\global\ATOlabelnametrue}
\makeatother
\lstnewenvironment{lstcpplabel}[1][]{
  \lstset{style=cpp,#1} % #1 allows to add new options with [] same as for normal lstlistings environment
  \ifATOlabelname
    \renewcommand{\tmpname}{\lstlistingname}
    \renewcommand{\tmplistingname}{\lstlistlistingname}
    \renewcommand{\lstlistingname}{\ATOlabelname}% Listing -> labelname
    \renewcommand{\lstlistlistingname}{List of \lstlistingname s}% List of Listings -> List of labelname
  \fi
}{
  \renewcommand{\lstlistingname}{\tmpname}
  \renewcommand{\lstlistlistingname}{\tmplistingname}
  \lstset{style=cpp}
}
\definecolor{solarized@base03}{HTML}{002B36}
\definecolor{solarized@base02}{HTML}{073642}
\definecolor{solarized@base01}{HTML}{586e75}
\definecolor{solarized@base00}{HTML}{657b83}
\definecolor{solarized@base0}{HTML}{839496}
\definecolor{solarized@base1}{HTML}{93a1a1}
\definecolor{solarized@base2}{HTML}{EEE8D5}
\definecolor{solarized@base3}{HTML}{FDF6E3}
\definecolor{solarized@yellow}{HTML}{B58900}
\definecolor{solarized@orange}{HTML}{CB4B16}
\definecolor{solarized@red}{HTML}{DC322F}
\definecolor{solarized@magenta}{HTML}{D33682}
\definecolor{solarized@violet}{HTML}{6C71C4}
\definecolor{solarized@blue}{HTML}{268BD2}
\definecolor{solarized@cyan}{HTML}{2AA198}
\definecolor{solarized@green}{HTML}{859900}
\definecolor{darkred}{HTML}{550003}
\definecolor{darkgreen}{HTML}{00AA00}
\definecolor{orchid}{HTML}{AF06F5}

\newcommand\YAMLstringstyle{\footnotesize\color{solarized@green}\mdseries}
\newcommand\YAMLkeystyle{\footnotesize\color{solarized@blue}\ttfamily}
\newcommand\YAMLvaluestyle{\footnotesize\color{blue}\mdseries}
\newcommand\ProcessThreeDashes{\llap{\color{cyan}\mdseries-{-}-}}
\newcommand\CPPcommentstyle{\color{solarized@violet}\footnotesize\ttfamily}
\newcommand\CPPdirectivestyle{\color{solarized@magenta}\footnotesize\ttfamily}
\newcommand\termplainstyle{\footnotesize\ttfamily}

\newcommand\YAMLcommentstyle{\color{solarized@orange}\ttfamily}

\newcommand\processLongMacroDelimiter
{%
\CPPdirectivestyle%
\#define%
}

\lstdefinestyle{cpp}
{
  language=C++,
  basicstyle=\footnotesize\ttfamily,
  basewidth={0.53em,0.44em}, %Ben: experimenting a bit with the fixed-width width (first argument); feels a bit more readable to me with the slightly smaller width (was 0.6em by default)
  numbers=none,
  tabsize=2,
  breaklines=true,
  escapeinside={@}{@},
  showstringspaces=false,
  numberstyle=\tiny\color{solarized@base01},
  keywordstyle=\color{solarized@orange},
  stringstyle=\color{solarized@red}\ttfamily,
  identifierstyle=\color{solarized@blue},
  commentstyle=\CPPcommentstyle,
  directivestyle=\CPPdirectivestyle,
  emphstyle=\color{solarized@green},
  frame=single,
  rulecolor=\color{solarized@base2},
  rulesepcolor=\color{solarized@base2},
  literate={~} {\customtilde}1,
  moredelim=*[directive]\ \ \#,
  moredelim=*[directive]\ \ \ \ \#
}

\lstdefinestyle{cppalt}
{
  language=C++,
  basicstyle=\footnotesize\ttfamily,
  basewidth={0.53em,0.44em}, %Ben: experimenting a bit with the fixed-width width (first argument); feels a bit more readable to me with the slightly smaller width (was 0.6em by default)
  numbers=none,
  tabsize=2,
  breaklines=true,
  escapeinside={*@}{@*},
  showstringspaces=false,
  numberstyle=\tiny\color{solarized@base01},
  keywordstyle=\color{solarized@orange},
  stringstyle=\color{solarized@red}\ttfamily,
  identifierstyle=\color{solarized@blue},
  commentstyle=\CPPcommentstyle,
  directivestyle=\CPPdirectivestyle,
  emphstyle=\color{solarized@green},
  frame=single,
  rulecolor=\color{solarized@base2},
  rulesepcolor=\color{solarized@base2},
  literate={~}{\customtilde}1,
  moredelim=**[is][\processLongMacroDelimiter]{BeginLongMacro}{EndLongMacro} %special delimiter for long macros that go over several lines
}

\lstdefinestyle{cppnum}
{
  language=C++,
  basicstyle=\footnotesize\ttfamily,
  basewidth={0.53em,0.44em}, %Ben: experimenting a bit with the fixed-width width (first argument); feels a bit more readable to me with the slightly smaller width (was 0.6em by default)
  numbers=none,
  tabsize=2,
  breaklines=true,
  escapeinside={@}{@},
  numberstyle=\tiny\color{solarized@base01},
  showstringspaces=false,
  keywordstyle=\color{solarized@orange},
  stringstyle=\color{solarized@red}\ttfamily,
  identifierstyle=\color{solarized@blue},
  commentstyle=\CPPcommentstyle,
  directivestyle=\CPPdirectivestyle,
  emphstyle=\color{solarized@green},
  frame=single,
  rulecolor=\color{solarized@base2},
  rulesepcolor=\color{solarized@base2},
  literate={~} {\customtilde}1,
  moredelim=*[directive]\ \ \#,
  moredelim=*[directive]\ \ \ \ \#
}

\lstdefinestyle{python}
{
  language=Python,
  basicstyle=\footnotesize\ttfamily,
  basewidth={0.53em,0.44em},
  numbers=none,
  tabsize=2,
  breaklines=true,
  escapeinside={@}{@},
  showstringspaces=false,
  numberstyle=\tiny\color{solarized@base01},
  keywordstyle=\color{blue},
  stringstyle=\color{orange}\ttfamily,
  identifierstyle=\color{darkred},
  commentstyle=\color{purple},
  emphstyle=\color{green},
  frame=single,
  rulecolor=\color{solarized@base2},
  rulesepcolor=\color{solarized@base2},
  literate = {~}{\customtilde}1
             {\ as\ }{{\color{blue}\ as\ \color{black}}}3
             {.set}{{\color{black}.}{\color{darkred}set}}4
}

\lstdefinestyle{fortran}
{
  language=Fortran,
  basicstyle=\footnotesize\ttfamily,
  basewidth={0.53em,0.44em},
  numbers=none,
  tabsize=2,
  breaklines=true,
  escapeinside={@}{@},
  showstringspaces=false,
  numberstyle=\tiny\color{solarized@base01},
  keywordstyle=\color{blue},
  stringstyle=\color{orange}\ttfamily,
  identifierstyle=\color{Periwinkle},
  commentstyle=\color{purple},
  emphstyle=\color{green},
  morekeywords={and, or, true, false},
  frame=single,
  rulecolor=\color{solarized@base2},
  rulesepcolor=\color{solarized@base2},
  literate={~}{\customtilde}1
}

\lstdefinestyle{terminal}
{
  language=bash,
  basicstyle=\termplainstyle,
  numbers=none,
  tabsize=2,
  breaklines=true,
  escapeinside={@}{@},
  frame=single,
  showstringspaces=false,
  numberstyle=\tiny\color{solarized@base01},
  keywordstyle=\color{solarized@orange},
  stringstyle=\color{solarized@red}\ttfamily,
  identifierstyle=\color{black},
  commentstyle=\color{solarized@violet},
  emphstyle=\color{solarized@green},
  frame=single,
  rulecolor=\color{solarized@base2},
  rulesepcolor=\color{solarized@base2},
  morekeywords={gambit, cmake, make, mkdir, gum, python, wget, tar, cp, pippi, mpirun},
  deletekeywords={test},
  literate = {/gambit}{{/}{\color{black}}gambit}6
             {gambit/}{{\color{black}}gambit{/}}6
             {gum/}{{\color{black}}gum{/}}4
             {/include}{{/}{\color{black}}include}8
             {cmake/}{{\color{black}}cmake/}6
             {.cmake}{{.}{\color{black}}cmake}6
             {.gum}{{.}{\color{black}}gum}6
             {.tar}{{.}{\color{black}}tar}4
             {source/}{{\color{black}}source{/}}7
             { type}{{\color{black}}{}type}5
             {~}{\customtilde}1
             {math}{{\color{solarized@orange}}math}4
}

\lstdefinestyle{terminalalt}
{
  language=bash,
  basicstyle=\footnotesize\ttfamily,
  numbers=none,
  tabsize=2,
  breaklines=true,
  escapeinside={*@}{@*},
  frame=single,
  showstringspaces=false,
  numberstyle=\tiny\color{solarized@base01},
  keywordstyle=\color{solarized@orange},
  stringstyle=\color{solarized@red}\ttfamily,
  identifierstyle=\color{black},
  commentstyle=\color{solarized@violet},
  emphstyle=\color{solarized@green},
  frame=single,
  rulecolor=\color{solarized@base2},
  rulesepcolor=\color{solarized@base2},
  morekeywords={gambit, cmake, make, mkdir},
  deletekeywords={test},
  literate = {\ gambit}{{\ }{\color{black}}gambit}7
             {/gambit}{{/}{\color{black}}gambit}6
             {gambit/}{{\color{black}}gambit{/}}6
             {/include}{{/}{\color{black}}include}8
             {cmake/}{{\color{black}}cmake/}6
             {.cmake}{{.}{\color{black}}cmake}6
             {~}{\customtilde}1
}

\lstdefinestyle{text}
{
  language={},
  basicstyle=\footnotesize\ttfamily,
  identifierstyle=\color{black},
  numbers=none,
  tabsize=2,
  breaklines=true,
  escapeinside={*@}{@*},
  showstringspaces=false,
  frame=single,
  rulecolor=\color{solarized@base2},
  rulesepcolor=\color{solarized@base2},
  literate={~}{\customtilde}1
}

\lstdefinestyle{yaml}
{
  language=bash,
  escapeinside={@}{@},
  keywords={true,false,null},
  otherkeywords={},
  keywordstyle=\color{solarized@base0}\bfseries,
  basicstyle=\footnotesize\color{black}\ttfamily,
  identifierstyle=\YAMLkeystyle,
  sensitive=false,
  commentstyle=\YAMLcommentstyle,
  morecomment=[l]{\#},
  morecomment=[s]{/*}{*/},
  stringstyle=\YAMLstringstyle\ttfamily,
  moredelim=**[s][\YAMLkeystyle]{,}{:},   % switch to value style at : but back to key style at,
  moredelim=**[l][\YAMLvaluestyle]{:},    % switch to value style at :
  morestring=[b]',
  morestring=[b]",
  literate =    {---}{{\ProcessThreeDashes}}3
                {>}{{\textcolor{solarized@red}\textgreater}}1
                {gtr}{\textgreater}1
                {grt}{\textgreater}1
                {|}{{\textcolor{solarized@red}\textbar}}1
                {\ -\ }{{\mdseries\color{black}\ -\ \negmedspace}}3
                {\}}{{{\color{black} \}}}}1
                {\{}{{{\color{black} \{}}}1
                {[}{{{\color{black} [}}}1
                {]}{{{\color{black} ]}}}1
                {~}{\customtilde}1,
  breakindent=0pt,
  breakatwhitespace,
  columns=fullflexible
}

\lstdefinestyle{gum}
{
  language=bash,
  escapeinside={@}{@},
  keywords={true,false,null,all},
  otherkeywords={},
  keywordstyle=\color{solarized@base02}\bfseries,
  basicstyle=\footnotesize\color{black}\ttfamily,
  identifierstyle=\color{solarized@magenta},
  sensitive=false,
  commentstyle=\color{solarized@cyan}\ttfamily,
  morecomment=[l]{\#},
  morecomment=[s]{/*}{*/},
  stringstyle=\footnotesize\color{solarized@base01}\mdseries\ttfamily,
  moredelim=**[l][\footnotesize\color{solarized@base02}\mdseries]{:},    % switch to value style at :
  morestring=[b]',
  morestring=[b]",
  literate =    {---}{{\ProcessThreeDashes}}3
                {grt}{{\textcolor{solarized@magenta}\textgreater}}1
                {gtr}{{\textcolor{solarized@base02}\textgreater}}1
                {/>}{{\textcolor{solarized@magenta}\textgreater}}1
                {/<}{{\textcolor{solarized@magenta}\textless}}1
                {lss}{{\textcolor{solarized@base02}\textless}}1
                {pls}{{\textcolor{solarized@magenta}+}}1
                {mns}{{\textcolor{solarized@magenta}-}}1
                {|}{{\textcolor{solarized@base02}\textbar}}1
                {\ -\ }{{\mdseries\color{solarized@base02}\ -\ \negmedspace}}3
                {\}}{{{\color{solarized@base02} \}}}}1
                {\{}{{{\color{solarized@base02} \{}}}1
                {[}{{{\color{solarized@base02} [}}}1
                {]}{{{\color{solarized@base02} ]}}}1
                {~}{\customtilde}1,
  breakindent=0pt,
  breakatwhitespace,
  columns=fullflexible
}

\lstdefinestyle{mathematica}
{
  language={Mathematica},
  basicstyle=\footnotesize\ttfamily,
  basewidth={0.53em,0.44em},
  numbers=none,
  tabsize=2,
  breaklines=true,
  postbreak=,
  escapeinside={@}{@},
  numberstyle=\tiny\color{black},
  showstringspaces=false,
  numberstyle=\tiny\color{solarized@base01},
  keywordstyle=\color{solarized@orange},
  stringstyle=\color{solarized@red}\ttfamily,
  identifierstyle=\color{solarized@orange}\ttfamily,
  commentstyle=\color{solarized@gray}\ttfamily,
  directivestyle=\color{solarized@orange}\ttfamily,
  emphstyle=\color{solarized@green},
  frame=single,
  rulecolor=\color{solarized@base2},
  rulesepcolor=\color{solarized@base2},
  literate={~} {\customtilde}1,
  moredelim=*[directive]\ \ \#,
  moredelim=*[directive]\ \ \ \ \#,
  mathescape=false
}

\newcommand{\doublecross}[2]{\hyperref[#2]{\textbf{#1}}}
\newcommand{\doublecrosssf}[2]{\hyperref[#2]{\textbf{\textsf{#1}}}}

\newcommand{\startglossary}{\section{Glossary}\label{glossary}Here we explain some terms that have specific technical definitions in \GB.\begin{description}}
\newcommand{\finishglossary}{\end{description}}

\newcommand{\bcode}{\begin{lstlisting}}
\newcommand{\ecode}{\end{lstlisting}}

\newcommand{\ie}{i.e.\ }

\newcommand{\gambit}{\textsf{GAMBIT}\xspace}

\newcommand{\darkbit}{\textsf{DarkBit}\xspace}
\newcommand{\cosmobit}{\textsf{CosmoBit}\xspace}
\newcommand{\colliderbit}{\textsf{ColliderBit}\xspace}

\newcommand{\GB}{\gambit}

\newcommand{\delphes}{\textsf{Delphes}\xspace}

\newcommand{\pythia}{\textsf{Pythia}\xspace}

\newcommand{\ds}{\textsf{DarkSUSY}\xspace}
\newcommand{\darksusy}{\ds}

\newcommand{\micromegas}{\textsf{micrOMEGAs}\xspace}

\newcommand\nulike{\textsf{nulike}\xspace}
\newcommand\gamLike{\textsf{gamLike}\xspace}
\newcommand\gamlike{\gamLike}

\newcommand\ddcalc{\textsf{DDCalc}\xspace}
\newcommand\capgen{\textsf{Capt'n General}\xspace}
\newcommand{\gum}{\textsf{GUM}\xspace}

\newcommand{\fr}{\textsf{FeynRules}\xspace}

\newcommand{\CH}{\textsf{CalcHEP}\xspace}
\newcommand{\MG}{\textsf{MadGraph}\xspace}

\newcommand{\ufo}{\textsf{UFO}\xspace}

\newcommand{\ddm}{\textsf{DirectDM}\xspace}

\newcommand\Python{\textsf{Python}\xspace}
\newcommand\python{\Python}

\newcommand\YAML{\textsf{YAML}\xspace}

\newcommand\beq{\begin{equation}}
\newcommand\eeq{\end{equation}}

\newcommand{\subparagraph}{} %< workaround for svjour not defining subparagraph
\journalname{Eur.\ Phys.\ J.\ C}
\smartqed

\makeatletter
\patchcmd{\ttlh@hang}{\parindent\z@}{\parindent\z@\leavevmode}{}{}
\patchcmd{\ttlh@hang}{\noindent}{}{}{}
\makeatother

\usepackage[title]{appendix}

\newcommand{\utc}{\textsf{ufo\_to\_mdl}\xspace}

\newcommand{\Q}[2]{
  \if\relax\detokenize{#2}\relax
    \mathcal{Q}_{#1}
  \else
    \mathcal{Q}_{#1}^{(#2)}
  \fi
}

\newcommand{\C}[2]{
  \if\relax\detokenize{#2}\relax
    \mathcal{C}_{#1}
  \else
    \mathcal{C}_{#1}^{(#2)}
  \fi
}
\newcommand{\La}{{\rm \Lambda}}

\usepackage{amssymb}% http://ctan.org/pkg/amssymb
\usepackage{pifont}% http://ctan.org/pkg/pifont

\begin{document}

\preprintnumber{ADP-21-9/T1156, CERN-TH-2021-084, CP3-21-15, P3H-21-038, TTK-21-19, gambit-physics-21}

\title{Thermal WIMPs and the Scale of New Physics: \\ Global Fits of Dirac Dark Matter Effective Field Theories}

\author{The GAMBIT Collaboration:
Peter Athron\thanksref{monash,nanjing} \and
Neal Avis Kozar\thanksref{mcdonald,queens} \and \\
Csaba Bal{\'a}zs\thanksref{monash} \and
Ankit Beniwal\thanksref{louvain,a} \and
Sanjay Bloor\thanksref{imperial,uq,b} \and
Torsten Bringmann\thanksref{oslo} \and \\
Joachim Brod\thanksref{cincy} \and
Christopher Chang\thanksref{uq} \and
Jonathan M. Cornell\thanksref{wsu} \and \\
Ben Farmer\thanksref{bom} \and
Andrew Fowlie\thanksref{nanjing} \and
Tom\'{a}s E. Gonzalo\thanksref{monash,aachen,c} \and
Will Handley\thanksref{kicc,cambridge} \and
Felix Kahlhoefer\thanksref{aachen,d} \and
Anders Kvellestad\thanksref{oslo} \and
Farvah Mahmoudi\thanksref{lyon,cern} \and \\
Markus T. Prim\thanksref{bonn} \and
Are Raklev\thanksref{oslo} \and
Janina J.\ Renk\thanksref{imperial,okc} \and
Andre Scaffidi\thanksref{adelaide,infn} \and
Pat Scott\thanksref{imperial,uq} \and
Patrick St\"{o}cker\thanksref{aachen} \and
Aaron C. Vincent\thanksref{mcdonald,queens,perimeter} \and
Martin White\thanksref{adelaide} \and
Sebastian Wild\thanksref{desy} \and
Jure Zupan\thanksref{cincy}
}

\institute{
  \gi{monash}
  \gi{nanjing}
  \gi{mcdonald}
  \gi{queens}
  \gi{louvain}
  \gi{imperial}
  \gi{uq}
  \gi{oslo}
  \gi{cincy}
  \gi{wsu}
  \gi{bom}
  \gi{aachen}
  \gi{kicc}
  \gi{cambridge}
  \gi{lyon}
  \gi{cern}
  \gi{bonn}
  \gi{okc}
  \gi{adelaide}
  \gi{infn}
  \gi{perimeter}
  \last{desy}
}

\thankstext{a}{ankit.beniwal@uclouvain.be}
\thankstext{b}{sanjay.bloor12@imperial.ac.uk}
\thankstext{c}{gonzalo@physik.rwth-aachen.de}
\thankstext{d}{kahlhoefer@physik.rwth-aachen.de}

\titlerunning{Global Fits of Dirac Dark Matter Effective Field Theories}
\authorrunning{The GAMBIT Collaboration}

\date{Received: date / Accepted: date}

\maketitle

\begin{abstract}
We assess the status of a wide class of WIMP dark matter (DM) models in light of the latest experimental results using the global fitting framework \gambit. We perform a global analysis of effective field theory (EFT) operators describing the interactions between a gauge-singlet Dirac fermion and the Standard Model quarks, the gluons and the photon. In this bottom-up approach, we simultaneously vary the coefficients of 14 such operators up to dimension 7, along with the DM mass, the scale of new physics and several nuisance parameters. Our likelihood functions include the latest data from \emph{Planck}, direct and indirect detection experiments, and the LHC. For DM masses below 100 GeV, we find that it is impossible to satisfy all constraints simultaneously while maintaining EFT validity at LHC energies. For new physics scales around 1 TeV, our results are influenced by several small excesses in the LHC data and depend on the prescription that we adopt to ensure EFT validity. Furthermore, we find large regions of viable parameter space where the EFT is valid and the relic density can be reproduced, implying that WIMPs can still account for the DM of the universe while being consistent with the latest data.
\end{abstract}

\tableofcontents

%%%%%%%%%%%%%%%%%%%%%%%%%%%%%%%%%%%%%%%%%%%%%%%%%%%%%%%%%%%%%%%%%%%%%%%%%%%%%%%%%%%%%%%%%%%%%%%%%%%%%%%%%%%%%%%%%%%%%%%%
\section{Introduction}

Despite years of searching, the identity of dark matter (DM) remains a mystery. Nevertheless, the large number of past, present and future probes of its particle interactions makes it essential to regularly revisit the constraints on the most popular theoretical candidates, in order to guide future searches.

A favoured paradigm for the particle nature of dark matter is that of Weakly Interacting Massive Particles (WIMPs), due to the fact that it allows for a simple thermal mechanism to produce DM with the cosmologically-observed abundance~\cite{Lee:1977ua}. Such models have also attracted attention due to the large number of possible signals they predict, none of which have been definitively observed so far. Although this has led some to make claims of the demise of WIMPs \cite{Arcadi:2017kky}, others have argued that such predictions are premature \cite{Leane:2018kjk}.

A relatively agnostic approach to WIMP model building is to pursue a bottom-up, Effective Field Theory (EFT) approach, in which one enumerates all of the allowed higher dimensional operators which lead to interactions between DM and Standard Model (SM) particles. Any result described by an EFT can in general be explained by many high-energy theories.  In this way, the EFT description is a model-independent one, as it does not depend on the Ultraviolet (UV) completion that describes an effective operator. This is, however, a double-edged sword: because an effective operator does not encode any information about the UV completion, it has no constraining power in distinguishing between the range of UV theories that can map to it -- nor can all UV-complete theories be mapped to an EFT description for the energies we are interested in here.

In spite of these limitations, the bottom-up approach is well-advised given the lack of direct evidence pointing to the properties of DM. The EFT approach in particular is highly suitable for low-velocity environments such as direct detection~\cite{Fan:2010gt,Agrawal:2010fh,Fitzpatrick:2010br,Crivellin:2014gpa,DEramo:2016gos,Hoferichter:2016nvd,Kahlhoefer:2016eds} and indirect detection~\cite{Goodman:2010qn,Beltran:2008xg,Cheung:2011nt,Harnik:2008uu,DeSimone:2013gj,Karwin:2016tsw,Carpenter:2016thc}.  At higher energy scales, the EFT approach starts breaking down, such that simplified models have become the theories of choice for the interpretation of LHC searches~\cite{Abdallah:2015ter,Kahlhoefer:2017dnp} (see also Refs.~\cite{Alanne:2017oqj,Alanne:2020xcb} for a hybrid approach called ``Extended Dark Matter EFT''). %(see Sec.~\ref{sec:DMEFT_simplified_models})
Nevertheless, there is an extensive literature on EFTs at colliders~\cite{Bai:2010hh,Dreiner:2013vla,Zhou:2013fla,Fox:2012ee,Rajaraman:2011wf,Goodman:2010ku,Fox:2011pm,Beltran:2010ww,Buchmueller:2013dya,Belyaev:2016pxe,Pobbe:2017wrj} including studies by ATLAS~\cite{ATLAS:2012ky} and CMS~\cite{Chatrchyan:2012me}, which may help to shed light on the nature of DM when interpreted with care.

A common approach to the analysis of EFTs for DM in the literature has been to consider a single operator at a time~\cite{Buckley:2011kk,Cheung:2012gi,MarchRussell:2012hi,Zheng:2010js,Belyaev:2018pqr,Bertuzzo:2017lwt,DelNobile:2013sia} and compare experimental bounds on the new physics scale $\La$ with the values implied by the observed DM relic density. This method, however, severely limits the scope of the analysis and potentially leads to overly-aggressive exclusions, not only because it neglects (potentially destructive) interferences between different operators~\cite{Kumar:2013iva}, but also because the relic density constraint can be considerably relaxed when several operators contribute to the DM annihilation cross-section. The first global study of EFTs for scalar, fermionic and vector DM taking interference effects into account was performed in Ref.~\cite{Balazs:2014rsa}, but no collider constraints were included in the analysis and no couplings to gluons were considered. More recently, Ref.~\cite{Liem:2016xpm} applied Bayesian methods to perform a global analysis of scalar DM, for which only a small number of effective operators need to be considered and collider constraints can be neglected.  Examples of global studies considering subspaces of a general DM EFT include Refs.~\cite{Matsumoto:2014rxa,Blennow:2015yca,Matsumoto:2016hbs}.

In the present work, we exploit the computational power of the \gambit framework~\cite{gambit} to perform the first global analysis of a very general set of effective operators up to dimension 7 that describe the interactions between a Dirac fermion DM particle (or a DM sub-component) and quarks or gluons. Such a set-up arises for example in many extensions of the SM gauge group, such as gauged baryon number~\cite{Duerr:2014wra} or other anomaly-free gauge extensions that require additional stable fermions~\cite{Dudas:2013sia,Bauer:2018egk}. Our novel approach of considering many operators simultaneously enables us to study parameter regions where several types of DM interactions need to be combined in order to satisfy all constraints. Our analysis substantially improves upon the previous state-of-the-art in both the statistical rigour with which the DM EFT parameter space is interrogated, and in the new combinations of constraints that are simultaneously applied. We also increase the level of detail with which individual constraints are modelled, summarised as follows.

First, we include a much improved calculation of direct detection constraints using the \gambit module \darkbit~\cite{DarkBit}. We consider the renormalization group (RG) evolution of all effective operators from the electroweak to the hadronic scale and then match the relativistic operators onto the non-relativistic effective theory~\cite{Fitzpatrick:2012ix} relevant for DM-nucleon scattering. We then calculate event rates in direct detection experiments to leading order in the chiral expansion, including the contributions from operators that are naively suppressed in the non-relativistic limit, and determine the resulting constraints using detailed likelihood functions for a large number of recent experiments. In the process, we include a number of nuisance parameters to account for uncertainties in nuclear form factors and the astrophysical distribution of DM.

Second, we consider the most recent constraints on DM annihilations using gamma rays and the Cosmic Microwave Background (CMB). To include the latter, we employ the recently released \gambit module \cosmobit~\cite{CosmoBit}, which uses detailed spectra to calculate effective functions for the efficiency of the injected energy deposition and obtain constraints on the DM annihilation cross-section while varying cosmological parameters. For the calculation of annihilation cross-sections we make use of the new \gambit Universal Model Machine (\gum)~\cite{Gonzalo:2021cnq, GUM} to automatically generate the relevant code based on the EFT Lagrangian.

Third, we combine the above detailed astrophysical and cosmological constraints with a state-of-the-art implementation of LHC constraints on WIMP dark matter. A central concern for any study of EFTs is the range of validity of the EFT approach~\cite{Shoemaker:2011vi,Busoni:2013lha,Busoni:2014sya,Busoni:2014haa,Endo:2014mja,Bell:2016obu,Racco:2015dxa,Bruggisser:2016nzw}. This is particularly true when considering constraints from the LHC, which may probe energies above the assumed scale of new physics. A naive application of the EFT in such a case may lead to unphysical predictions, such as unitarity violation. Whenever this is the case it becomes essential to adopt some form of truncation to ensure that only reliable predictions are used to calculate experimental constraints.

In the present work we address these challenges in two key ways. First, we separate the scale of new physics $\La$ from the individual Wilson coefficients $\C{}{}$ (rather than scanning over a combination such as $\C{}{} / \La^2$), such that the former can be directly interpreted as the scale where the EFT breaks down and the latter can be constrained by perturbativity. Second, we check the impact of a phenomenological nuisance parameter that describes the possible modification of LHC spectra at energies beyond the range of EFT validity. The nuisance parameter smoothly interpolates between an abrupt truncation and no truncation at all.

Our analysis reveals viable parameter regions for general WIMP models across a wide range of new physics scales, including very small values of $\La$ ($\La < 200 \, \mathrm{GeV}$), where there are no relevant LHC constraints and very large values of $\La$ ($\La > 1.5 \,\mathrm{TeV}$), where LHC constraints are largely robust. Of particular interest are the intermediate values of $\La$ ($\La \sim 700\text{--} 900 \, \mathrm{GeV}$), for which our DM EFT partly accommodates several small LHC data excesses that could be interesting to analyse in more detail in the context of specific UV completions or simplified models. However, our analysis also reveals that there cannot be a large hierarchy between $\La$ and the DM mass $m_\chi$. In particular, even with the most general set of operators we consider, it is impossible to simultaneously have a small DM mass ($m_\chi \lesssim 100 \, \mathrm{GeV}$) and a large new physics scale ($\La > 200 \, \mathrm{GeV}$). In other words, for light DM to be consistent with all constraints, it is necessary for the new physics scale to be so low that the EFT approach breaks down for the calculation of LHC constraints. 
For heavier DM, on the other hand, thermal production of DM in the early universe would exceed the observed abundance
whenever $\La$ is more than one order of magnitude larger than $m_\chi$ (up to the unitarity bound
at a few hundred TeV~\cite{Griest:1989wd}, where the maximum possible value of $\La$ approaches $m_\chi$).

This work is organised as follows. We introduce the DM EFT description in Sec.~\ref{sec:model}. In Sec.~\ref{sec:constraints}, we discuss the constraints used in this study, and our methods for computing likelihoods and observables. We present our results in Sec.~\ref{sec:results}. Finally, we present our conclusions in Sec.~\ref{sec:summary}. The samples from our scans and the corresponding \gambit input files, and plotting scripts can be downloaded from \textsf{Zenodo}~\cite{Zenodo_DMEFT}.

%%%%%%%%%%%%%%%%%%%%%%%%%%%%%%%%%%%%%%%%%%%%%%%%%%%%%%%%%%%%%%%%%%%%%%%%%%%%%%%%%%%%%%%%%%%%%%%%%%%%%%%%%%%%%%%%%%%%%%%%
\section{Dark Matter Effective Field Theory}\label{sec:model}

In this study, we consider possible interactions of SM fields with a Dirac fermion DM field, $\chi$, that is a singlet
under the SM gauge group. For phenomenological reasons discussed in detail in Sec.~\ref{sec:constraints},
we focus on interactions between $\chi$
and the quarks or gluons of the SM.
We assume that the mediators that generate these interactions are
heavier than the scales probed by the experiments under consideration. Following the notation
of Refs.~\cite{Bishara:2017nnn,Brod:2017bsw}, the interaction Lagrangian for the theory can be written as
\begin{equation}
  \mathcal{L}_{\rm{int}} = \sum_{a,d} \dfrac{\C{a}{d}}{\La^{d-4}} \Q{a}{d}\,,
\end{equation}
where $\Q{a}{d}$ is the DM-SM operator, $d\geq 5$ is the mass dimension of the operator, $\C{a}{d}$ is the
dimensionless Wilson coefficient associated to $\Q{a}{d}$, and $\La$ is the scale of new physics (which
can be identified with the mediator mass). The full Lagrangian for the theory is then
\begin{equation}
  \mathcal{L} = \mathcal{L}_{\rm{SM}} + \mathcal{L}_{\rm{int}} + \overline{\chi}\left(i\slashed{\partial}-m_\chi\right)\chi\,,
\end{equation}
such that the free parameters of the theory are the DM mass $m_\chi$, the scale of new physics $\La$, and
the set of dimensionless Wilson coefficients $\{ \C{a}{d} \}$.

For sufficiently large $\La$, the phenomenology at small
energies is dominated by the operators of lowest dimension, and we therefore limit ourselves to $d \leq 7$. However, even this
leaves a relatively large set of operators. The DM EFT that is valid below the electroweak (EW) scale (with the Higgs, $W$, $Z$ and the top quark integrated out) contains 2 dimension five, 4 dimension six, and 22 dimension seven operators (not counting flavour multiplicities), while the DM EFT above the EW scale for a singlet Dirac fermion DM has 4 dimension five, 12 dimension six, and 41 dimension seven operators (again, not counting flavour multiplicities) \cite{Brod:2017bsw}. The large set of possible operators poses a challenge for a global statistical analysis where
bounds on $\La$ and $\{ \C{a}{d} \}$ are derived from experimental observations (see Sec.~\ref{sec:constraints} for details). An added complexity is that we consider both processes where the typical energy transfer is above the EW scale (such as collider searches and indirect detection) as well as processes in which the energy release is small (direct detection). The consistent implementation of these bounds requires the combination of both DM EFTs, together with the appropriate matching conditions between the two.

To make the problem tractable we focus in our numerical analysis on a subset of DM EFT operators - the dimension six operators involving DM, $\chi$, and SM quark fields, $q$,
\begin{align} \label{dim6efts}
  \Q{1,q}{6} &= (\overline\chi \gamma_\mu \chi)(\overline{q} \gamma^\mu q)\,, \\
  \Q{2,q}{6} &= (\overline\chi \gamma_\mu \gamma_5 \chi)(\overline{q} \gamma^\mu q)\,, \\
  \Q{3,q}{6} &= (\overline\chi \gamma_\mu \chi)(\overline{q} \gamma^\mu \gamma_5 q)\,, \\
  \Q{4,q}{6} &= (\overline\chi \gamma_\mu \gamma_5 \chi)(\overline{q} \gamma^\mu \gamma_5 q)\,.
  \label{dim6efts:end}
\end{align}
The difference between the DM EFT below the EW scale and the DM EFT above the EW scale is in this case very simple: above the EW scale the quark flavours run over all SM quarks, including the top quark, while below the EW scale the top quark is absent.

While the above set of operators does not span the full dimension six bases of the two DM EFTs, it does collect the most relevant operators. The full dimension six operator basis contains operators where quarks are replaced by the SM leptons. These are irrelevant for the collider and direct detection constraints we consider, and are thus omitted for simplicity. The basis of dimension six operators for the DM EFT above the EW scale contains, in addition, operators that are products of DM and Higgs currents. These are expected to be tightly constrained by direct detection to have very small coefficients such that they are irrelevant in other observables, and are thus also dropped for simplicity.

To explore to what extent the numerical analyses would change, if the set of considered DM EFT operators were enlarged, we also perform global fits including, in addition to the dimension six operators~\eqref{dim6efts}-\eqref{dim6efts:end}, a set of dimension seven operators that comprise interactions with the gluon field either through the QCD field strength tensor $G^a_{\mu\nu}$ or its dual $\widetilde{G}_{\mu\nu}=\frac12\epsilon_{\mu\nu\rho\sigma}G^{\rho\sigma}$, as well as operators constructed from scalar, pseudoscalar and tensor bilinears:
\begin{align}
  \Q{1}{7} &= \frac{\alpha_s}{12\pi}(\overline\chi \chi)G^{a\mu\nu}G^a_{\mu\nu}\,, \\
  \Q{2}{7} &= \frac{\alpha_s}{12\pi}(\overline\chi i\gamma_5 \chi)G^{a\mu\nu}G^a_{\mu\nu}\,, \\
  \Q{3}{7} &= \frac{\alpha_s}{8\pi}(\overline\chi \chi)G^{a\mu\nu}\widetilde{G}^a_{\mu\nu}\,, \\
  \Q{4}{7} &= \frac{\alpha_s}{8\pi}(\overline\chi i\gamma_5 \chi)G^{a\mu\nu}\widetilde{G}^a_{\mu\nu}\,, \\
  \Q{5,q}{7} &= m_q(\overline\chi \chi)( \overline{q} q )\,, \\
  \Q{6,q}{7} &= m_q(\overline\chi i\gamma_5 \chi)( \overline{q} q )\,, \\
  \Q{7,q}{7} &= m_q(\overline\chi \chi)( \overline{q} i\gamma_5 q )\,, \\
  \Q{8,q}{7} &= m_q(\overline\chi i\gamma_5 \chi)( \overline{q} i\gamma_5 q )\,, \\
  \Q{9,q}{7} &= m_q(\overline\chi \sigma^{\mu\nu} \chi)( \overline{q} \sigma_{\mu\nu} q )\,, \\
  \Q{10,q}{7} &= m_q(\overline\chi i\sigma^{\mu\nu}\gamma_5 \chi)( \overline{q} \sigma_{\mu\nu} q )\,.
  \label{dim7efts}
\end{align}
The definition of the operators describing interactions with the gluons, $\Q{1\text{--}4}{7}$, includes a loop factor since in most new physics models these operators are generated at one loop.
Similarly, the couplings to scalar and tensor quark bilinears, $\Q{5\text{--}10,q}{7}$, include a conventional factor of the quark mass $m_q$, since they have the same flavour structure as the quark mass terms (coupling left-handed and right-handed quark fields). The $m_q$ suppression of these operators is thus naturally encountered in new physics models that satisfy low energy flavour constraints, such as minimal flavour violation and its extensions. Note that, unless explicitly stated otherwise, $m_q$ always refers to the running mass in the modified minimal subtraction ($\overline{\rm{MS}}$) scheme.

The complete dimension-seven basis below the EW scale contains eight additional operators with derivatives acting on the DM fields \cite{Brod:2017bsw}. To simplify the discussion we do not include these operators in our analysis, partially because they do not lead to new chiral structures in the SM currents.  Moreover, the direct detection constraints on these additional operators are expressible in terms of the operators that we do include in the global fits due to the non-relativistic nature of the scattering process.

Note that the operators $\Q{5\text{--}10,q}{7}$ are not invariant under EW gauge transformations, and are thus replaced in the DM EFT above the EW scale by operators of the form $(\bar{\chi}\chi)(\bar{q}_Lq_R)H$, where $H$ is the Higgs doublet. In all the processes we consider, $H$ can be replaced by its vacuum expectation value - either because the emission of the Higgs boson is phase-space suppressed or suppressed by small Yukawa couplings, or both. This means that, up to renormalization group effects (to be discussed in Sec.~\ref{sec:RG:mix}), the operators $\Q{5\text{--}10,q}{7}$ can also be used in our fitting procedure above the EW scale.

In principle, analogous operators to $\Q{1\text{--}10,q}{7}$ exist for leptons instead of quarks~\cite{Kopp:2009et,Fox:2011fx} and weak gauge bosons instead of gluons~\cite{Weiner:2012gm,Frandsen:2012db,Paz:2020pbc}.\footnote{Furthermore, there are two additional dimension-6 operators describing DM-photon interactions: the anapole moment and the charge radius~\cite{Kavanagh:2018xeh}. For a recent discussion of LHC constraints on these operators we refer to Ref.~\cite{Arina:2020mxo}.}~In general, these play a much smaller role in the phenomenology and will not be considered here. Similarly, throughout this work the Wilson coefficients of any dimension five operators are set to zero at the UV scale.

The Wilson coefficients of the operators defined above depend implicitly on the energy scale of the
process under consideration. In our fits, all Wilson coefficients are specified at the new physics scale $\La$.
If this scale is larger than the top mass, $\La > m_t$, all six quarks are active degrees of freedom and the Wilson coefficients need to be specified for $q = u, d, s, c, b, t$. For $\La < m_t$, the top quarks are integrated out, and only the Wilson coefficients for $q = u, d, s, c, b$ need to be specified.
This is done automatically in our fitting procedures, such that effectively both EFTs are used in the fit, according to the numerical value of the scale $\La$.

Although, a priori, the Wilson coefficients for each quark flavour are independent, we will restrict
ourselves to the assumption of minimal flavour violation (which implies
$\C{i,d}{d} = \C{i,s}{d} = \C{i,b}{d}$ and $\C{i,u}{d} = \C{i,c}{d} = \C{i,t}{d}$), and the assumption of isospin invariance (which implies $\C{i,d}{d} = \C{i,u}{d}$).\footnote{%
These constraints also ensure that the dimension-six operators do not explicitly break
electroweak symmetry, which requires $\C{1,u}{6} - \C{3,u}{6} = \C{1,d}{6} - \C{3,d}{6}$~\cite{Haisch:2016usn}.
}~Hence, each operator comes with only one free parameter in addition to the global parameters $\La$ and $m_\chi$. Under these assumptions, the two EFTs above and below the EW scale have the same number of free parameters.

\subsection{Running and mixing}\label{sec:RG:mix}

For many applications, the RG running of the Wilson coefficients (i.e.\ their dependence on the energy scale $\mu$)
can be neglected. In fact, the operators $\Q{1,q}{6}$, $\Q{2,q}{6}$, $\Q{5,q}{7}$ and $\Q{6,q}{7}$ have
vanishing anomalous dimension, while $\Q{3,q}{6}$, $\Q{4,q}{6}$, $\Q{7,q}{7}$, $\Q{8,q}{7}$ as well as
$\Q{1\text{--}4}{7}$ exhibit no running at one-loop order in QCD~\cite{Hill:2014yxa}. Nevertheless, there
are two cases when the effects of running can be important:
\begin{enumerate}
 \item \textbf{Mixing:} Different operators can mix with each other under RG evolution,
 such that operators assumed negligible at one scale may give a relevant contribution at a different scale. This
 is particularly important in the context of direct detection, because for certain operators the DM-nucleon
 scattering cross-section is strongly suppressed in the non-relativistic limit. In such a case, the dominant
 contribution to direct detection may arise from operators induced only at the loop level~\cite{Bishara:2018vix,Brod:2018ust}. In our case, the dominant effects arise from the top quark Yukawa and are discussed below.
 \item \textbf{Threshold corrections:} Whenever the scale $\mu$ drops below the mass of one of the quarks,
 the number of active degrees of freedom is reduced and a finite correction to various operators arises. In our context, the only effect is the matching of the operators $\Q{5\text{--}8,q}{7}$ onto the operators
 $\Q{1\text{--}4}{7}$ at the heavy quark thresholds, which is given by
\begin{align}
  \C{1}{7} &= \C{1}{7} - \C{5,q}{7}\, , \nonumber \\[1.5mm]
    \C{2}{7} &= \C{2}{7} - \C{6,q}{7}\, , \nonumber \\[1.5mm]
  \C{3}{7} &= \C{3}{7} + \C{7,q}{7}\,,  \nonumber \\[1.5mm]
  \C{4}{7} &= \C{4}{7} + \C{8,q}{7}\, .
 \label{eq:threshold}
\end{align}
Mixing of the tensor operators $\Q{9,q}{7}, \Q{10,q}{7}$ above the EW scale and subsequent matching gives rise to the dimension-five dipole operators
\begin{align}
  \Q{1}{5} &= \frac{e}{8\pi^2} (\overline\chi \sigma_{\mu\nu} \chi) F^{\mu\nu} \,, \\
  \Q{2}{5} &= \frac{e}{8\pi^2} (\overline\chi i \sigma_{\mu\nu} \gamma_5 \chi) F^{\mu\nu} \, ,
\end{align}
where $F_{\mu\nu}$ is the electromagnetic field strength tensor and $e$ is the electromagnetic charge.
These operators give an important contribution to direct detection experiments and are thus kept.\footnote{Note that as per our assumptions the Wilson coefficients $\C{1}{5}, \C{2}{5}$ are taken to be zero at scale $\La$  and are only generated by the RG effects.}
\end{enumerate}

In the present work we include these effects as follows. To calculate the Wilson coefficients at the hadronic scale
$\mu = 2\,\mathrm{GeV}$ (relevant for direct detection) we make use of the public code \ddm \textsf{v2.2.0} \cite{Bishara:2017nnn,Brod:2017bsw},
which calculates the RG evolution of the operators defined above, including threshold corrections
and mixing effects. The code furthermore performs a matching of the resulting operators at
$\mu = 2\,\mathrm{GeV}$ onto the basis of non-relativistic effective operators relevant for DM direct detection
(see Sec.~\ref{sec:dd}).

\ddm currently requires as input the Wilson coefficients in the five-flavour scheme given at the scale $m_Z = 91.1876\,\mathrm{GeV}$. For $\La < m_t$ (five-flavour EFT), we can therefore directly pass the Wilson coefficients defined above to \ddm.
For $\La > m_t$ (six-flavour EFT), there are three additional effects that are considered. First, as pointed out in Ref.~\cite{Haisch:2013uaa}, the operators $\Q{9,10,t}{7}$
give a contribution to the dipole operators $\Q{1,2}{5}$ at the one-loop level, which is given by\footnote{For historical reasons, in the numerical code $\log(m_t^2/\La^2)$ instead of $\log(m_Z^2/\La^2)$ was used. The effect on the numerical results is negligible.}
\begin{equation}\label{eq:MLL}
\C{1,2}{5}(m_Z) = \frac{4 \, m_t^2}{\La^2} \log \left( \frac{m_Z^2}{\La^2} \right) \, \C{9,10;t}{7}(\La)\,.
\end{equation}
Second, as pointed out first in Ref.~\cite{Crivellin:2014qxa}, the operator with an axial-vector top-quark current $\Q{3,t}{6}$ mixes into the operators $\Q{1,q}{6}$ with light quark vector currents. The relevant effects are given by~\cite{Bishara:2018vix}
\begin{equation}\label{eq:yt}
\begin{split}
\C{1,u/d}{6} & (m_Z) = \C{1,u/d}{6}(\La) \\[0.5em]
  & + \frac{2s_w^2 \mp (3 - 6s_w^2)}{8\pi^2} \frac{m_t^2}{v^2} \log \left( \frac{m_Z^2}{\La^2} \right) \, \C{3,t}{6}(\La)\,,
\end{split}
\end{equation}
after integrating out the $Z$ boson at the weak scale. Here, $s_w\equiv\sin\theta_w$ with $\theta_w$ the weak mixing angle, and $v=246\,$GeV is the Higgs field vacuum expectation value.
The flavour universal UV contributions $\C{1,u/d}{6}(\La)$ largely compensate the mixing effect in the fit; the remnant effect, due to the isospin-breaking $Z$ couplings, is small.

Third, in order to match the EFT with six active quark flavours onto the five-flavour scheme, we need to
integrate out the top quark and apply the top quark threshold corrections given in Eq.~(\ref{eq:threshold}).
We neglect any other effects of RG evolution between the scales $\La$ and $m_Z$, i.e.~all Wilson
coefficients other than $\C{1,2}{5}$ and $\C{1\text{--}4}{7}$ are directly passed to \ddm.%
\footnote{Small remnant effects of the bottom and charm Yukawa coupling are taken into account below the EW scale via double weak insertions~\cite{Brod:2018ust} that are included in  the \ddm code.}

For the purpose of calculating the LHC constraints, we neglect the effects of running and do not consider
loop-induced mixing between different operators, which is a good approximation for the operators
$\Q{1\text{--}4,q}{6}$ and $\Q{1\text{--}4}{7}$. For the operators $\Q{5\text{--}10,q}{7}$ mixing effects
are known to be important in principle~\cite{Haisch:2012kf}, but these operators are currently
unconstrained by the LHC in the parameter region where the EFT is valid (see Sec.~\ref{sec:validity}). Likewise we also
calculate DM annihilation cross-sections at tree level. In particular, in these calculations we neglect the running of the strong coupling $(\alpha_s)$ and use the pole quark masses $(m_q^\text{pole})$ instead of the running quark masses. Moreover, we neglect a small loop-level
contribution from the operators $\Q{5\text{--}8,q}{7}$ to the operators $\Q{1\text{--}4}{7}$.

\subsection{EFT validity}
\label{sec:validity}

A central concern when employing an EFT to capture the effects of new physics is that the scale of new
physics must be sufficiently large compared to the energy scales of interest for the EFT description to be valid.
Unfortunately, the point at which the EFT breaks down is difficult to determine from the low-energy theory alone.
Considerations of unitarity violation make it possible to determine the scale where the EFT becomes
unphysical, but in many cases the EFT description already fails at lower energies, in particular if the UV
completion is weakly coupled.

\begin{figure}[t]
	\centering
	\includegraphics[width=\columnwidth]{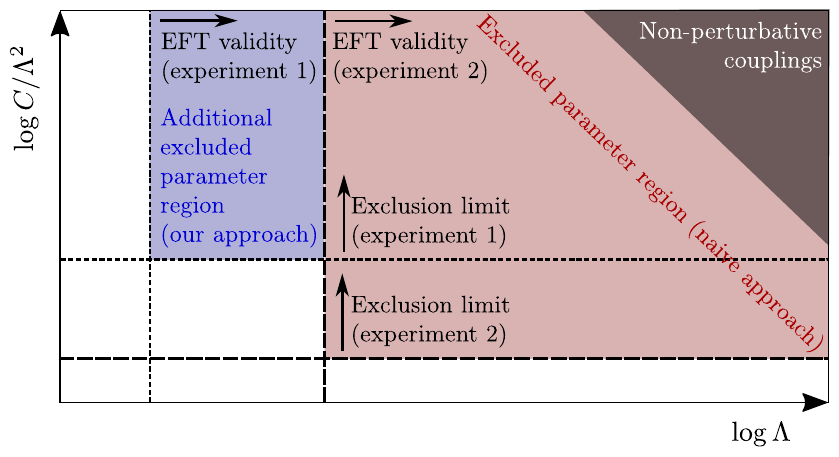}
	\caption{Illustration of our approach for studying DM EFTs compared to a more naive approach, in which one only uses the experiment that yields the strongest bound on $C / \La^2$. The resulting exclusion is indicated by the red shaded region. By independently varying $\La$, we can include additional information from experiments that give weaker bounds on $C / \La^2$ but for which the EFT has a larger range of validity. The additional exclusion obtained in this way is indicated by the blue shaded region. The region of parameter space that corresponds to the non-perturbative values of Wilson coefficient $C$ is excluded in either approach (shaded brown). }
	\label{fig:EFT_illustration}
\end{figure}

To address this issue in the present study, we simultaneously vary the overall scale $\La$, which corresponds
to the energy where new degrees of freedom become relevant and the EFT description breaks down, and
the Wilson coefficients $\C{a}{d}$ for each operator. Doing so introduces a degeneracy, because cross
sections are invariant under the rescaling $\La \to \alpha \La$ and $\C{a}{d} \to \alpha^{d-4} \C{a}{d}$.
However, the advantage of this approach is that the parameter $\La$ can be used to determine which
constraints can be trusted in the EFT limit. This is illustrated in Fig.~\ref{fig:EFT_illustration}, which compares our approach of varying $\La$ and $\C{a}{d}$ separately to the naive approach where only $\C{a}{d} / \La^{d-4}$ is constrained.

We emphasize that this approach assumes the same new-physics scale for all effective operators, even though they may be generated through different mechanisms, and hence at different scales, in the UV. In practice, one should think of $\La$ as the minimum of all of these scales, i.e.\ the energy at which new degrees of freedom first become relevant. These new degrees of freedom may not contribute to all processes, such that some effective operators may provide an accurate description even at energies above $\La$. Whether or not this is the case cannot be determined from the low-energy viewpoint, such that we conservatively limit the EFT validity to energies below $\La$.

For the purpose of direct detection constraints, the only requirement on $\La$ is that it is larger than the
hadronic scale, so that the effective operators can be written in terms of free quarks and gluons. This is
the case for $\La \gtrsim 2\,\mathrm{GeV}$, which will always be satisfied in the present study. However,
in order to evaluate direct detection constraints, it is necessary to determine the relic abundance of DM
particles, which depends on the cross-sections for the processes $\chi \chi \to q q$ or $\chi \chi \to g g$,
just as in the case of indirect detection constraints
(see Sec.~\ref{sec:id}). For this calculation to be meaningful in the EFT framework, we require $\La > 2 m_\chi$.
Parameter points with smaller values of $\La$ will thus be invalidated.
A dedicated study of direct detection constraints for $\La < 2 m_\chi$ will be left for future work.

In the context of LHC searches for DM, EFT validity requires that the invariant mass of the DM pair
produced in a collision satisfies $m_{\chi\chi} < \La$~\cite{Berlin:2014cfa}. To obtain robust constraints, only events with smaller
energy transfer should be included in the calculation of likelihoods. The problem with this prescription is
that $m_{\chi\chi}$ does not directly correspond to any observable quantity (such as the missing energy
$\slashed{E}_T$ of the event) and hence the impact of varying $\La$ on predicted LHC spectra is difficult
to assess. One possible way to address this issue would be to generate new LHC events for each parameter
point and include only those events with small enough $m_{\chi\chi}$ in the likelihood calculation, but this
is not computationally feasible in the context of a global scan.

In the present work, we adopt the following simpler approach: Rather than comparing $\La$
to the invariant mass of the DM pair, we compare it to the typical overall energy scale of the event, which
can be estimated by the amount of missing energy produced. In other words, we do not modify the missing energy spectrum for $\slashed{E}_T < \La$ and only apply the EFT validity requirement for larger values of $\slashed{E}_T$. This approach is less conservative than the one advocated, for instance in
Refs.~\cite{Racco:2015dxa,Bruggisser:2016nzw}, where the energy scale of the event is taken to be
the partonic centre-of-mass energy $\sqrt{\hat{s}}$, but it has the crucial advantage that it can be applied
\emph{after} event generation, since the differential cross-section with respect to missing energy
$\mathrm{d}\sigma / \mathrm{d}\slashed{E}_T$ is exactly the quantity that is directly compared to data.\footnote{We emphasize that $m_{\chi\chi}$ and $\slashed{E}_T$ are not strongly correlated in the sense that there are events with both $\slashed{E}_T \ll m_{\chi\chi}$ (if the DM pair is emitted approximately in the longitudinal direction) and $\slashed{E}_T \gg m_{\chi\chi}$ (if the two DM particles are light and approximately collinear). Since our approach does not modify the spectrum for $\slashed{E}_T < \La$, we risk overestimating the differential cross-section in this regime. However, the sensitivity of the LHC to DM EFTs typically stems from events with large $\slashed{E}_T$, where our prescription is more appropriate.}

In the following, we will consider two different prescriptions for how to impose the EFT validity. The first one is to introduce a hard cut-off, i.e.\ to set $\mathrm{d}\sigma / \mathrm{d}\slashed{E}_T = 0$ for $\slashed{E}_T > \La$. The second, more realistic, prescription is to introduce a smooth cut-off that leads to a non-zero but steeply falling missing energy spectrum above $\La$. For this we make the replacement
\begin{equation}
\frac{\mathrm{d}\sigma}{\mathrm{d}\slashed{E}_T} \to \frac{\mathrm{d}\sigma}{\mathrm{d}\slashed{E}_T} \left(\frac{\slashed{E}_T}{\La}\right)^{-a},
\end{equation}
for $\slashed{E}_T > \La$. Here $a$ is a free parameter that depends on the specific UV completion. The limits $ a \to 0 $ and $a \to \infty$ correspond to no truncation and an abrupt truncation above the cut-off, respectively. For the case that the EFT results from the exchange of an $s$-channel mediator with mass close to $\La$, one finds $a \approx 2$~\cite{Buchmueller:2013dya}. Rather than taking inspiration from a specific UV completion, we will instead keep $a$ as a free parameter in the interval $[0,4]$ and find the value that gives the best fit to data at each parameter point.  This approach typically leads to conservative LHC bounds in the sense that much stronger exclusions may be obtained in specific UV completions, if the heavy particles that generate the effective DM interactions can be directly produced at the LHC. However, this truncation procedure can lead to unrealistic spectral shapes with sharp features that may be tuned to fit fluctuations in the data. As will be discussed in more detail in Sec.~\ref{sec:results}, any explanation of data excesses through this approach must be interpreted with care.
% 
%\bigskip

Without upper bounds on the Wilson coefficients, any requirement on EFT validity could be satisfied by
making both $\La$ and the Wilson coefficients arbitrarily large. We therefore require $\C{a}{d} < 4\pi$,
which is necessary for a perturbative UV completion and ensures that there is no unitarity violation in
the validity range of the EFT~\cite{Bell:2016obu}.

One drawback of this prescription is that the EFT validity requirement depends on the normalisation of
the effective operators. For example, we have written $\Q{1\text{--}2}{7}$ with a prefactor $\alpha_s / (12\pi)$
and $\Q{3\text{--}4}{7}$ with a prefactor $\alpha_s / (8\pi)$ to reflect the fact that in many UV completions,
these operators would be generated at the one-loop level. If these operators are instead generated at
tree level (e.g.\ from a strongly interacting theory), it would be more appropriate to write the prefactor
as $4\pi\alpha_s$. With the latter convention any constraint on the new physics scale $\La$ becomes
stronger by a factor $(48\pi^2)^{1/3} \approx 5.3$ for $\Q{1\text{--}2}{7}$ and by a factor
$(32\pi^2)^{1/3} \approx 4.6$ for $\Q{3\text{--}4}{7}$, meaning that much larger values of $\La$ are
experimentally testable and the range of EFT validity is substantially increased.
We have confirmed explicitly that the results presented in Sec.~\ref{sec:results} do not depend on the specific definition of the Wilson coefficients for $\Q{1\text{--}4}{7}$.\footnote{We note that the explicit factor of $m_q$ in the definition of $\Q{5\text{--}10,q}{7}$ not only affects the EFT validity but also directly the resulting phenomenology. Hence our results cannot be easily translated to operators with non-trivial flavour structure.}

\subsection{Parameter ranges}

In this study we focus on the following parameter regions. In order to be able to neglect QCD resonances in
the process $\chi \bar{\chi} \to q \bar{q}$, we restrict ourselves to $m_\chi > 5\,\mathrm{GeV}$. In order to
have a sufficiently large separation of scales between the new physics scale $\La$ and the hadronic scale,
we also require $\La > 20\,\mathrm{GeV}$. As discussed in Sec.~\ref{sec:validity}, we furthermore impose the bound
$|\C{a}{d}| < 4\pi$ on all Wilson coefficients and the bound $\La > 2 m_\chi$. The upper bounds on $m_\chi$ and $\La$ depend on the details of the scans that we perform and will be discussed in Sec.~\ref{sec:results}.

%%%%%%%%%%%%%%%%%%%%%%%%%%%%%%%%%%%%%%%%%%%%%%%%%%%%%%%%%%%%%%%%%%%%%%%%%%%%%%%%%%%%%%%%%%%%%%%%%%%%%%%%%%%%%%%%%%%%%%%%
\section{Constraints}\label{sec:constraints}

In this section we describe the constraints relevant for our model. A summary of all likelihoods included in our scans is provided in Table~\ref{tab:experiments}. For each likelihood that directly constrains the interactions of the DM particle we also quote the background-only log-likelihood $\ln \mathcal {L}^\text{bg}$ obtained when setting all Wilson coefficients to zero. For the remaining likelihoods we instead quote the maximum achievable value of the log-likelihood $\ln \mathcal {L}^\text{max}$. The sum of all these contributions, $\ln \mathcal{L}^\text{ideal} = -105.3$ will be used to calculate log-likelihood differences below.

%%%%%%%%%%%%%%%%%%%%%%%%%%%%%%%%%%%%%%%%%%%%%%%%%%%%%%%%%%%%%%%
\begin{table}[t]
  \centering
  \begin{tabular}{lcc}
    \toprule
    \textbf{Experiment} & \textbf{$\ln \mathcal {L}^\text{bg}$} & \textbf{$\ln \mathcal {L}^\text{max}$}    \\ \midrule
	CDMSlite~\cite{Agnese:2015nto} & $-16.68$ \\
	CRESST-II~\cite{Angloher:2015ewa} & $-27.59$ \\
	CRESST-III~\cite{Abdelhameed:2019hmk} & $-27.22$ \\
	DarkSide 50~\cite{Agnes:2018fwg} & $-0.09$ \\
	LUX 2016~\cite{LUXrun2} & $-1.47$ \\
	PICO-60~\cite{Amole:2017dex,Amole:2019fdf} & $-1.496$ \\
	PandaX~\cite{Tan:2016zwf,Cui:2017nnn} & $-3.436$ \\
	XENON1T~\cite{Aprile:2018dbl} & $-3.651$
	\\[2mm]
	ATLAS monojet~\cite{Aad:2021egl} & $0$ \\
	CMS monojet~\cite{Sirunyan:2017jix} & $0$
	\\[2mm]
	\emph{Fermi}-LAT~\cite{LATdwarfP8} & $-33.245$ \\
	IceCube 79-string~\cite{IC79_SUSY} & $0$ \\
	\emph{Planck} 2018: $p_\text{ann}$~\cite{Aghanim:2018eyx} & $-1.507$ \\[2mm]
	\emph{Planck} 2018: $\Omega h^2$~\cite{Aghanim:2018eyx} & & $5.989$
	\\[2mm]
	Nuisances (see Table~\ref{tab:nuis_params}) & & $5.141$
	\\[1mm] \bottomrule
  \end{tabular}
  \caption{Likelihoods included in our scans and their respective values for the background-only hypothesis. For each likelihood that directly constrains the interactions of the DM particle we also quote the background-only log-likelihood $\ln \mathcal {L}^\text{bg}$ obtained when setting all Wilson coefficients to zero. For the remaining likelihoods we instead quote the maximum achievable value of the log-likelihood $\ln \mathcal {L}^\text{max}$.}
  \label{tab:experiments}
\end{table}
%%%%%%%%%%%%%%%%%%%%%%%%%%%%%%%%%%%%%%%%%%%%%%%%%%%%%%%%%%%%%%%

\subsection{Direct detection}\label{sec:dd}

Direct detection experiments search for the scattering of DM particles from the Galactic halo off nuclei in an ultra-pure target by measuring the energy $E_\mathrm{R}$ of recoiling nuclei. The differential event rate with respect to recoil energy is given by
\begin{equation}
\frac{\text{d}R}{\text{d}E_\text{R}} = \frac{\rho_0}{m_T \, m_\chi}  \int_{v_\text{min}}^\infty v f(v) \frac{\text{d} \sigma}{\mathrm{d} E_{\text{R}}}  \text{d}^3 v\; ,
\label{eq:dRdE}
\end{equation}
where $\rho_0$ is the local DM density,
%\AB{Local DM density is referred as $\rho_0$ in Table 3; shouldn't we keep consistency here too?},
 $m_T$ is the target nucleus mass, $f(v)$ is the local DM velocity distribution and
\begin{equation}\label{eq:vmin}
v_\text{min}(E_\text{R}) = \sqrt{\frac{m_T E_\text{R}}{2 \, \mu^2}}
\end{equation}
is the minimal DM velocity to cause a recoil carrying away a kinetic energy $E_\text{R}$,
where $\mu = m_T \, m_\chi / (m_T + m_\chi)$ is the reduced mass of the DM-nucleus system.

%%%%%%%%%%%%%%%%%%%%%%%%%%%%%%%%%%%%%%%%%%%%%%%%%%%%%%%%%%%%%%%
\begin{table*}[t]
  \centering
  \begin{tabular}{lccc}
    \toprule
    & \textbf{SI scattering} & \textbf{SD scattering} & \textbf{Annihilations} \\ \midrule
  \textbf{Dimension-6 operators} & & & \\[1mm]
  $\Q{1,q}{6} = (\overline\chi \gamma_\mu \chi)(\overline{q} \gamma^\mu q)$ & unsuppressed  & --- & $s$-wave \\[1.5mm]
  $\Q{2,q}{6} = (\overline\chi \gamma_\mu \gamma_5 \chi)(\overline{q} \gamma^\mu q)$ & suppressed & --- & $p$-wave \\[1.5mm]
    $\Q{3,q}{6} = (\overline\chi \gamma_\mu \chi)(\overline{q} \gamma^\mu \gamma_5 q)$ & --- & suppressed & $s$-wave \\[1.5mm]
    $\Q{4,q}{6} = (\overline\chi \gamma_\mu \gamma_5 \chi)(\overline{q} \gamma^\mu \gamma_5 q)$ & --- & unsuppressed & $s$-wave $\propto m_q^2 / m_\chi^2$ \\ \midrule
  \textbf{Dimension-7 operators} & & & \\[1mm]
  $\Q{1}{7} = \dfrac{\alpha_s}{12\pi}(\overline\chi \chi)G^{a\mu\nu}G^a_{\mu\nu}$ & unsuppressed & --- & $p$-wave \\[2mm]
  $\Q{2}{7} = \dfrac{\alpha_s}{12\pi}(\overline\chi i\gamma_5 \chi)G^{a\mu\nu}G^a_{\mu\nu}$ & suppressed & --- & $s$-wave \\[2mm]
    $\Q{3}{7} = \dfrac{\alpha_s}{8\pi}(\overline\chi \chi)G^{a\mu\nu}\widetilde{G}^a_{\mu\nu}$ & --- & suppressed & $p$-wave \\[2mm]
  $\Q{4}{7} = \dfrac{\alpha_s}{8\pi}(\overline\chi i\gamma_5 \chi)G^{a\mu\nu}\widetilde{G}^a_{\mu\nu}$ & --- & suppressed & $s$-wave \\[2mm]
    $\Q{5,q}{7} = m_q(\overline\chi \chi)( \overline{q} q )$ & unsuppressed & --- &  $p$-wave $\propto m_q^2 / m_\chi^2$ \\[1.5mm]
    $\Q{6,q}{7} = m_q(\overline\chi i\gamma_5 \chi)( \overline{q} q )$ & suppressed & --- & $s$-wave $\propto m_q^2 / m_\chi^2$ \\[1.5mm]
    $\Q{7,q}{7} = m_q(\overline\chi \chi)( \overline{q} i\gamma_5 q )$ & --- & suppressed & $p$-wave $\propto m_q^2 / m_\chi^2$ \\[1.5mm]
  $\Q{8,q}{7} = m_q(\overline\chi i\gamma_5 \chi)( \overline{q} i\gamma_5 q )$ & --- & suppressed &  $s$-wave $\propto m_q^2 / m_\chi^2$ \\[1.5mm]
    $\Q{9,q}{7} = m_q(\overline\chi \sigma^{\mu\nu} \chi)( \overline{q} \sigma_{\mu\nu} q )$ & loop-induced & unsuppressed &  $s$-wave $\propto m_q^2 / m_\chi^2$  \\[1.5mm]
  $\Q{10,q}{7} = m_q(\overline\chi i\sigma^{\mu\nu}\gamma_5 \chi)( \overline{q} \sigma_{\mu\nu} q )$ & loop-induced & suppressed &  $s$-wave $\propto m_q^2 / m_\chi^2$ \\
  \bottomrule
  \end{tabular}
  \caption{A full list of dimension-6 and 7 operators included in this study, and the types of interactions they induce. For the DM-nucleon scattering cross-section, we distinguish between spin-independent (SI) and spin-dependent (SD) interactions, with the former receiving a large coherent enhancement and the latter vanishing for nuclei with zero spin. We use ``unsuppressed'' (``suppressed'') to denote tree-level contributions that do not vanish (that vanish) in the zero-velocity limit, while ``loop-induced'' implies that an unsuppressed interaction is induced at the one-loop level. For the annihilation cross-section we use ``$s$-wave'' (``$p$-wave'') to denote annihilations that do not vanish (that vanish) in the zero-velocity limit. Note that if the $s$-wave contribution is helicity suppressed (i.e.\ proportional to $m_q^2 / m_\chi^2$), the $p$-wave contribution may dominate in the relic density calculation.}
  \label{tab:op_suppression}
\end{table*}
%%%%%%%%%%%%%%%%%%%%%%%%%%%%%%%%%%%%%%%%%%%%%%%%%%%%%%%%%%%%%%%

The local DM density and velocity distribution are not very well known and introduce sizeable uncertainties in the prediction of experimental signals (see the discussion of nuisance parameters in Sec.~\ref{sec:nuisance}). Nevertheless, the greatest challenge in the present context is the calculation of the differential scattering cross-section $\mathrm{d}\sigma / \mathrm{d}E_\mathrm{R}$. For this purpose, one needs to map the effective interactions between relativistic DM particles and quarks or gluons defined above onto effective interactions between non-relativistic DM particles and nucleons $N = p,n$. The EFT of non-relativistic interactions can be written as
\begin{equation}
 \mathcal{L}_\text{NR} = \sum_{i,N} c_i^N(q^2) \, \mathcal{O}^N_i \; ,
 \label{eq:NREFT}
\end{equation}
where the operators $\mathcal{O}^N_i$ depend only on the DM spin $\vec{S}_\chi$, the nucleon spin $\vec{S}_N$, the momentum transfer $\vec{q}$ and the DM-nucleon relative velocity $\vec{v}$~\cite{Fitzpatrick:2012ix,Anand:2013yka,Fan:2010gt}.

The non-relativistic operators can be divided into four categories according to whether or not they depend on the nucleon spin $\vec{S}_N$, such that scattering is suppressed for nuclei with vanishing spin, and whether or not they depend on $\vec{q}$ and/or $\vec{v}$, such that scattering is suppressed in the non-relativistic limit. Specifically, $\mathcal{O}^N_1$ leads to spin-independent (SI) unsuppressed scattering, $\mathcal{O}^N_4$ leads to spin-dependent (SD) unsuppressed scattering, $\mathcal{O}^N_5$, $\mathcal{O}^N_8$, $\mathcal{O}^N_{11}$ lead to SI momentum-suppressed scattering and $\mathcal{O}^N_6$, $\mathcal{O}^N_7$, $\mathcal{O}^N_9$, $\mathcal{O}^N_{10}$, $\mathcal{O}^N_{12}$ lead to SD momentum-suppressed scattering, which is typically unobservable.
For the relativistic operators included in this study we give the dominant type of interaction they induce in the non-relativistic limit in Table~\ref{tab:op_suppression}.

The coefficients $c_i^N(q^2)$ can be directly calculated from the Wilson coefficients of the relativistic operators at $\mu = 2\,\mathrm{GeV}$. The explicit dependence on the momentum transfer $q = \sqrt{2 m_T E_\text{R}}$ is a result of two effects. First, under RG evolution some of the effective DM-quark operators mix into the DM dipole operators $\Q{1,2}{5}$ (see Eq.~(\ref{eq:MLL})). These operators then induce long-range interactions, i.e.\ contributions to the $c_i^N(q^2)$ that scale as $q^{-2}$. Since the momentum transfer can be very small in direct detection experiments, these contributions can be important in spite of their loop suppression. Second, the coefficients include nuclear form factors, obtained by evaluating expectation values of quark currents like $\langle N' | \overline{q} \gamma^\mu q | N \rangle$. These form factors can be calculated in chiral perturbation theory and exhibit a pion pole for axial and pseudoscalar currents, i.e.\ a divergence for $q \to m_\pi$~\cite{Bishara:2016hek,Bishara:2017pfq}.

%%%%%%%%%%%%%%%%%%%%%%%%%%%%%%%%%%%%%%%%%%%%%%%%%%%%%%%%%%%%%%%
\begin{table*}
  \centering
  \begin{tabular}{lrcclr}
  \toprule
  \textbf{Parameter} & \textbf{Value} & & & \textbf{Parameter} & \textbf{Value} \\ \midrule
   $\sigma_{\pi N} $ & $50(15)$~MeV \cite{Bishara:2017pfq} & & & $\mu_p$ & $2.793$ \cite{PDG20} \\[1.5mm]
   $B c_5 (m_d-m_u) $ & $-0.51(8)$ MeV \cite{Crivellin:2013ipa} & & & $\mu_n$ & $-1.913$ \cite{PDG20} \\[1.5mm]
   $g_A$ & $1.2756(13)$ \cite{PDG20} & & & $\mu_s$ & $-0.036(21)$ \cite{Djukanovic:2019jtp, Sufian:2016pex} \\[1.5mm]
   $m_G$ & $836(17){\rm~MeV}$ \cite{Bishara:2017nnn} & & & $g_T^u$ & $0.784(30)$ \cite{Gupta:2018lvp} \\[1.5mm]
   $\sigma_s$ & $52.9(7.0)\rm{~MeV}$ \cite{Aoki:2019cca} & & & $g_T^d$ & $-0.204(15)$ \cite{Gupta:2018lvp} \\[1.5mm]
   $\Delta u + \Delta d$  & $0.440(44)\,$ \cite{Liang:2018pis} & & & $g_T^s$ & $-27(16) \cdot 10^{-3}$ \cite{Gupta:2018lvp} \\[1.5mm]
   $\Delta s$ & $-0.035(9)$ \cite{Liang:2018pis} & & & $B_{T,10}^{u/p}$ & $3.0(1.5)$ \cite{Pasquini:2005dk} \\[1.5mm]
   $B_0 m_u$ & $0.0058(5)\rm{~GeV}^2$ \cite{Bishara:2017nnn} & & & $B_{T,10}^{d/p}$ & $0.24(12)$ \cite{Pasquini:2005dk} \\[1.5mm]
   $B_0 m_d$ & $0.0124(5)\rm{~GeV}^2$ \cite{Bishara:2017nnn} & & & $B_{T,10}^{s/p}$ & $0.0(2)$ \cite{Pasquini:2005dk} \\[1.5mm]
   $B_0 m_s$ & $0.249(9)\rm{~GeV}^2$  \cite{Bishara:2017nnn} & & & $r_s^2 $ & $ -0.115(35){\rm~GeV}^{-2}$ \cite{Sufian:2016pex,Djukanovic:2019jtp} \\ \bottomrule
 \end{tabular}
 \caption{The hadronic input parameters as used in \ddm \textsf{v2.2.0}. }
 \label{tab:hadr:inputs}
\end{table*}
%%%%%%%%%%%%%%%%%%%%%%%%%%%%%%%%%%%%%%%%%%%%%%%%%%%%%%%%%%%%%%%

All of these effects are fully taken into account in $\ddm$, which calculates the coefficients $c_i^N(q^2)$ for given Wilson coefficients $\C{a}{d}$ at a higher scale (see App.~\ref{app:directdm}). These coefficients are then passed onto \textsf{DDCalc v2.2.0}~\cite{DarkBit,HP}, which calculates the differential cross-section for each operator $\mathcal{O}^N_i$ (including interference) and target element of interest. \textsf{DDCalc} also performs the velocity integrals needed for the calculation of the differential event rate, and the convolution with energy resolution and detector acceptance needed to predict signals in specific experiments:
\begin{equation}
 N_p = M \, T_\text{exp} \int \phi(E_\mathrm{R}) \, \frac{\mathrm{d}R}{\mathrm{d}E_\mathrm{R}} \, \mathrm{d}E_\mathrm{R} \, ,
\end{equation}
where $M$ is the detector mass, $T_\text{exp}$ is the exposure time and $\phi(E_\mathrm{R})$ is the acceptance function.

By combining \ddm and \ddcalc, we can obtain likelihoods for a wide range of direct detection experiments. In the present analysis, we include constraints from the most recent XENON1T analysis \cite{Aprile:2018dbl}, LUX 2016 \cite{LUXrun2}, PandaX 2016 \cite{Tan:2016zwf} and 2017 \cite{Cui:2017nnn} analyses, CDMSlite \cite{Agnese:2015nto}, CRESST-II \cite{Angloher:2015ewa} and CRESST-III~\cite{Abdelhameed:2019hmk}, PICO-60 2017 \cite{Amole:2017dex} and 2019~\cite{Amole:2019fdf}, and DarkSide-50 \cite{Agnes:2018fwg}.

The hadronic inputs to \ddm \textsf{v2.2.0} \cite{Bishara:2017nnn} were updated with the most recent $N_f=2+1$ lattice QCD results, following the FLAG quality requirements~\cite{Aoki:2019cca}, see Table \ref{tab:hadr:inputs}. All the inputs are evaluated at $\mu=2$ GeV. The hadronic matrix elements for protons and neutrons are related using isospin conservation.

For operators with vector quark currents, the least well known are the hadronic matrix elements involving the strange quark, while the matrix elements for operators with $u,d$ quark vector currents have negligible errors to the precision we are working with. Since the strange quark vector current vanishes at $q^2=0$, the first non-vanishing contribution is  obtained only at next-to-leading order in the chiral expansion, and depends on the strange quark charge radius, $r_s^2 = -0.0045(14)\,$fm$^2$~\cite{Sufian:2016pex,Djukanovic:2019jtp}. For the nuclear magnetic moment induced by the strange quark, $\mu_s= -0.036(21)$~\cite{Djukanovic:2019jtp, Sufian:2016pex}, we inflate the errors according to the Particle Data Group prescription.

The scalar form factors at zero recoil are obtained from expressions in Ref.~\cite{Crivellin:2013ipa}, namely
\beq\label{eq:sigmaudN}
  \sigma_{u(d)}^N = \frac{\sigma_{\pi N}}{2} (1 \mp \xi)\pm Bc_5 \, (m_d-m_u)\left(1 \mp \frac{1}{\xi}\right),
\eeq
where the upper (lower) sign is for the proton (neutron).  We use a rather conservative estimate $\sigma_{\pi N}=(50\pm15)$ MeV \cite{Bishara:2017pfq} that covers the spread between the lattice QCD \cite{Horsley:2011wr, Durr:2015dna, Yang:2015uis, Abdel-Rehim:2016won, Bali:2016lvx, Alexandrou:2019brg, Yamanaka:2018uud, Borsanyi:2020bpd} and pionic atom determinations \cite{Alarcon:2011zs, Hoferichter:2015dsa, Dmitrasinovic:2016hup, RuizdeElvira:2017stg, Friedman:2019zhc, Yang:2015uis, Abdel-Rehim:2016won, Yamanaka:2018uud}. The other two parameters are $\xi \equiv (m_d-m_u)/(m_d +  m_u)= 0.36\, \pm 0.04$ and $B c_5 \, (m_d-m_u)=(-0.51\pm0.08)$ MeV~\cite{Crivellin:2013ipa}.

The matrix elements of tensor currents are described by three sets of form factors, but only two, $g_{T}^{q}$ and $B_{T,10}^{q/N} (0)$, enter the chirally leading expressions. For $g_{T}^{q}$, the only $N_f=2+1$ result from Ref.~\cite{Yamanaka:2018uud} does not satisfy the FLAG quality requirements, so we use the $N_f=2+1+1$ results from Ref.~\cite{Gupta:2018lvp} instead; the difference between the $N_f=2+1$ and $N_f=2+1+1$ results is expected to be small. For $B_{T,10}^{q/N}(0)$, we use the results from the constituent quark model in Ref.~\cite{Pasquini:2005dk}.

\subsection{Relic abundance of DM}\label{sec:rd}

The Early Universe time evolution of the number density of the $\chi$ particles, $n_\chi$, is governed by the Boltzmann equation~\cite{Gondolo:1990dk}
\begin{align} \label{eq:Boltzmann}
  \frac{dn_\chi}{dt} + 3Hn_\chi = -\langle\sigma v_\textrm{rel}\rangle \left(n_\chi n_{\bar\chi}-n_{\chi,\textrm{eq}}n_{\bar\chi,\textrm{eq}}\right) \, ,
\end{align}
where $n_{\chi,\textrm{eq}}$ is the number density in equilibrium, $H(t)$ is the Hubble rate and $\langle\sigma v_{\textrm{{rel}}}\rangle$ is the thermally averaged cross-section times the relative (Møller) velocity, given by
\begin{align} \label{eq:sigmavthermal_def}
  \langle\sigma v_\textrm{rel}\rangle = \int^{\infty}_{4m_\chi^2} \!ds \, \frac{\sqrt{s-4m_\chi^2}(s-2m_\chi^2)\,K_1 \left(\sqrt{s}/T\right)}{8m_\chi^4 \,T K_2^2\left(m_\chi/T\right)} \,\sigma v_{\rm lab} \, ,
\end{align}
where $K_{1,2}$ are the modified Bessel functions and $v_{\rm lab}$ is the velocity of one of the annihilating
(anti-)DM particles in the rest frame of the other (for a discussion, see also Ref.~\cite{Binder:2017rgn}).
We stress that there is no additional factor of $1/2$ in the above equations. However, the fact that DM
consists of Dirac particles implies that the total contribution to the observed DM density is given by
$n_\chi+n_{\bar\chi}=2n_\chi$ %. In the following, we will disregard the
(disregarding the possibility of an initial asymmetry~\cite{Kaplan:2009ag}).
% and therefore set $n_\chi=n_{\bar\chi}$.

We compute tree-level annihilation cross-sections using \CH \textsf{v3.6.27} \cite{Pukhov:2004ca,Belyaev:2012qa}, where
the implementation of the four-fermion interactions is generated by \gum~\cite{Gonzalo:2021cnq,GUM} from \ufo files via the
tool \utc (described in App.~\ref{app:ufo_to_mdl}). To ensure the EFT picture is valid, we invalidate points
where $\La \leq 2 m_\chi$. We obtain the relic density of $\chi$ by numerically solving
Eq.~\eqref{eq:Boltzmann} at each parameter point, assuming the standard cosmological history\footnote{For a recent review on the effects of non-standard cosmological scenarios, see Ref.~\cite{Arbey:2021gdg}.} and using the routines implemented
in \darksusy \textsf{v6.2.2} \cite{darksusy,darksusy4} via \darkbit.
We then compare the prediction to the relic density constraint from \emph{Planck} 2018: $\Omega_{\textrm{DM}}\,h^2 = 0.120 \pm 0.001$~\cite{Aghanim:2018eyx}.
We include a $1\%$ theoretical error on the computed values of the relic density, which we combine in quadrature with the observed error
on the \emph{Planck} measured value.
More details on this prescription can be found in Refs.~\cite{gambit,DarkBit}.

We note that our uncertainty estimate does not include uncertainties in the calculation of the annihilation cross-section very close to quark thresholds, which may be considerably larger. Moreover, our approach does not capture the potential effect of additional degrees
of freedom on $\langle\sigma v_\textrm{rel}\rangle$ during freeze-out. The resulting effects, such as resonances or coannihilations could both
 increase and decrease the resulting value of $\Omega_\chi$ (see e.g.\ Ref.~\cite{Bell:2013wua,Baker:2015qna}), so the relic density
 constraint should be interpreted with care for $\La\sim2m_\chi$, i.e.~close to the EFT validity boundary (see Sec.~\ref{sec:validity}).

The very nature of the EFT construction implies additional degrees of freedom above the energy scale
$\La$.
Given the potential for a rich dark sector containing $\chi$, and in particular, 
the possibility of additional DM candidates not
captured by the EFT, we will by default \emph{not} demand that the particle $\chi$ constitutes all of the
observed DM, \ie we allow for the possibility of other DM species to contribute to the observed relic density.
In practice, this means that we modify the relic density constraint in such a way that the likelihood is flat if the predicted value is smaller than the observed one.
In this case, we rescale all predicted direct and indirect detection signals by
\begin{equation}
f_\chi \equiv (\Omega_\chi + \Omega_{\bar\chi})/\Omega_{\mathrm{DM}} = 2 \Omega_\chi / \Omega_{\mathrm{DM}}
\label{eq:fDM}
\end{equation}
and $f_\chi^2$, respectively. 
In doing so, we assume that the fraction $f_\chi$ is the same in all astrophysical systems and that any additional DM population does \emph{not} contribute to
signals in these experiments.
In a second set of scans we then impose a stricter requirement, namely that the DM particle under consideration saturates the DM relic abundance ($f_\chi \approx 1$) rather than imposing the relic density as an upper bound ($f_\chi \leq 1$).\footnote{Note that, since we include uncertainties in both the relic density calculation and the \emph{Planck} measurement, $\Omega_\chi h^2$ can deviate slightly from 0.120 even when we require that the DM relic abundance is saturated. In this case we set $f_\chi = \text{min}(\Omega_\chi h^2 / 0.120, 1)$, which can therefore slightly deviate from (but never exceed) unity.}

\subsection{Indirect detection with gamma rays}\label{sec:id}

If DM is held in thermal equilibrium in the early universe via collisions with SM particles, then it can still
annihilate today,
especially in regions of high DM density.  As with the relic
abundance calculation, in order for the effective picture to hold for DM annihilation, we must impose
$\La > 2 m_\chi$.

Gamma rays from dwarf spheroidal galaxies (dSphs) are a particularly robust way of constraining annihilation signals from
DM~\cite{Bringmann:2012ez}.
In general, for a given energy bin $i$, the DM-induced $\gamma$-ray flux from target $k$ can be written
in the factorised form $\Phi_i \cdot J_k$, where details of the particle physics processes are encoded in
$\Phi_i$, and details of the astrophysics are encoded in $J_k$. See the \darkbit manual~\cite{DarkBit} for more details.

In general, only operators that lead to $s$-wave annihilation ($\Q{1,3,4,q}{6},\Q{2,4,6,8,9,10,q}{7}$) give rise to observable gamma-ray signals; see for instance, Table~\ref{tab:op_suppression}. For the operators $\Q{2,q}{6}$ and $\Q{1,3,5,7,q}{7}$, the leading contribution to the annihilation cross-section is $p$-wave suppressed, i.e.\ proportional to $v_\mathrm{rel}^2$. As DM in dSphs is extremely cold, with $\langle v^2\rangle^{1/2}\sim10^{-4}$, this factor is very small, and the resulting limits are exceedingly weak.  We therefore neglect
$p$-wave contributions to all annihilation processes here.

For $s$-wave annihilation, one obtains
\begin{align}
 \Phi_i &= \frac{f_\chi^2}{4} \sum_{j} \frac{(\sigma v)_{0,j}}{4\pi m_\chi^2}\int_{\Delta E_i} dE \, \frac{dN_{\gamma,j}}{dE} \, ,
\end{align}
where $f_\chi$ is the DM fraction defined in eq.~(\ref{eq:fDM}), $(\sigma v)_{0,j}$ denotes the zero-velocity limit of the cross-section for $\chi\bar\chi\to j$ and $N_{\gamma,j}$
is the number of photons, per annihilation, resulting from the final state channel $j$. The prefactor $1/4$ accounts
for the Dirac nature of the DM particles (under the assumption that $n_\chi=n_{\bar\chi}$). Again, we use \CH
to compute annihilation cross-sections, with the \CH model files generated by
\utc via \gum (see App.~\ref{app:ufo_to_mdl}). The photon yields  ${dN_{\gamma,j}}/{dE}$ used in
\darkbit are based on tabulated \pythia runs, as provided by \ds.

The $J$-factor for each dSph $k$ is simply the line-of-sight integral over the DM distribution assuming an NFW density profile and the solid
angle $\Omega$,
\begin{align}
  J_k &= \int_{\Delta\Omega_k} d\Omega \int_{\mathrm{l.o.s.}} ds \, \rho_{\rm DM}^2\simeq
  D_k^{-2} \int d^3x\,\rho_{\rm DM}^2\,, \label{eq:def_jk}
\end{align}
where $D_k$ is the distance to the dSph.
In our analysis we use the \texttt{Pass-8} combined analysis of 15 dSphs after 6 years of
\emph{Fermi}-LAT data~\cite{LATdwarfP8}. We use the \gamlike~\textsf{v1.0.1} interface within
\darkbit~\cite{DarkBit} to compute the likelihood for the gamma-ray observations, $\ln \mathcal{L}_{\rm{exp}}$,
constructed from the product $\Phi_i \cdot J_k$ and summed over all targets and energy bins,
\begin{align}
 \ln \mathcal{L}_{\rm{exp}} = \sum^{\rm{N_{dSphs}}}_{k=1} \sum^{\rm{N_{eBins}}}_{i=1} \ln \mathcal{L}_{ki}\left(\Phi_i \cdot J_k\right) \, .
\end{align}
We also include a contribution from profiling over the $J$-factors of each dSph, $\ln \mathcal{L}_J = \sum_k \ln \mathcal{L}(J_k)$~\cite{DarkBit,LATdwarfP8}, such that the full likelihood reads
\begin{align}
 \ln \mathcal{L}_{\rm{dSphs}}^{\rm{prof.}} = \underset{\{J_k\}}{\textrm{max}}\left(\ln\mathcal{L}_{\rm{exp}} + \ln\mathcal{L}_J \right) \, .
\end{align}

Gamma rays from the Galactic centre region provide a promising complementary way of constraining
a signal from annihilating DM. While the $J$-factor is expected to be significantly higher than for dSphs,
however, this conclusion is largely based on the result of numerical simulations of gravitational clustering rather
than on the direct analysis of kinematical data. The reason for this is that the gravitational potential within
the solar circle is dominated by baryons, not by DM, which adds additional uncertainty due to
a dominant component of astrophysical gamma rays from this target region.
As a result, Galactic centre observations with \emph{Fermi}-LAT are somewhat less competitive than the dSph
limits discussed above~\cite{TheFermi-LAT:2017vmf}. The upcoming Cherenkov Telescope Array (CTA), on the other
hand, has a good chance of probing thermally produced DM up to particle masses of several TeV~\cite{Acharyya:2020sbj}.
We will not include the projected CTA likelihoods in our scans, but indicate the reach of CTA when discussing our
results.

\subsection{Other indirect detection constraints}
\label{sec:id2}

\subsubsection*{Solar capture}

The presence of non-zero elastic scattering cross-sections with nuclei combined with self-annihilation to heavy SM states leads to an additional, unique signature of DM in the form of high-energy neutrinos from the Sun. If Milky Way DM scatters with solar nuclei and loses enough kinetic energy to fall below the local escape velocity, it will become gravitationally bound. As long as it is above the evaporation mass threshold $\simeq 4$ GeV, captured DM will thermalize in a small region near the solar centre, and annihilate to SM products which then produce neutrinos via regular decay processes. These are distinct from the neutrinos from Solar fusion, as they are expected to have much higher energies than the $\sim$ MeV scales of fusion processes. Leading constraints have been obtained by Super-Kamiokande down to a few GeV \cite{Choi:2015ara}, and by the IceCube South Pole Neutrino Observatory, between 20 and 10$^4$ GeV \cite{Aartsen:2016zhm}. For typical annihilation cross-sections, the captured DM population reaches an equilibrium that is determined by the capture rate.
For each likelihood evaluation, we obtain the non-relativistic effective operators (Eq.~\eqref{eq:NREFT}) as described in Sec. \ref{sec:dd}, using \ddm to obtain the non-relativistic Wilson coefficients. These are passed to the public code \capgen \cite{Kozar:2021iur}, which computes the DM capture rate via the integral over the solar radius $r$ and DM halo velocity $u$:
\begin{equation}
C_\odot(t) = 4\pi \int_0^{R_\odot} r^2 \int_0^\infty \frac{f(u)}{u} \, w \Omega(w,r) \, d u \, d r \, ,
\end{equation}
where $w(r) = \sqrt{u^2 + v^2_{\text{esc},\odot}(r)}$ is the DM velocity at position $r$, and
\begin{align}
 \Omega(w) &= w \sum_i n_i(r,t) \frac{\mu_{i,+}^2}{\mu_i}\Theta\left(\frac{\mu_i v^2}{\mu_{i,-}^2} - u^2 \right) \nonumber \\
 &\hspace{5mm} \times \int_{m_\chi u^2/2}^{m_\chi w^2 \mu_i/2\mu_{i,+}^2} \frac{d\sigma_{i}}{dE_\text{R}} \, d E_\text{R} \,,
\end{align}
is the probability of scattering from velocity $w$ to a velocity less than the local Solar escape velocity $v_{\mathrm{esc},\odot}(r)$, ${d\sigma_{i}}/{dE_\text{R}}$ is the DM-\textit{nucleus} scattering cross-section, $n_i(r)$ is the number density of species $i$ with atomic mass $m_{N,i}$, and $\mu_i = m_\chi/m_{N,i}$. Version 2.1 of \capgen uses the method described in detail in Ref.~\cite{Catena:2015uha}, separating the DM-nucleus cross-section into factors proportional to non-relativistic Wilson coefficients, powers of $w$ and exchanged momentum $q$, and operator-dependent \textit{nuclear response functions} computed in Ref.~\cite{Catena:2015uha} for the 16 most abundant elements in the Sun. Solar parameters are based on the Barcelona Group's AGSS09ph Standard Solar Model \cite{Vinyoles:2016djt,AGSS}.

Annihilation cross-sections are computed as described in Sec.~\ref{sec:rd}, via \CH. Once the equilibrium population of DM in the Sun has been obtained, cross-sections and annihilation rates are passed to \ds, which computes the neutrino yields as a function of energy.  These are finally passed to \nulike \textsf{v1.0.9} \cite{IC79,IC22Methods}, which computes event-level likelihoods based on a re-analysis of the 79-string IceCube search for DM annihilation in the Sun \cite{IC79_SUSY}.

\subsubsection*{Cosmic Microwave Background}

Additional constraints on the DM annihilation cross-section arise from the early universe, more specifically from observations of the Cosmic Microwave Background (CMB). Annihilating DM particles inject energy into the primordial plasma, which affects the reionisation history and alters the optical depth $\tau$. The magnitude of this effect depends on the specific annihilation channel and how efficiently the injected energy is deposited. These details can be encoded in an effective efficiency coefficient $f_\text{eff}$, which depends on the injected yields of photons, electrons and positrons, and thus on the DM mass and its branching ratios into different final states~\cite{Slatyer15a}. The CMB is then sensitive to the following parameter combination:

\begin{equation}
 p_\text{ann} \equiv f_\chi^2 f_\text{eff} \frac{\langle \sigma v_\mathrm{rel} \rangle}{m_\chi} \; ,
 \label{eq:CMB_energy_injection}
\end{equation}
where $\langle \sigma v_\mathrm{rel} \rangle \approx (\sigma v)_0$ to a very good approximation during recombination;
we thus also neglect $p$-wave contributions to all annihilation processes here.

In order to calculate $p_\text{ann}$ for a given parameter point, one first needs to calculate the injected spectrum of photons, electrons and positrons and then convolve the result with suitable transfer functions that link the energy injection rate to the energy deposition rate~\cite{Slatyer:2015kla}. The first part of this calculation has been automated within \ds and is accessible via \darkbit. The second part relies on \textsf{DarkAges}~\cite{Stocker:2018avm} (which is part of the \textsf{ExoCLASS} branch of \textsf{CLASS}) and is accessible via \cosmobit \cite{CosmoBit}, see App.~\ref{app:energy_injection} for further details.\footnote{It is noteworthy that \textsf{DarkAges} calculates $ f_\text{eff} (z) $ as a redshift-dependent function instead of a single redshift-independent coefficient $ f_{\rm eff} $, as it is implicitly assumed in Eq.~\eqref{eq:CMB_energy_injection}. In order to compress the function $ f_\text{eff} (z) $ into this coefficient, it is convolved with a weighting function $ W(z) $ that encodes the CMB sensitivity to energy injection through $s$--wave annihilation as a function of redshift~\cite{Slatyer15a}.}

As the \emph{Planck} collaboration only quotes the 95\% credible interval for $p_\text{ann}$~\cite{Aghanim:2018eyx}, the remaining challenge is to obtain a likelihood for $p_\text{ann}$ from cosmological data. Although this likelihood can, in principle, be calculated for each parameter point individually using the \textsf{CosmoBit} interface to \textsf{CLASS} and the \emph{Planck} likelihoods, carrying out such a large number of calculations would be prohibitively slow, in particular if the cosmological parameters of the $\La$CDM model are to be varied simultaneously. In the present work, we therefore adopt a simpler approach, where we first calculate the likelihood when varying $p_\text{ann}$  (while profiling over the $\La$CDM and cosmological nuisance parameters). This approach yields
\begin{equation}
 \mathcal{L}(p_\text{ann}) = \mathcal{L}_0 \exp\left[ - \left(\frac{p_\text{ann}^{28} + 0.48}{2.45} \right)^2 \right] \; ,
\end{equation}
where $p_\text{ann}^{28} \equiv p_\text{ann} / \left(10^{-28} \, \mathrm{cm^3 \, s^{-1} \, GeV^{-1}} \right)$. 
In arriving at this result, we have included the \emph{Planck} TT,TE,EE+lowE+lensing likelihoods (using the `lite' likelihood for multipoles $ \ell\geq30 $, which only require one additional nuisance parameter~\cite{Aghanim:2019ame}), as well as the BAO data of 6dF~\cite{2011MNRAS.416.3017B}, SDSS DR7 MGS~\cite{2015MNRAS.449..835R}, and the SDSS BOSS DR12 galaxy sample~\cite{Alam:2016hwk}. This profile likelihood, which reproduces the 95\% credible interval obtained by the \emph{Planck} collaboration~\cite{Aghanim:2018eyx}, can then be used in all subsequent scans, so that only $p_\text{ann}$ needs to be calculated for each parameter point and it is no longer necessary to call \textsf{CLASS} or \textsf{plc}.

\subsubsection*{Charged cosmic rays}

Finally, DM particles annihilating in the Galactic halo also produce positrons, antiprotons and, to a lesser degree,
heavier anti-nuclei that could in principle be observed in the spectrum of charged cosmic rays. Positrons
quickly lose their energy through synchrotron radiation, and are thus a robust probe of exotic contributions
from the {\it local} Galactic environment; the resulting bounds on DM annihilating to quarks or gluons are,
however, much weaker than the other indirect detection constraints discussed here~\cite{%Ellis:1988qp,
Kopp:2013eka}.\footnote{%
We note that this conclusion would be radically different for unsuppressed direct annihilation to
leptons~\cite{Bergstrom:2013jra,Ibarra:2013zia},
which would result from leptonic operators analogous to $\Q{1,3}{6}$.
}
Anti-nuclei, on the other hand, probe a significant fraction of the entire Galactic halo because energy
losses are much less efficient in this case.
For antiprotons, this generally leads to competitive constraints on DM annihilation
signals~\cite{Bergstrom:1999jc,Bringmann:2006im,Cuoco:2016eej},
but it also means that such bounds necessarily strongly depend on uncertainties relating to modelling the production and propagation of cosmic rays in the Galactic halo. In addition to the dozen (or more) free parameters in the diffusion-reacceleration equations, there exist significant uncertainties on the energy dependence of  the nuclear cross-sections responsible for the conventional antiproton flux  \cite{Heisig:2020nse} and possible correlated systematics \cite{Boudaud:2019efq}. 
 A full statistical analysis, which would require a treatment of the large number of (effective) propagation parameters
as nuisance parameters in our scans,
is prohibitive in terms of computational costs  \cite{Johannesson:2016rlh} and
hence beyond the scope of this work.

\subsection{Collider physics}\label{sec:collider}
The effective operators defined in Sec.~\ref{sec:model} allow for the pair production of WIMPs in the proton--proton collisions at the LHC. If one of the incoming partons radiates a jet through initial state radiation (ISR), one can observe the process $pp \rightarrow \chi \chi j$ as a single jet associated with missing transverse energy ($\slashed{E}_T$). In this study, we include the CMS~\cite{Sirunyan:2017jix} and ATLAS~\cite{Aad:2021egl} monojet analyses based on $36\,\rm{fb}^{-1}$ and $139\,\rm{fb}^{-1}$ of data from Run II, respectively. ATLAS and CMS have performed a number of further searches for other types of ISR, leading for example to mono-photon signatures, but these are known to give weaker bounds on DM EFTs than monojet searches~
\cite{Bauer:2017fsw,Zhou:2013fla,Brennan:2016xjh}.

The expected number of events in a given bin of the $\slashed{E}_T$ distribution is
$$N = L\times\sigma \times(\epsilon A)\;,$$
where $L =36\,\text{fb}^{-1}$ or $139\,\text{fb}^{-1}$ is the total integrated luminosity, $\sigma$ the total production cross-section and the factor $(\epsilon A)$ is the efficiency times acceptance for passing the kinematic selection requirements for the analysis. Both $\sigma$ and $(\epsilon A)$ can be obtained via Monte Carlo simulation, but given the dimensionality of the DM EFT parameter space it is computationally too expensive to perform these simulations on the fly during the parameter scan, as would be the standard approach to collider simulations within \colliderbit in \gambit.

Starting from \textsf{UFO} files generated using \fr \textsf{v2.0}~\cite{Alloul:2013bka}, we have therefore produced separate interpolations of $\sigma$ and $\epsilon A$ based on the output of Monte Carlo simulations with \textsf{MadGraph\_aMC@NLO} \textsf{v2.6.6}~\cite{Alwall:2011uj} (\textsf{v2.9.2}) for the CMS (ATLAS) analysis, interfaced to \pythia \textsf{v8.1} \cite{Sjostrand:2007gs} for parton showering and hadronization. The matching between \MG and \pythia is performed according to the CKKW prescription, and the detector response is simulated using \delphes \textsf{v3.4.2} \cite{DELPHES3}. The \colliderbit code extension that enables $\sigma$ and $(\epsilon A)$ interpolations to be used as an alternative to direct Monte Carlo simulation will be generalised and documented in the next major version of \colliderbit.

We only include the dimension-6 and 7 EFT operators ($\C{i}{6}$ and $\C{i=1,...,4}{7}$) which are relevant for collider searches. Other operators give a negligible contribution due to either being suppressed by the parton distribution functions (in the case of heavy quarks), or by a factor of the fermion mass (small in the case of light quarks).

To reduce the computation time for our study, we generate events in discrete grids of the Wilson coefficients and DM mass. Separate grids are defined for each set of operators that do not interfere, such that the total number of events will simply be the sum of the contributions calculated from each grid. At dimension-6, there is interference between operators $\Q{1,q}{6}$/$\Q{4,q}{6}$ and $\Q{2,q}{6}$/$\Q{3,q}{6}$. For these Wilson coefficients, we parametrize the tabulated grids in terms of a mixing angle $\theta$, defined via $\C{1,2}{6}=\sin\theta$ and $\C{3,4}{6} = \cos\theta$.

The CMS and ATLAS analyses have 22 and 13 exclusive signal regions, respectively, corresponding to the individual bins in the missing transverse energy distributions. As discussed below, the publicly available information makes it possible to combine all signal regions for the CMS analysis, while for the ATLAS analysis, only a single signal region can be used at once. To maximize the sensitivity of the ATLAS analysis, we combine the three highest missing energy bins, for which systematic uncertainties in the background estimation (and hence their correlations) are negligible, such that the highest bin in our analysis corresponds to all events with $\slashed{E}_T > 1000 \, \mathrm{GeV}$.\footnote{We note that this combination also reduces the impact of a local $\sim 2.5\sigma$ excess in the third-highest bin, which would otherwise strongly bias our analysis.} Once the predicted yields for all bins have been evaluated, taking into account the EFT validity constraint as described in Sec.~\ref{sec:validity}, we compute a likelihood for each analysis as follows.

For the CMS analysis, we follow the ``simplified likelihood'' method \cite{Collaboration:2242860}, since the required covariance matrix was published by CMS. In this approach, the full experimental likelihood function is approximated by a standard convolved Poisson--Gaussian form, with the systematic uncertainties on the background predictions treated as correlated Gaussian distributions:
\begin{equation}
  \label{eq:simplike}
  \begin{split}
    \mathcal{L}_{\text{CMS}}(\bm{s}, \bm{\gamma})
    =& \prod_{i=1}^{22} \left[ \frac{(s_i + b_i + \gamma_i)^{n_i} \, e^{-(s_i + b_i + \gamma_i)}}{n_i!} \right]\\
    & \hphantom{\int} \times \frac{1}{\sqrt{\det2\pi\Sigma}} e^{-\frac{1}{2} \bm{\gamma}^T \bm{\Sigma^{-1}} \bm{\gamma}} \, .
  \end{split}
\end{equation}
For each signal region $i$, the observed yield, expected signal yield and expected background yield are given by $n_i$, $s_i$ and $b_i$, respectively. The deviation from the nominal expected yield due to systematic uncertainties is given by $\gamma_i$. The correlations between the different $\gamma_i$ are encoded in the covariance matrix $\bm{\Sigma}$ provided by CMS, where we also add the signal yield uncertainties in quadrature along the diagonal. We follow the procedure in Ref.~\cite{Collaboration:2242860} in treating the $\gamma_i$ nuisance parameters as linear corrections to the expected yields. For every point in our scans of the DM EFT parameter space, we profile Eq.~(\ref{eq:simplike}) over the 22 nuisance parameters in $\bm\gamma$ to obtain a likelihood solely in terms of the set of DM EFT signal estimates $\bm{s}$:
\begin{equation}
  \mathcal{L}_{\text{CMS}}(\bm{s}) \equiv \mathcal{L}_{\text{CMS}}(\bm{s}, \hat{\hat{\bm{\gamma}}}).
\end{equation}

In the case of the ATLAS analysis, for which such a covariance matrix is not available, the conservative course of action is to calculate a likelihood using only the signal region with the best expected sensitivity. The ATLAS likelihood is therefore given by
\begin{equation}
  \mathcal{L}_{\text{ATLAS}}(s_i) \equiv \mathcal{L}_{\text{ATLAS}}(s_i, \hat{\hat{\gamma_i}}) \, ,
\end{equation}
where $\mathcal{L}_{\text{ATLAS}}(s_i, \hat{\hat{\gamma_i}})$ is the single-bin equivalent of Eq.~(\ref{eq:simplike}), and $i$ refers to the signal region with the best expected sensitivity, i.e.\ the signal region that would give the lowest likelihood in the case $n_i = b_i$.

The total LHC log-likelihood is then given by $\ln \mathcal{L}_{\text{LHC}} = \ln \mathcal{L}_{\text{CMS}} + \ln \mathcal{L}_{\text{ATLAS}}$. However, due to the per-point signal region selection required in the evaluation of $\ln \mathcal{L}_{\text{ATLAS}}$, the variation in typical yields between the different signal regions would manifest as a large variation in the effective likelihood normalization between different parameter points. To avoid this we follow the standard approach in \colliderbit of using the log-likelihood difference
\begin{equation}
  \Delta \ln \mathcal{L}_{\text{LHC}} = \ln \mathcal{L}_{\text{LHC}}(\bf{s}) - \ln \mathcal{L}_{\text{LHC}}(\bf{s}=\bf{0})
  \label{eq:LHC_loglike}
\end{equation}
as the LHC log-likelihood contribution in the parameter scan \cite{ColliderBit}.

%%%%%%%%%%%%%%%%%%%%%%%%%%%%%%%%%%%%%%%%%%%%%%%%%%%%%%%%%%%%%%%
\begin{table*}[t]
  \centering
  \begin{tabular}{lcr}
    \toprule
    \textbf{Nuisance parameter} & & \textbf{Value} ($\pm$\,$3\sigma$\,\textbf{range}) \\ \midrule
    Local DM density & $\rho_0$ & $0.2$--$0.8$\,GeV\,cm$^{-3}$ \\[1mm]
    Most probable speed & $v_{\rm{peak}}$ & $240\,(24)$\,km s$^{-1}$\\[1mm]
    Galactic escape speed &$v_{\textrm{esc}}$ & $528\,(75)$\,km s$^{-1}$ \\[1mm] \midrule
    Running top mass ($\overline{\rm{MS}}$ scheme)  & $m_t (m_t)$ & $162.9\,(6.0)$\,GeV \\[1mm] \midrule
  Pion-nucleon sigma term & $\sigma_{\pi N} $ & $50\,(45)$~MeV \\[1mm]
  Strange quark contrib. to nucleon spin & $\Delta s$ & $-0.035\,(0.027)$ \\[1mm]
  Strange quark nuclear tensor charge & $g_T^s$ & $-0.027\,(0.048)$ \\[1mm]
  Strange quark charge radius of the proton & $r_s^2 $ & $ -0.115\,(0.105){\rm~GeV}^{-2}$ \\[1mm] \bottomrule
  \end{tabular}
  \caption{A list of nuisance parameters that are varied simultaneously with the DM EFT model parameters in our scans (the hadronic parameters are given at $\mu=2$ GeV). All parameters are scanned over their $3\sigma$ range using flat parametrisation. For more details on the respective nuisance likelihoods, see Sec.~\ref{sec:nuisance}.  }
  \label{tab:nuis_params}
\end{table*}
%%%%%%%%%%%%%%%%%%%%%%%%%%%%%%%%%%%%%%%%%%%%%%%%%%%%%%%%%%%%%%%

When presenting the results of a global fit we identify the maximum-likelihood point $\bf{\Theta}_\text{best-fit}$ in the DM EFT parameter space and map out the $1\sigma$ and $2\sigma$ confidence regions defined using the likelihood ratio $\mathcal{L}(\bf{\Theta}) / \mathcal{L}(\bf{\Theta}_\text{best-fit})$.
Thus, in cases where some region of the DM EFT parameter space can accommodate a modest excess in the collider data, other DM EFT parameter regions that might still perform better than the SM, or that are experimentally indistinguishable from SM, can appear as excluded. While this is perfectly reasonable, given that the comparison is to the best-fit DM EFT point and not to the SM expectation, it is also interesting to study the global fit results under the assumption that mild excesses in the collider data indeed do not originate from a true new physics signal. A simple and pragmatic approach is then to replace $\Delta \ln \mathcal{L}_{\text{LHC}}$ with a capped version,
\begin{align}
  \label{eq:LHC_loglike_capped}
  \Delta \ln \mathcal{L}_{\text{LHC}}^\text{cap}(\bf{s}) =
  \min[\mathit{\Delta} \ln \mathcal{L}_{\text{LHC}}(\bf{s}),
       \mathit{\Delta} \ln \mathcal{L}_{\text{LHC}}(\bf{s}=\bf{0})].
\end{align}
This will assign the same log-likelihood value, $\Delta \ln \mathcal{L}_{\text{LHC}}^\text{cap} = 0$, for all DM EFT parameter points whose prediction fit the collider data as well as, or better than, the SM prediction ($\bf{s} = \bf{0}$) does. Thus, analogous to how exclusion limits from LHC searches are constructed to only exclude new physics scenarios that predict too \textit{many} signal events, the capped likelihood only penalizes parameter points for performing worse than the background-only scenario. The result obtained from using $\Delta \ln \mathcal{L}_{\text{LHC}}^\text{cap}$ in a fit is therefore close to the result one would obtain by constructing a joint exclusion limit for the LHC searches, and applying this limit as a hard cut on the parameter space favoured by the other observables. The main difference is that the capped LHC likelihood incorporates a continuous likelihood penalty.\footnote{A practical benefit of having a continuous likelihood penalty rather than a hard cut is that it helps guide the parameter sampler towards the viable regions in the high-dimensional DM EFT parameter space.} A more detailed introduction to the capped likelihood construction can be found in Ref.~\cite{EWMSSM}.

Below we will present some results using this capped LHC likelihood, and some using the full LHC likelihood in Eq.\ (\ref{eq:LHC_loglike}). In light of the discussion above, the two sets of results should be interpreted as answering slightly different questions: The fit results with the full LHC likelihood show what DM EFT scenario is in best agreement with the complete set of current data, and how much worse other DM EFT scenarios perform in comparison. The results with the capped LHC likelihood map out the DM EFT parameter space that is preferred by the non-collider observables and not excluded by a combination of the LHC searches.

\subsection{Nuisance parameter likelihoods} \label{sec:nuisance}

In our scans we also vary a set of relevant nuisance parameters related to the DM observables and SM measurements. Most of these nuisance parameters are directly constrained by dedicated measurements, which we include through appropriate likelihood functions. In some cases, however, several conflicting measurements exist, indicating additional systematic uncertainties in the methodology. In these cases we constrain the nuisance parameters through effective likelihoods intended to give a conservative constraint on the allowed ranges. The nuisance parameters and $3\sigma$ ranges used in this study are summarised in Table~\ref{tab:nuis_params}. We briefly cover each nuisance likelihood in turn below.

We follow the default prescription in \darkbit for the local DM density $\rho_0$, where the likelihood is given by a log-normal distribution with central value $\rho_0 = 0.40$\,GeV\,cm$^{-3}$ and error $\sigma_{\rho_0}=0.15$\,GeV\,cm$^{-3}$. We scan over an asymmetric range in $\rho_0$ to reflect the log-normal distribution -- see Ref.~\cite{DarkBit} for more details.

We follow the same treatment of the Milky Way halo as in the \GB Higgs portal study~\cite{HP}. We utilise Gaussian likelihoods for parameters describing the Maxwell-Boltzmann velocity distribution, specifically the peak of the distribution $v_{\rm{peak}} = 240\, \pm \,8$\,km\,s$^{-1}$ \cite{Reid:2014boa}, and the galactic escape velocity $v_{\rm{esc}} = 528 \pm 25$\,km\,s$^{-1}$, based on the \emph{Gaia} data \cite{Deason:2019kgj}.

We employ a Gaussian likelihood for the running top quark mass in the $\overline{\textrm{MS}}$ scheme with a central value $m_t (m_t) = 162.9$\,GeV and an error $2.0$\,GeV \cite{Aad:2019mkw}.\footnote{This is based on taking an average of the asymmetric uncertainty $m_t (m_t) = 162.9^{+2.3}_{-1.6}$\,GeV; see table 2 in Ref.~\cite{Aad:2019mkw}.}~The top pole mass $(m_t^\textrm{pole})$ is then computed using the following formula:
\begin{equation}
  m_t^{\textrm{pole}} = m_t (m_t) \left[ 1 + \dfrac{4}{3} \dfrac{\alpha_s (m_Z)}{\pi} \right] \,.
\end{equation}
We use only the one-loop QCD corrections in this shift in order to be consistent with the procedure carried out in Ref.~\cite{Aad:2019mkw}. We have checked that the above expression gives the expected result for the top pole mass and matches well with Ref.~\cite{Aad:2019mkw}.

For direct detection, we employ nuisance parameter likelihoods for a number of hadronic input parameters that are used to evaluate form factors at the nuclear scale. Specifically, we use a product of four Gaussian likelihoods to include the constraints on $\sigma_{\pi N}$, $\Delta s$, $g_T^s$ and $r_s^2$ quoted in Table~\ref{tab:hadr:inputs}. The remaining hadronic input parameters are fixed to the central values given in Table~\ref{tab:hadr:inputs}.

%%%%%%%%%%%%%%%%%%%%%%%%%%%%%%%%%%%%%%%%%%%%%%%%%%%%%%%%%%%%%%%%%%%%%%%%%%%%%%%%%%%%%%%%%%%%%%%%%%%%%%%%%%%%%%%%%%%%%%%%
\section{Results}\label{sec:results}

We now present the results obtained from comprehensive scans of the parameter space introduced above. These scans were carried out with the differential evolution sampler \textsf{Diver v1.4.0} \cite{ScannerBit} using a population of $5 \times 10^4$ and a convergence threshold of either $10^{-5}$ or $3 \times 10^{-5}$. As we will analyse our scan results using profile likelihood maps, the sole aim of the scans is to map out the likelihood function in sufficient detail across the high-likelihood regions of parameter space. In particular, no statistical interpretation is associated with the density of parameter samples, and we can therefore combine samples from scans that use different metrics on the parameter space. To ensure that all parameter regions are properly explored, we perform two different types of scans:
\begin{itemize}
 \item \textbf{Full:} We explore DM masses up to the unitarity bound ($5 \, \mathrm{GeV} < m_\chi < 150\,\mathrm{TeV}$ and $20 \, \mathrm{GeV} < \La < 300 \, \mathrm{TeV}$).\footnote{We note that for the largest values of $m_\chi$ and $\La$ that we consider in these scans our approach of specifying all operators in the broken phase of electroweak symmetry and ignoring the effects of running between $\mu = \La$ and $\mu = m_Z$ becomes questionable. The constraints that we obtain above the TeV scale are therefore only approximate and should be interpreted with care.}~In these scans, $m_\chi$ and $\La$ are scanned on a logarithmic scale, while the Wilson coefficients are scanned on both a linear and a logarithmic scale (i.e.\ we combine the samples from both scanning strategies to achieve a thorough exploration of the whole parameter space).
 \item \textbf{Restricted:} We consider the parameter region where experimental constraints are most relevant ($m_\chi < 500\,\mathrm{GeV}$ and $\La < 2 \, \mathrm{TeV}$). In these scans the DM mass is scanned on a linear scale, the scale $\La$ on a logarithmic scale and the Wilson coefficients on a scale that is logarithmic on $[-4\pi,-10^{-6}]$, linear on $[-10^{-6},10^{-6}]$ and logarithmic on $[10^{-6},4\pi]$. This approach was found to achieve the optimum resolution of the LHC constraints while simultaneously ensuring that enough viable samples are also found for small $\La$ when some or all of the Wilson coefficients are tightly constrained.
\end{itemize}
All nuisance parameters are scanned on a linear scale. In the first set of scans, we fix the Wilson coefficients for all dimension-7 operators to zero, so that there are 6 model parameters and 8 nuisance parameters. The second set of scans then includes all 14 Wilson coefficients, bringing the total number of parameters up to 24. 

\begin{figure*}[t]
	\centering
	\includegraphics[width=\columnwidth]{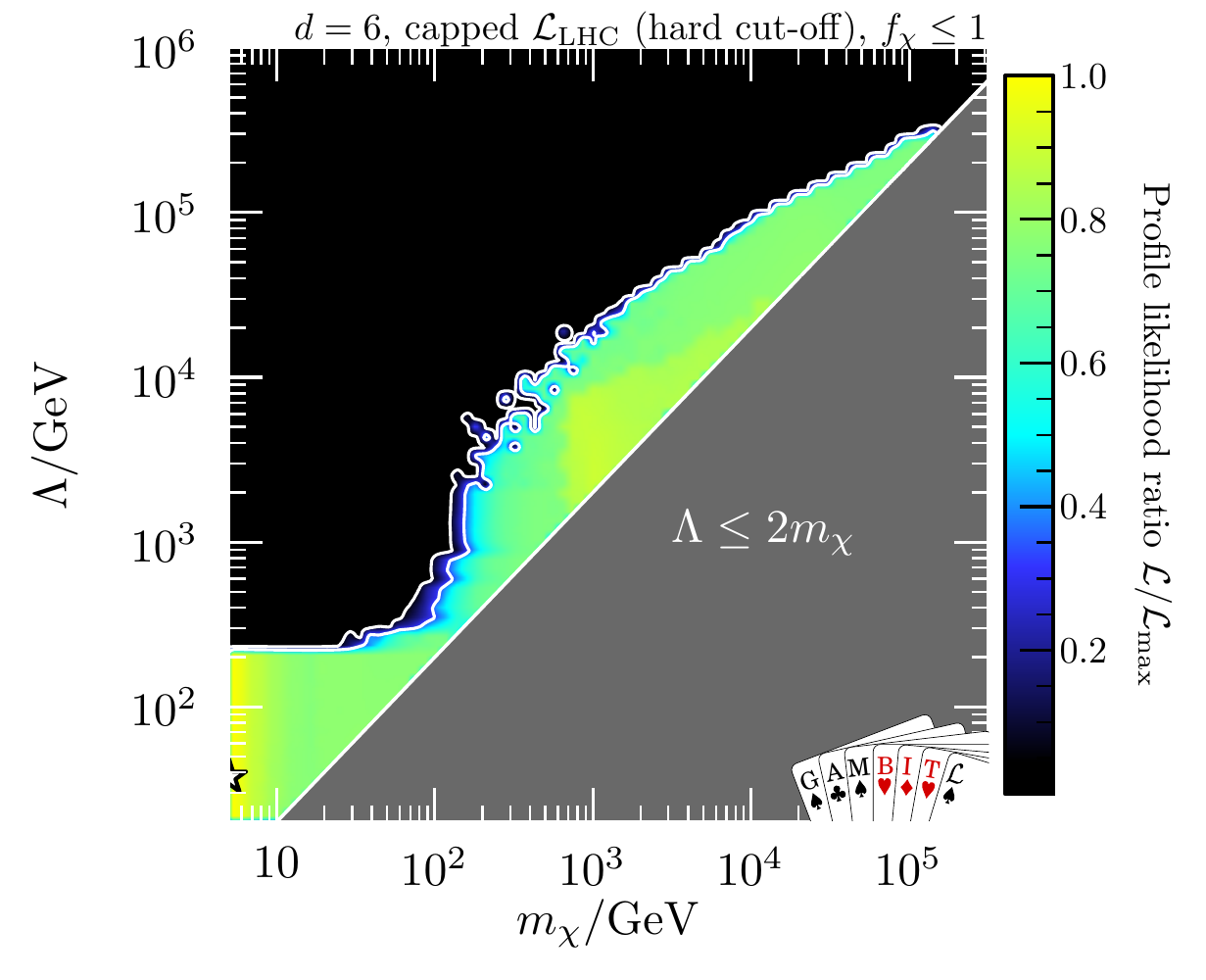}
	\includegraphics[width=\columnwidth]{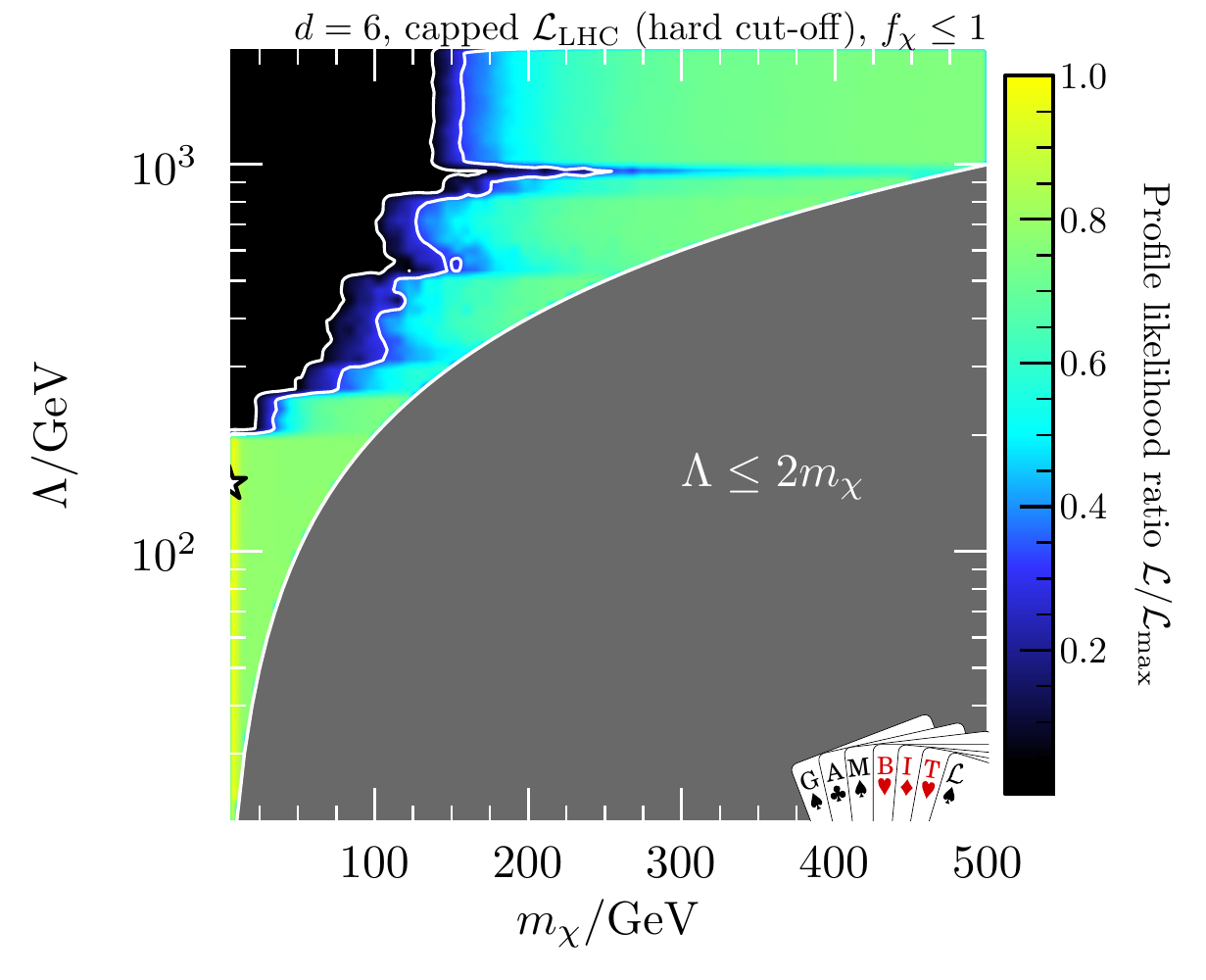}
	\caption{Profile likelihood in the $m_\chi$--$\La$ plane when considering only dimension-6 operators and capping the LHC likelihood at the value of the background-only hypothesis. The white contours indicate the $1\sigma$ and $2\sigma$ confidence regions and the best-fit point is indicated by the white star. The shaded region (corresponding to $\La \leq 2 m_\chi$) is excluded by the EFT validity requirement. In the right panel, the parameter ranges have been restricted to the most interesting region. Note that the position of the best-fit points in the two panels is somewhat arbitrary, as there is a degeneracy between $\La$ and $\mathcal{C}_{3,4}^{(6)}$ and hence the likelihood is essentially constant across the entire yellow region (see also Fig.~\ref{fig:dim_6_capped_coefficients}).}
	\label{fig:dim_6_capped_main}
\end{figure*}

We furthermore consider a number of variations in the constraints that we include in our scans:
\begin{itemize}
 \item We perform scans where the DM particle is allowed to be a sub-component ($f_\chi \leq 1$) and scans where we require that the DM relic density be saturated ($f_\chi \approx 1$), see Sec.~\ref{sec:rd};
 \item We perform scans with both the capped LHC likelihood and the full LHC likelihood (see Sec.~\ref{sec:collider});
 \item When considering the full LHC likelihood, we furthermore apply two different prescriptions for imposing the EFT validity: a hard cut-off and a smooth cut-off (see Sec.~\ref{sec:validity}).
\end{itemize}
Unless explicitly stated otherwise, our default choices for the discussion below are to allow a DM sub-component and consider the capped LHC likelihood with a hard cut-off.

\subsection{Capped LHC likelihood}

Let us begin with the case that the LHC likelihood is capped, i.e.\ it cannot exceed the likelihood of the background-only hypothesis. We first consider only dimension-6 operators with different requirements for the DM relic density, and then also include dimension-7 operators.

\subsubsection*{Dimension-6 operators only (relic density upper bound)}

Our main results for this case are shown in Fig.~\ref{fig:dim_6_capped_main} in terms of the DM mass and the new physics scale $\La$. The left panel corresponds to the full parameter range, whereas the right panel provides a closer look at the most interesting parameter region. We find a large viable parameter space but also a number of notable features. For large values of $m_\chi$ and $\La$, the allowed parameter space is determined by the EFT validity requirement $\La > 2 m_\chi$ and the relic density requirement which, combined with the perturbativity bound on the Wilson coefficients, implies an upper bound on $\La$ for given $m_\chi$. These different constraints are compatible only for $m_\chi < 150 \, \mathrm{TeV}$, implying an upper bound on the scale of new physics of $\La < 300 \, \mathrm{TeV}$. This limit corresponds to the well-known unitarity bound for thermal freeze-out~\cite{Griest:1989wd}.

The zoomed-in version in the right panel reveals a number of additional features. In the top-left corner (small $m_\chi$, large $\La$), there are strong constraints from the LHC, which make it impossible to satisfy the relic density requirement. These constraints become weaker as $\La$ decreases and the EFT can only be trusted for smaller values of $\slashed{E}_T$. The various sharp features correspond to the points where $\La$ crosses the boundary of a specific $\slashed{E}_T$ bin, leading to a jump in the likelihood. In our conservative approach, LHC constraints are completely absent for $\La < 200 \, \mathrm{GeV}$. Finally, we find that there is a slight upward fluctuation in \emph{Fermi}-LAT data, which can be fitted for $m_\chi = 5.0 \,\mathrm{GeV}$ and $f_\chi^2 \langle \sigma v \rangle_0 = 1.1 \times 10^{-27} \, \mathrm{cm^3 \, s^{-1}}$.\footnote{We emphasize that, although the best-fit point lies close to the boundary of the parameter space, there is no preference for even smaller values of the DM mass and hence our findings would not change when extending the scan range.}

We emphasize that a great advantage of our approach is that we treat the new-physics scale $\La$ as an independent parameter, which is kept explicit in Fig.~\ref{fig:dim_6_capped_main} (rather than being profiled out like the individual Wilson coefficients). This makes it possible in a straight-forward way to distinguish those parameter regions where the EFT predictions can be considered robust and those parameter regions where additional constraints may apply. As discussed in Sec.~\ref{sec:validity}, the EFT is expected to be valid if $\La$ is sufficiently greater than the largest $p_T$ bin considered in the LHC analyses, i.e.\ $\La > 1.3 \, \mathrm{TeV}$. Conversely, for $\La < 200 \, \mathrm{GeV}$ we conservatively suppress constraints from the LHC, such that the viable parameter regions found in this range must be interpreted with great care. For intermediate values of $\La$, LHC constraints are being applied but may depend on the specific UV completion. Which of these parameter regions is considered most interesting depends on the specific context and is left to the reader.

\begin{figure}[p]
	\centering

	\includegraphics[width=\columnwidth]{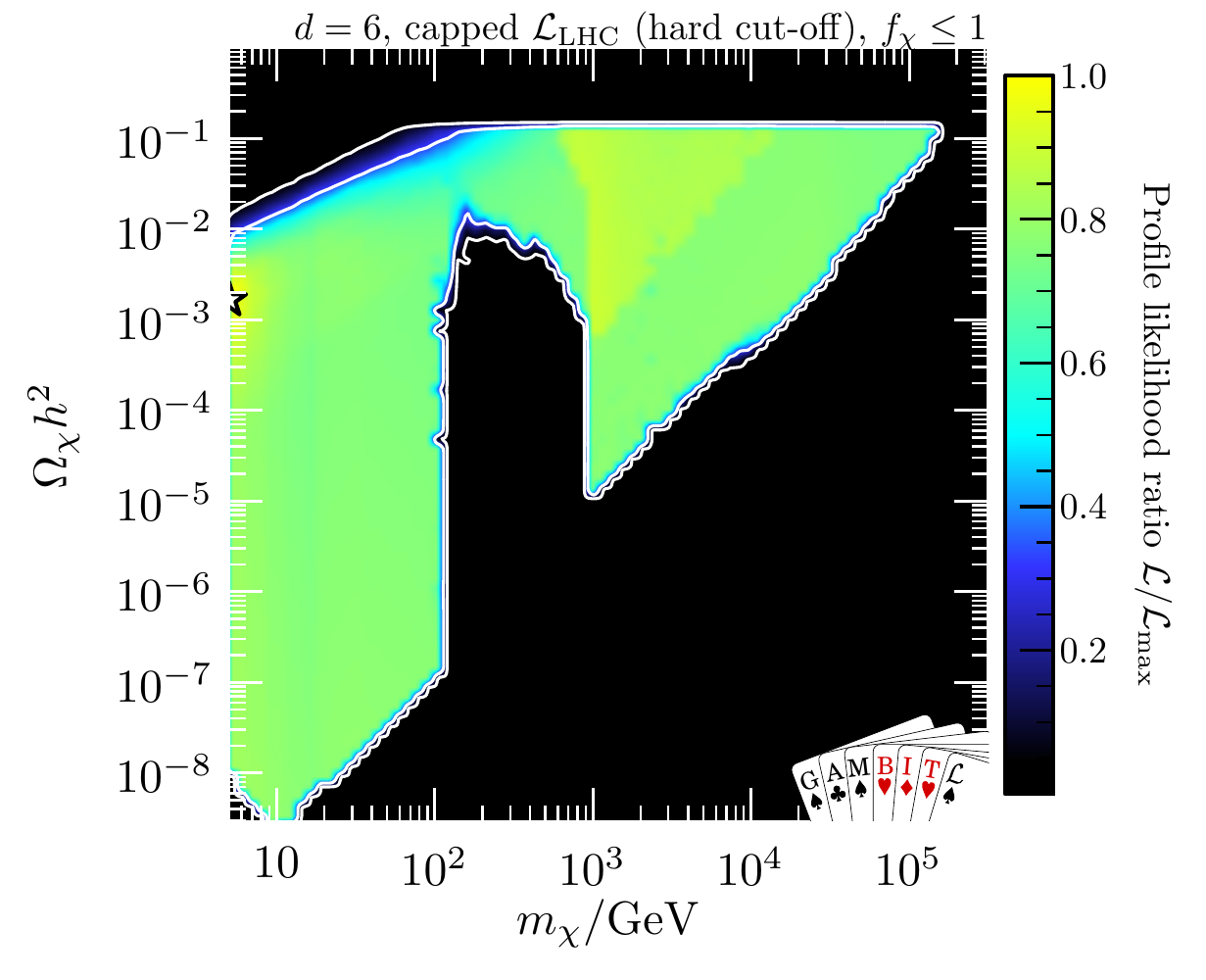}

	\includegraphics[width=\columnwidth]{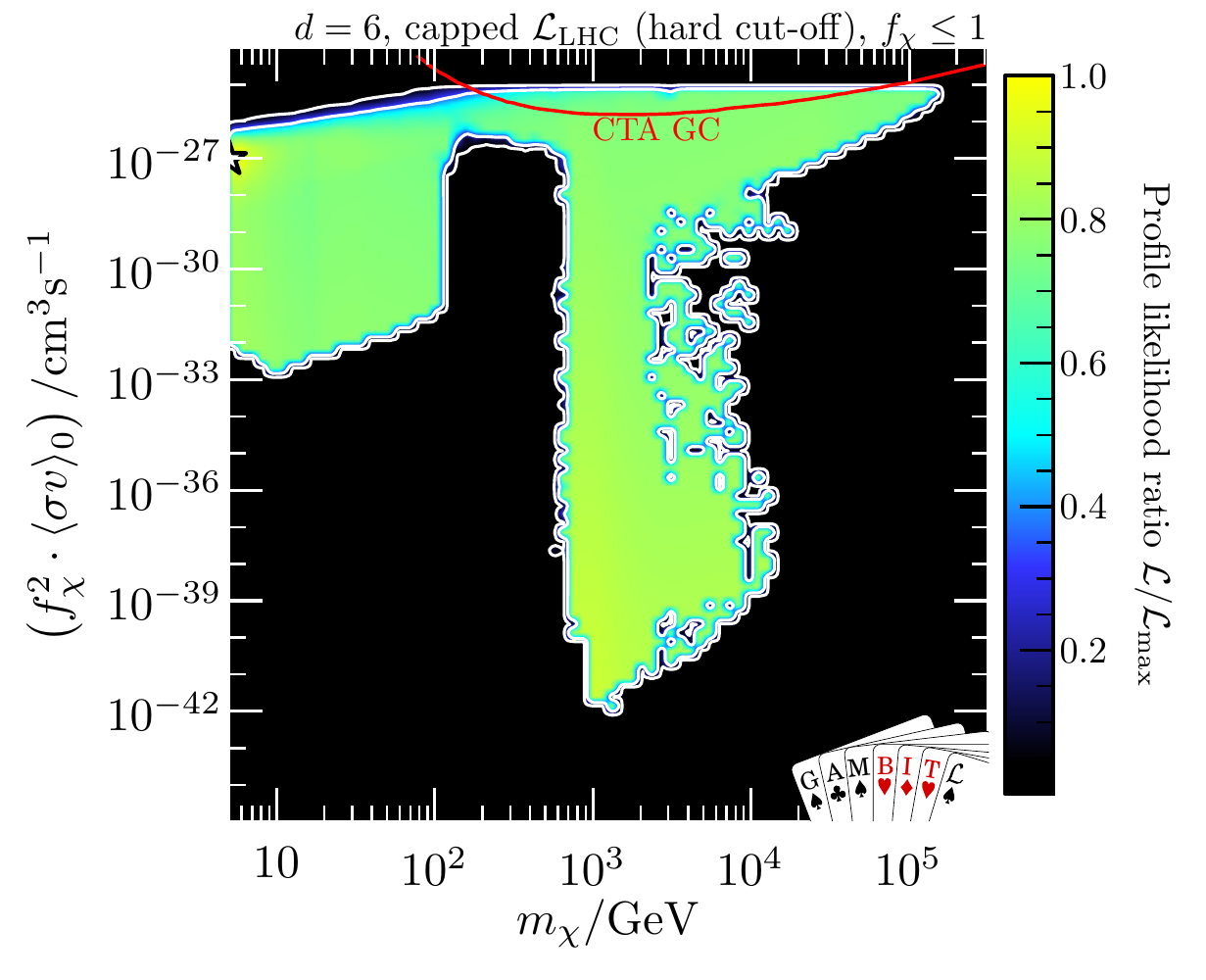}

	\includegraphics[width=\columnwidth]{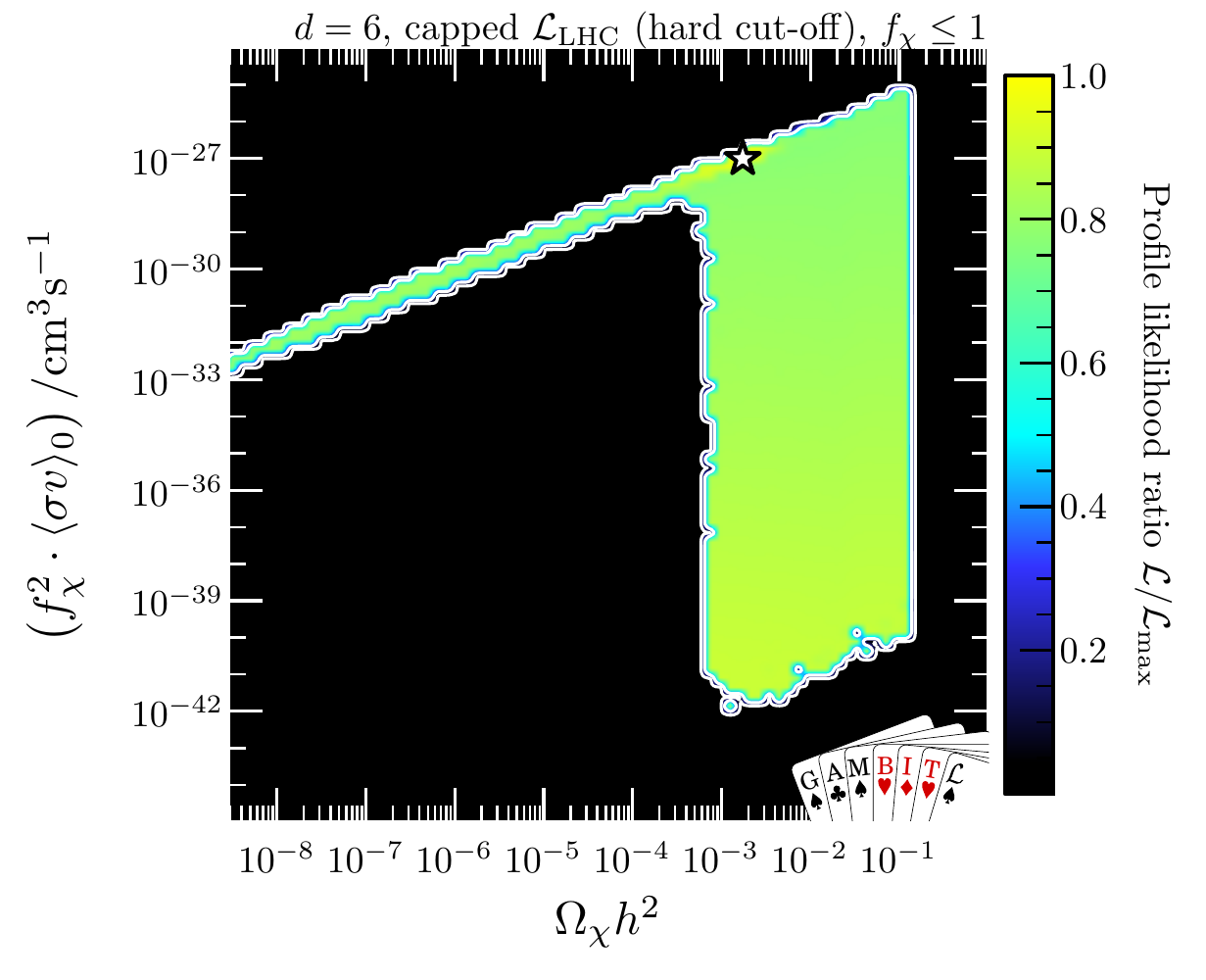}

	\caption{Profile likelihood in terms of the DM mass, the relic density and the rescaled annihilation cross-section. As in Fig.~\ref{fig:dim_6_capped_main}, we consider only dimension-6 operators and cap the LHC likelihood at the value of the background-only hypothesis. The solid red line in the middle panel denotes the ``initial construction'' projection sensitivity of Cherenkov Telescope Array (CTA) towards the Galactic Centre (GC) for the $b\bar{b}$ final state~\cite{Acharyya:2020sbj}.}
	\label{fig:dim_6_capped_relic}
\end{figure}

\begin{figure*}[t]
	\centering
	\includegraphics[width=\columnwidth]{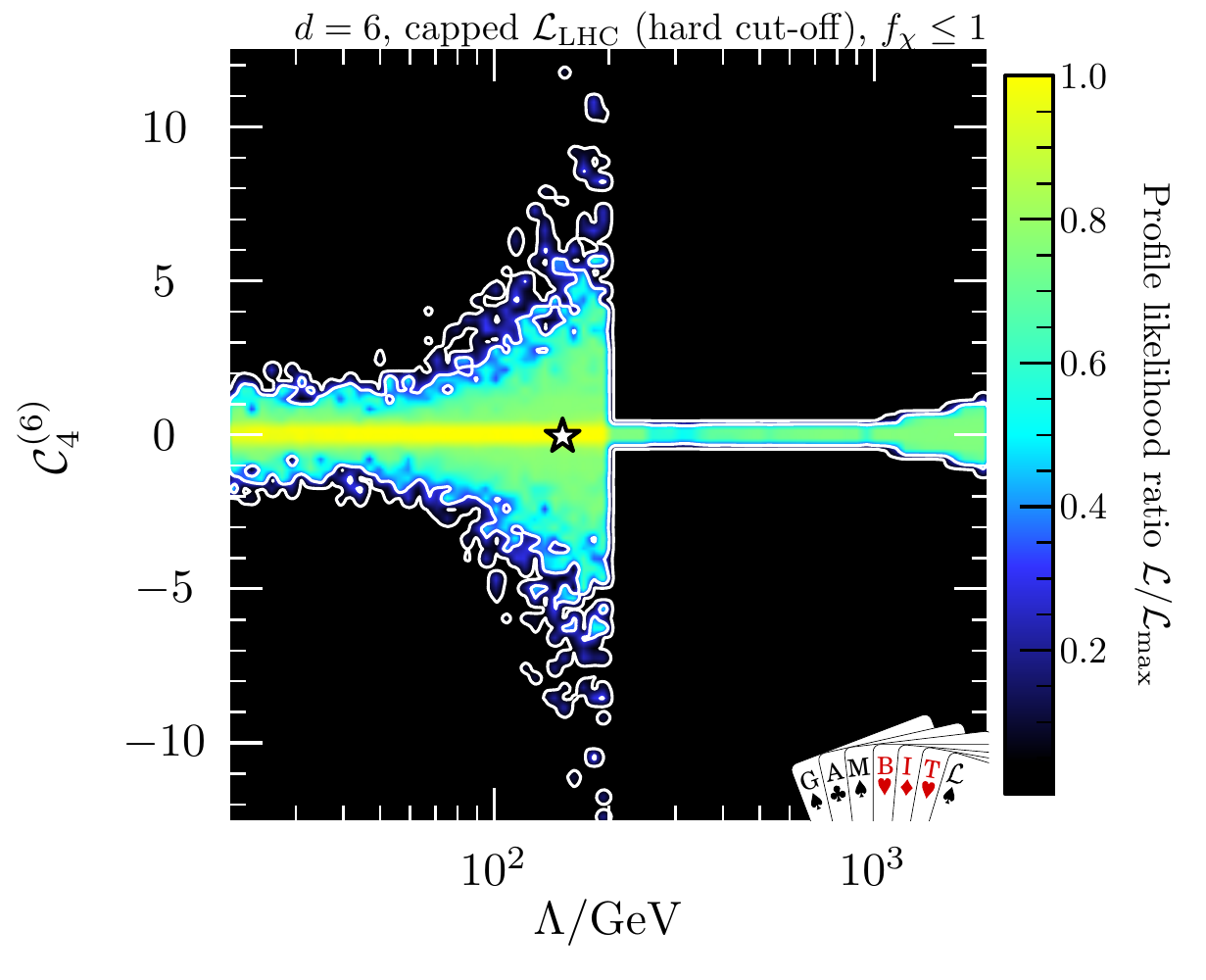}\hfill
	\includegraphics[width=\columnwidth]{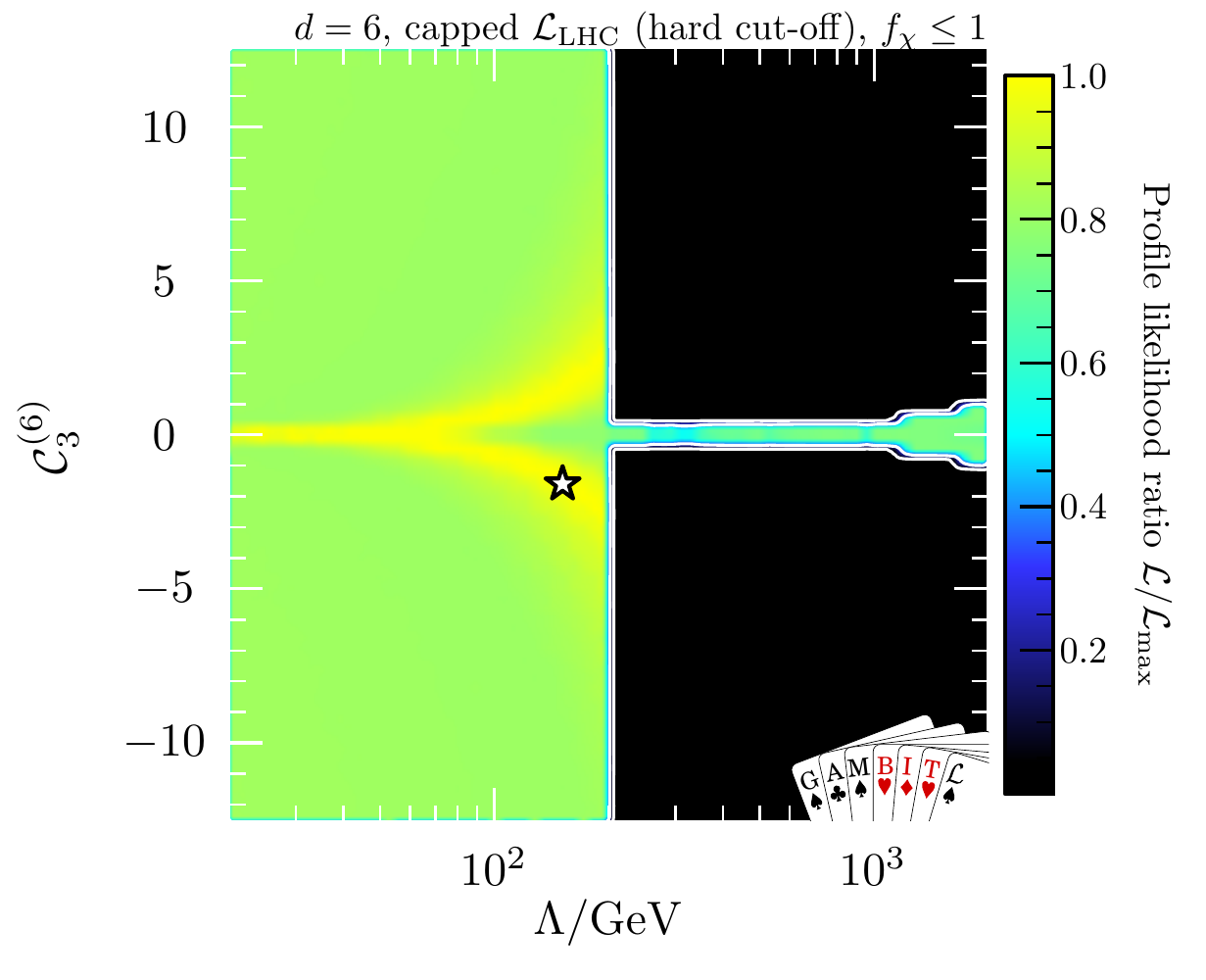}
	\caption{Profile likelihood in the $\La$--$\mathcal{C}^{(6)}_4$ plane (left) and the $\La$--$\mathcal{C}^{(6)}_3$ plane (right) for the restricted parameter ranges. As in Fig.~\ref{fig:dim_6_capped_main}, we only consider dimension-6 operators and cap the LHC likelihood at the value of the background-only hypothesis. The contour lines show the $1\sigma$ and $2\sigma$ confidence regions. Note that the position of the best-fit points (white stars) in the two panels is somewhat arbitrary, as there is a degeneracy between $\La$ and $\mathcal{C}_{3,4}^{(6)}$ and hence the likelihood is essentially constant across the entire yellow region.} 
	\label{fig:dim_6_capped_coefficients}
\end{figure*}

A complementary perspective is provided in Fig.~\ref{fig:dim_6_capped_relic}, which shows the allowed parameter regions in terms of the DM mass, the relic density and the rescaled annihilation cross-section.
A number of additional features become apparent in these plots. First, for $m_\chi \lesssim 100\,\mathrm{GeV}$ it is impossible to saturate the observed DM relic density, $\Omega_\text{DM} h^2 = 0.12$, due to the combined constraints from direct and indirect detection experiments. However, these constraints are suppressed for DM sub-components, such that it is possible to have very small relic densities in this mass region. For $m_\chi > 100 \, \mathrm{GeV}$ (corresponding to $\La > 200 \, \mathrm{GeV}$), on the other hand, constraints from the LHC become relevant, which are not suppressed for DM sub-components. These constraints are then again relaxed for $m_\chi \gtrsim 1 \, \mathrm{TeV}$ as the LHC energy becomes insufficient to produce a pair of DM particles.

For $m_\chi \lesssim 1 \, \mathrm{TeV}$, we find that there is a direct correspondence between $\Omega_\chi h^2$ and the rescaled annihilation cross-section $f_\chi^2 \langle \sigma v \rangle_0$. This is because the operators that induce $p$-wave annihilations (in particular $\mathcal{C}^{(6)}_2$) are strongly constrained by the LHC and direct detection experiments, and the annihilation cross-section is therefore always dominated by the $s$-wave contribution. For larger DM masses, it becomes possible for the $p$-wave contribution to dominate the relic density calculation, such that the total annihilation cross-section is velocity-dependent and becomes tiny in the present universe.
While indirect detection experiments presently cannot probe the relevant parameter space for TeV-scale DM,
it is worth stressing that CTA will be able to do so for operators that induce $s$-wave annihilation.
We illustrate this in Fig.~\ref{fig:dim_6_capped_relic} by indicating the sensitivity of CTA to a DM signal from the
galactic center~\cite{Acharyya:2020sbj} (for simplicity based on the assumption of $b\bar{b}$ final states,
noting that {\it any} hadronic DM annihilation channel results in very similar gamma-ray spectra
at these energies). We note that the CTA sensitivity indicated in Fig.~\ref{fig:dim_6_capped_relic} is based
on assuming a standard Einasto profile as expected for WIMP DM; if the DM density in the galactic center is
instead roughly constant, the sensitivity can worsen by up to about one order of
magnitude~\cite{Acharyya:2020sbj}.

Let us finally consider the allowed parameter space in terms of the Wilson coefficients. The coefficient $\mathcal{C}^{(6)}_1$ gives rise to spin-independent scattering, which is very strongly constrained by direct detection experiments. Thus, this coefficient is required to be so small that it cannot give a sizeable contribution to any other process. The coefficient $\mathcal{C}^{(6)}_4$, on the other hand, gives rise to spin-dependent interactions, for which constraints are significantly weaker. We show the allowed parameter regions for this coefficient in the left panel of Fig.~\ref{fig:dim_6_capped_coefficients}. The observed mirror symmetry results from the fact that all experimental predictions (and hence the likelihoods) are invariant under a global sign change of all Wilson coefficients. For $\La < 200\,\mathrm{GeV}$, all constraints are furthermore invariant under the rescaling $\mathcal{C} \to \alpha^2 \mathcal{C}$, $\La \to \alpha \La$, which explains why the allowed parameter region grows with increasing $\La$. For $\La > 200 \, \mathrm{GeV}$, 
LHC constraints become relevant and strongly constrain the magnitude of the coefficient. Very similar results are obtained for the coefficient $\mathcal{C}^{(6)}_2$, which gives rise to a momentum-suppressed spin-independent scattering (see Table~\ref{tab:op_suppression}).

In the right panel of Fig.~\ref{fig:dim_6_capped_coefficients}, we show the allowed parameter region in terms of $\mathcal{C}^{(6)}_3$, which induces scattering that is \emph{simultaneously} momentum-suppressed and spin-dependent, such that direct detection constraints are very weak. Correspondingly, we find
that this coefficient is largely unconstrained
for $\La < 200\,\mathrm{GeV}$. We also identify this coefficient as giving the main contribution for fitting the \emph{Fermi}-LAT excess. For larger values of $\La$, on the other hand, the constraints are very similar to the ones for $\mathcal{C}^{(6)}_{2,4}$ as the LHC only has limited sensitivity to distinguish the spin structure of the operators.

\subsubsection*{Dimension-6 operators only (relic density saturated)}

\begin{figure}[t]
	\centering
	\includegraphics[width=\columnwidth]{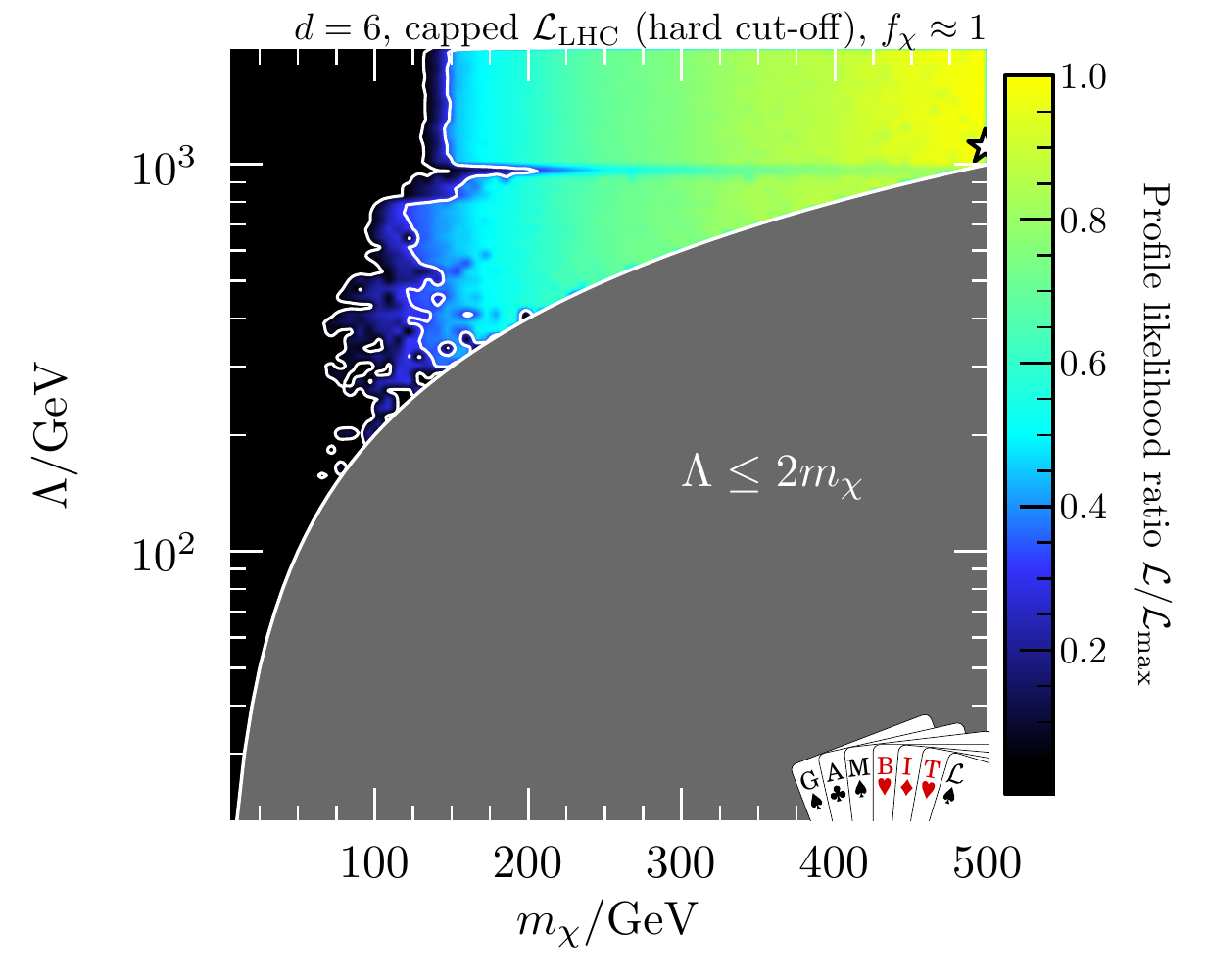}
	\caption{Same as the right panel of Fig.~\ref{fig:dim_6_capped_main} but requiring the DM relic density to be saturated (rather than imposing an upper bound only).}
	\label{fig:dim_6_capped_RF1}
\end{figure}

Next we consider the case where the relic density constraint is imposed not only as an upper limit but as an actual measurement, i.e.,\ the DM particle under consideration is required to account for all of the DM in the universe via the effective interactions that we consider. We show in Fig.~\ref{fig:dim_6_capped_RF1} the allowed parameter space in the restricted $m_\chi$--$\La$ plane when considering a capped LHC likelihood, i.e.\ the same likelihoods as in Fig.~\ref{fig:dim_6_capped_main} apart from the modified relic density requirement. As expected from the top row of Fig.~\ref{fig:dim_6_capped_relic}, it is not possible to saturate the observed relic density for $m_\chi \lesssim 100 \, \mathrm{GeV}$. The reason is that for such small DM masses the relic density requirement is incompatible with \emph{Fermi}-LAT and CMB bounds on the annihilation cross-section for operators that predict dominantly $s$-wave annihilation ($\Q{1,q}{6}$ and $\Q{3,q}{6}$), and incompatible with direct detection and LHC constraints for $\Q{2,q}{6}$ and $\Q{4,q}{6}$.

Constraints from direct and indirect detection experiments are also responsible for the preference for larger DM masses visible in Fig.~\ref{fig:dim_6_capped_RF1}. In particular, the \emph{Fermi}-LAT likelihood pushes the best-fit point towards the boundary $m_\chi = 500 \, \mathrm{GeV}$. We find the likelihood of the best-fit point to be slightly worse than for the background-only hypothesis: $2 \Delta \ln \mathcal{L} \equiv 2(\ln \mathcal{L}^{\mathrm{best-fit}} - \ln \mathcal{L}^\text{ideal}) = -0.5$. Extending the range of the scan to larger DM masses would allow the model to fully evade the \emph{Fermi}-LAT constraint. This would shift the best-fit point and the allowed parameter regions to slightly larger DM masses
without changing the remaining conclusions (see also Fig.~\ref{fig:dim_6_capped_relic}).

\begin{figure}[t]
	\centering
	\includegraphics[width=\columnwidth]{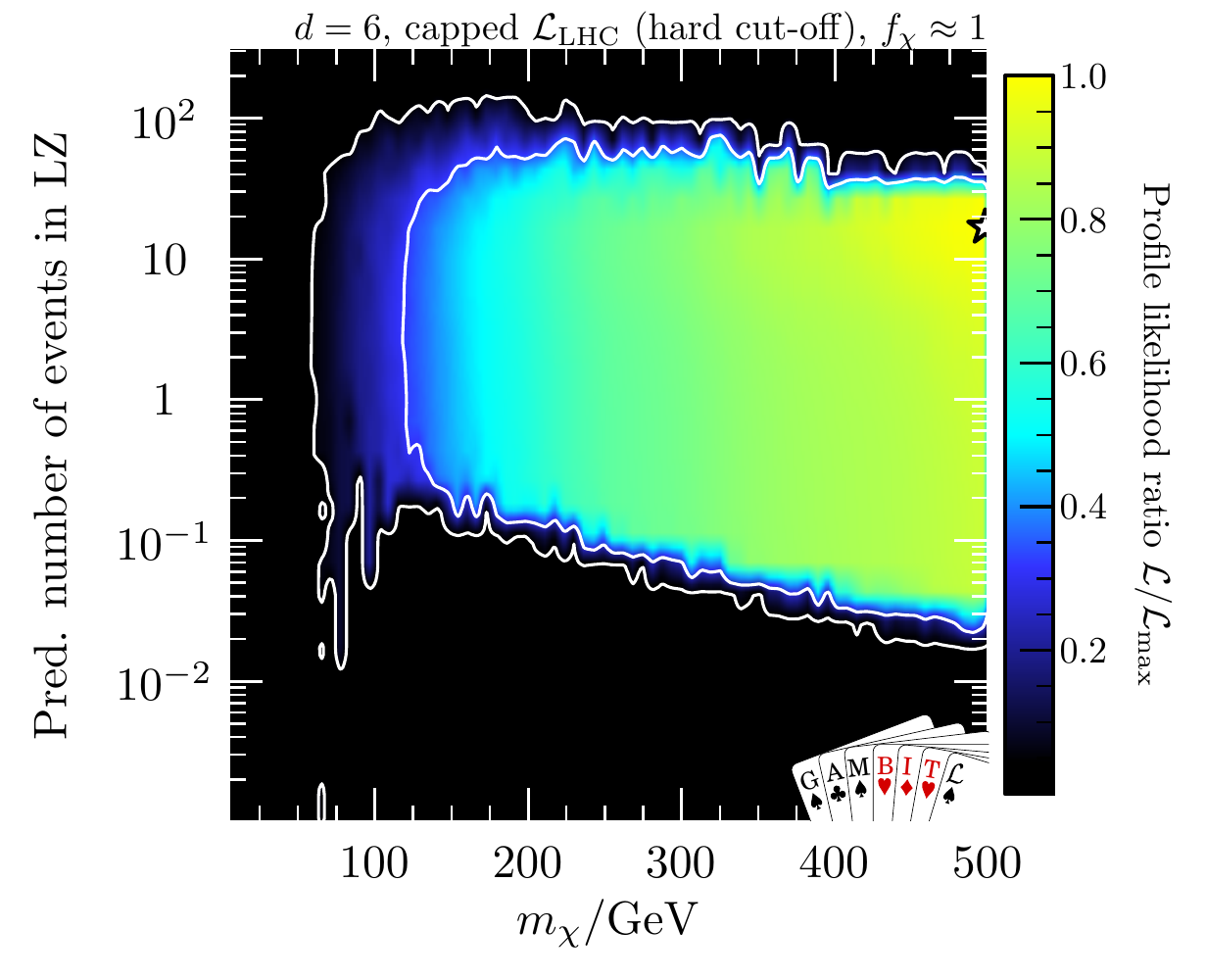}
	\caption{Profile likelihood in terms of $m_\chi$ and the predicted number of signal events in the LZ experiment when $\chi$ accounts for all of the observed DM abundance (as in Fig.~\ref{fig:dim_6_capped_RF1}).}
	\label{fig:dim_6_capped_RF1_LZ_events}
\end{figure}

For a complementary view of the parameter space, we show in Fig.~\ref{fig:dim_6_capped_RF1_LZ_events} the predicted number of signal events in the next-generation direct detection experiment LZ~\cite{Akerib:2018lyp} as a function of the DM mass. Due to the various different operators contributing to the DM-nucleus scattering, the predicted number of signal events is a more useful quantity to consider than the DM-nucleon scattering cross-section at zero momentum transfer. The predicted number of events corresponds to nuclear recoil energies in the search window $[6 \, \mathrm{keV}, 30 \, \mathrm{keV}]$ and assumes an exposure of $5.6 \times 10^6 \, \mathrm{kg\,days}$ and 50\% acceptance for nuclear recoils (see Ref.~\cite{HP} for details on our implementation of LZ).

We find that of the order of 10 events are predicted around the best-fit point, which requires a non-zero contribution from the operator $\Q{2}{6}$ leading to spin-independent (but momentum-suppressed) scattering. However, the predicted number of events varies significantly within the allowed region of parameter space and can be as small as 0.1 at 68\% confidence level. In this case the main contribution arises from the mixing of the operator $\Q{3}{6}$ into the operator $\Q{1}{6}$ as given in Eq.~(\ref{eq:yt}).\footnote{We note that this mixing effect could in principle be cancelled by contributions from additional effective operators not included in our analysis, such that even smaller event rates may be achievable.} While such an event number is too low to be detected with next-generation experiments, it is still well above the neutrino background and should be observable with more ambitious future detectors such as DARWIN~\cite{Aalbers:2016jon} or DarkSide-20k~\cite{Aalseth:2017fik}.

%\bigskip

Another interesting approach would be to not perform a relic density calculation at all and simply assume that $f_\chi = 1$ is achieved through some modification of early universe cosmology. In this case it would also be possible to consider $\La < 2 m_\chi$ since the calculation of the annihilation cross-section is unnecessary. However, since none of the other likelihoods that we consider give a strong preference for a DM signal, there would then be no lower bound on the interaction strength of the DM particle, i.e.\ it would be possible for all Wilson coefficients to vanish simultaneously. Hence we expect all combinations of $m_\chi$ and $\La$ to be viable in this approach, and we do not explore this direction further in the present work.

\begin{figure*}[t]
	\centering
	\includegraphics[width=\columnwidth]{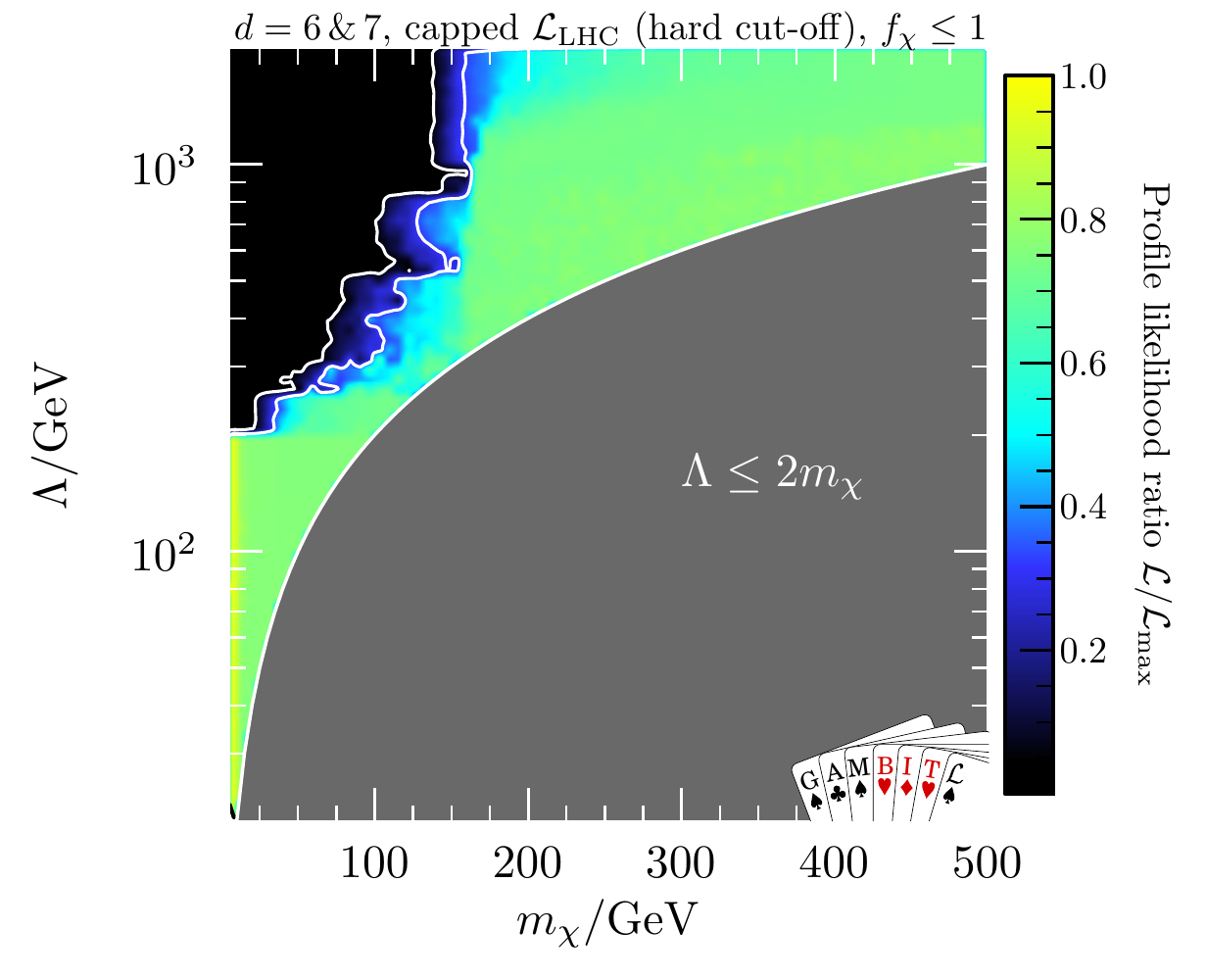} \hfill
	\includegraphics[width=\columnwidth]{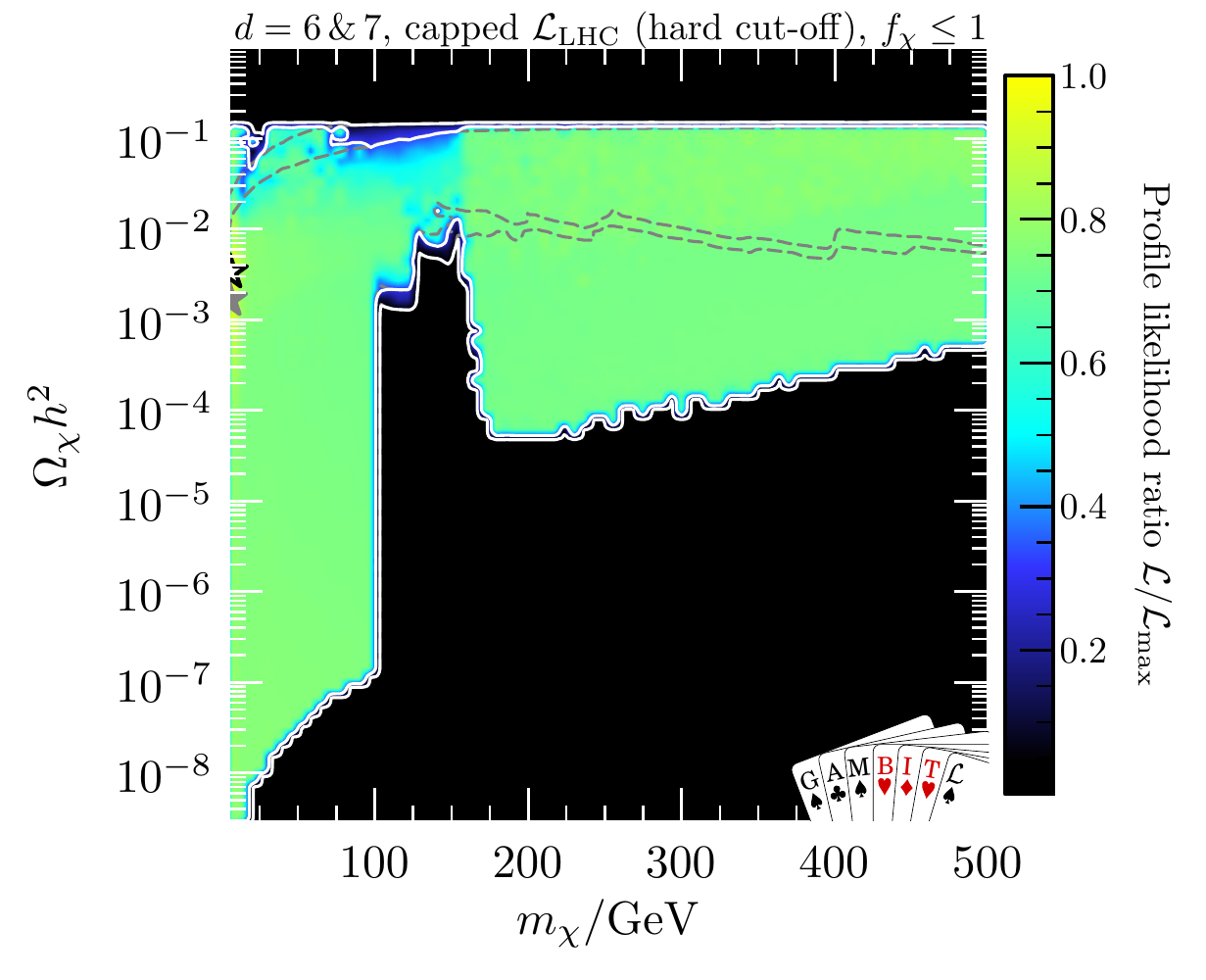}
	\caption{Profile likelihood in the $m_\chi$--$\La$ plane (left) and in terms of $m_\chi$ and the predicted relic density (right) when considering all dimension-6 and dimension-7 operators, and capping the LHC likelihood at the value of the background-only hypothesis. The white contours show the $1\sigma$ and $2\sigma$ confidence regions and the white star marks the best-fit point. For comparison, we also show the $1\sigma$ and $2\sigma$ confidence region contours (dashed grey lines) and best-fit point (grey star) for the case of dimension-6 operators only in the right panel (see also Fig.~\ref{fig:dim_6_capped_relic}).}
	\label{fig:dim_6_7_capped_main}
\end{figure*}

\subsubsection*{Operators up to dimension 7 (relic density upper bound)}

We now turn to the case where we simultaneously consider all dimension-6 operators as well as the dimension-7 operators involving DM particles and quarks or gluons introduced in Sec.~\ref{sec:model}.
We remind the reader that we neglect additional dimension-7 operators involving Higgs bosons that would arise in theories respecting unbroken electroweak symmetry (which are phenomenologically irrelevant) as well as operators with derivative interactions (which largely give redundant information). Even with these restrictions our analysis requires 24-dimensional (16 model + 8 nuisance) parameter scans.

In Fig.~\ref{fig:dim_6_7_capped_main}, we show the allowed regions in the $m_\chi$-$\La$ plane (left) and in the $m_\chi$--$\Omega_\chi h^2$ plane (right) when using the capped LHC likelihood. As before, we find that the parameter region at small $m_\chi$ and $\La$ can fit the slight \emph{Fermi}-LAT excess with best-fit values: $m_\chi = 5.5$\,GeV and $f_\chi^2 \langle \sigma v \rangle_0 = 1.9 \times 10^{-27}$\,cm$^{3}$\,s$^{-1}$.

As the inclusion of additional parameters can only increase the profile likelihood, we expect the allowed regions of parameter space to be larger than the ones found above. Interestingly, the differences between the left panel of Fig.~\ref{fig:dim_6_7_capped_main} and the right panel of Fig.~\ref{fig:dim_6_capped_main} are rather minimal. In other words, the inclusion of the 10 additional dimension-7 operators does not open up new parameter space in terms of $m_\chi$ and $\La$. This is of course expected for the parameter region with large $m_\chi$ and small $\La$ (bottom-right), which is excluded by the EFT validity constraint but surprising for the region with small $m_\chi$ and large $\La$ (top-left), which is excluded by the combination of the LHC constraints and the relic density requirement.

The reason why this parameter space remains inaccessible is that the gluon operators $\Q{1\text{--}4}{7}$ are strongly constrained by the LHC for $\La > 200 \, \mathrm{GeV}$ and can therefore not contribute significantly to the annihilation cross-section. The quark operators $\Q{5\text{--}10,q}{7}$, on the other hand, are unconstrained by the LHC, but for $m_\chi < m_t$, the resulting annihilation cross-section is suppressed by a factor
%proportional to
$m_b^2 m_\chi^2 / \La^6$, and therefore too small to produce a relic abundance that evades the upper bound from the relic density requirement given the perturbativity bound on the Wilson coefficients.

\begin{figure*}[t]
	\centering
	\includegraphics[width=\columnwidth]{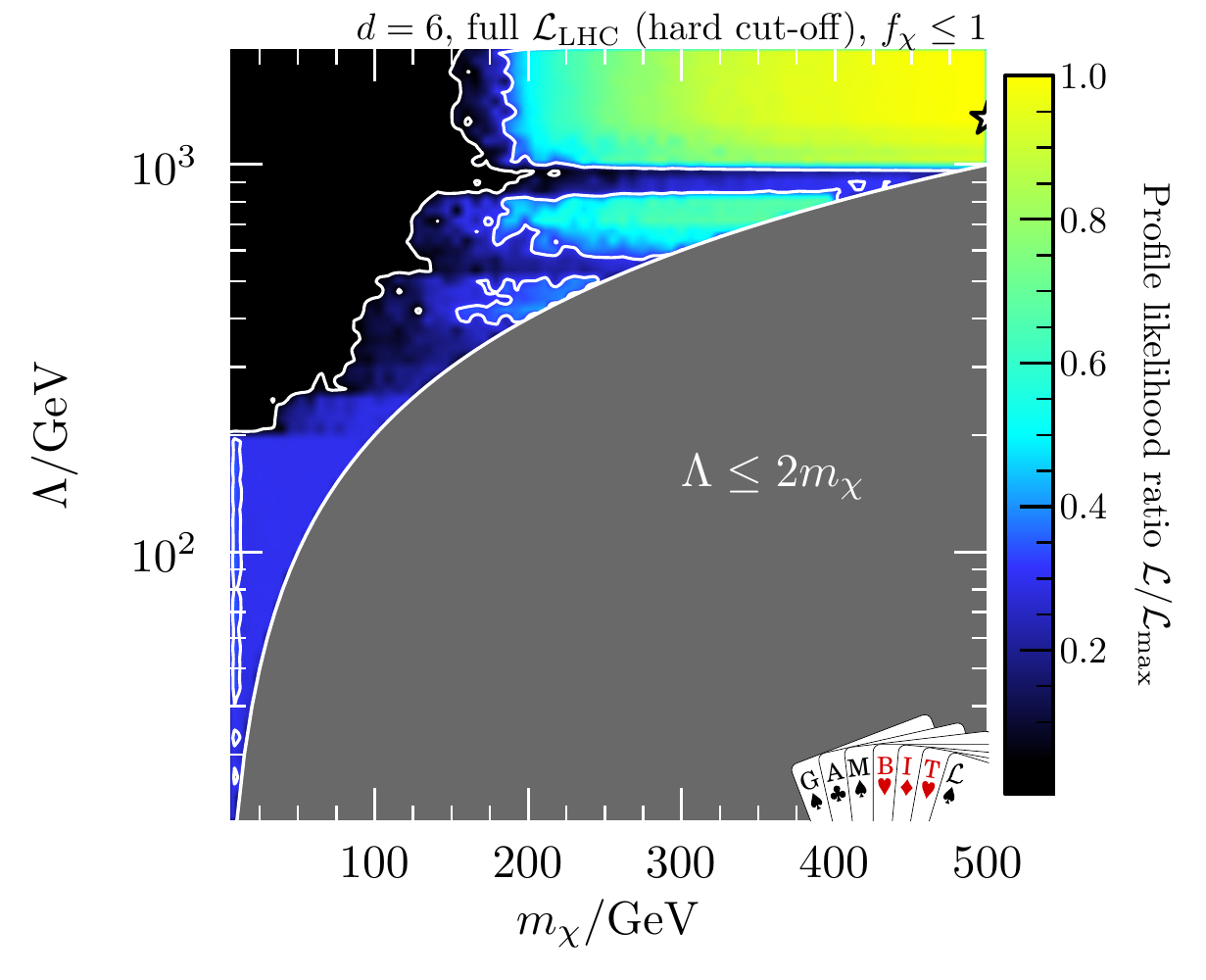} \hfill
	\includegraphics[width=\columnwidth]{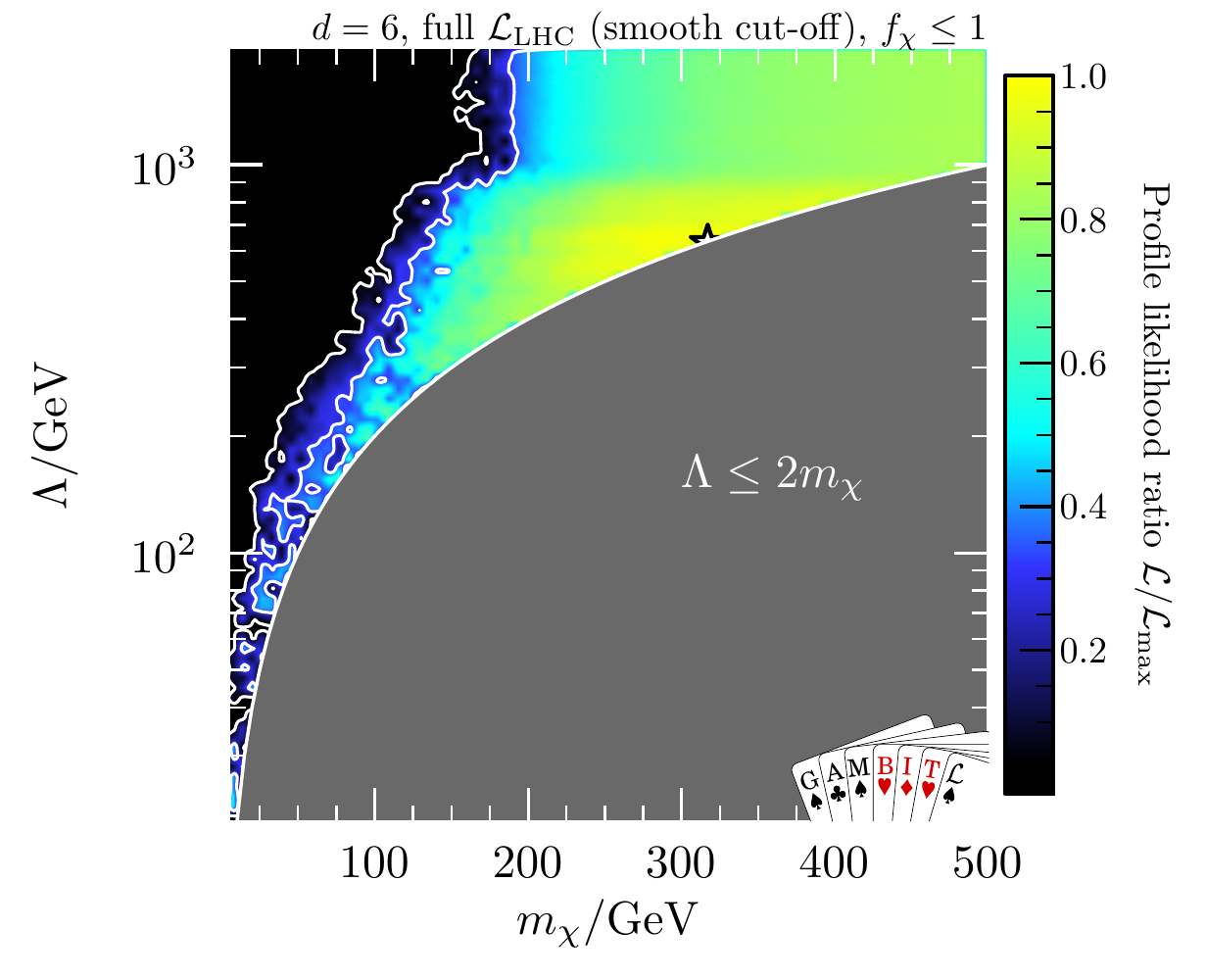}
	\caption{Profile likelihood in the $m_\chi$--$\La$ parameter plane when considering only dimension-6 operators and including the full LHC likelihood. In the left (right) panel, we impose a hard (smooth) cut-off in the predicted missing energy spectrum for $\slashed{E}_T > \La$ (see text for details).}
	\label{fig:dim_6_full_main}
\end{figure*}

\begin{figure*}[t]
	\centering
	\includegraphics[width=1.3\columnwidth]{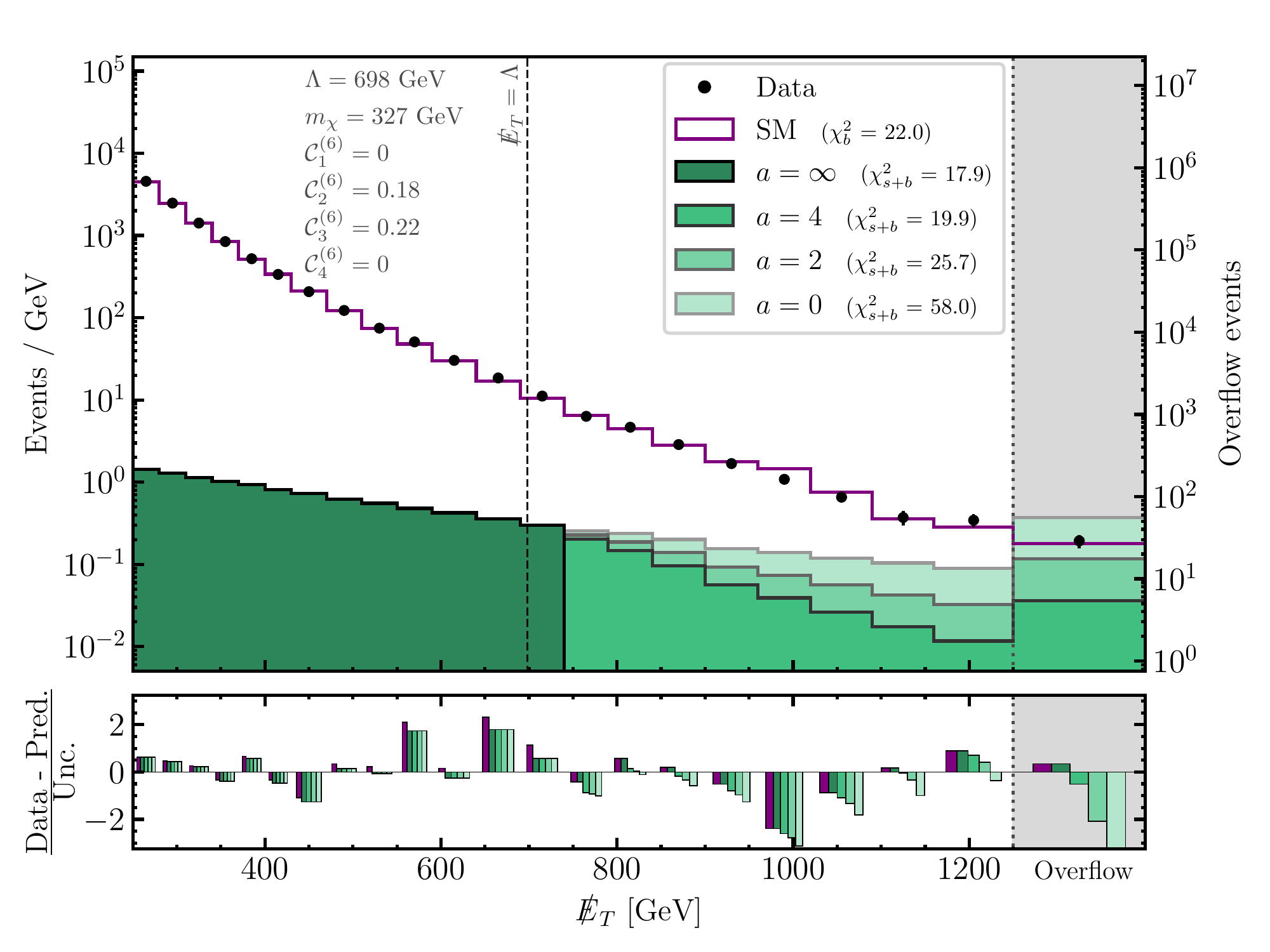} \hfill
	\caption{ \textit{Top panel:} Examples of missing energy spectra for the CMS monojet search~\cite{Sirunyan:2017jix}, illustrating different choices for imposing an EFT validity requirement on the signal prediction. For $\slashed{E}_T > \La$, we scale the $\slashed{E}_T$ signal spectrum with the factor $(\slashed{E}_T/\La)^{-a}$ as described in Sec.~\ref{sec:validity}. The green distributions show the resulting signal predictions for four different choices of $a$, from $a = 0$ (lightest green), corresponding to no modification of the spectrum, to $a \rightarrow \infty$ (darkest green), which removes any signal contribution in $\slashed{E}_T$ bins above $\slashed{E}_T = \La$. The SM background prediction (purple) and the observed event counts (black points) are taken from Ref.~\cite{Sirunyan:2017jix}. The last bin, starting at $\slashed{E}_T=1250\, \mathrm{GeV}$, contains any overflow and is thus not normalised to a given bin width.
    \textit{Bottom panel:} A small bar chart per bin showing the pulls, defined as $(\textrm{data} - \textrm{prediction}) / \textrm{uncertainty}$, resulting from adding the indicated signal prediction on top of the SM background prediction. The uncertainty includes the background uncertainty, signal uncertainty and statistical data uncertainty, added in quadrature. The purple bars show the pulls when only including the SM background prediction.
    The $\chi^2$ values in the top panel legend correspond to the sum of the squared pulls in each case. These values are intended for illustration only, i.e.\ they do not correspond directly to $-2\ln\mathcal{L}_{\text{CMS}}$ where $\mathcal{L}_{\text{CMS}}$ is defined in Eq.~\eqref{eq:simplike}.
	\label{fig:MET_spectrum}}
\end{figure*}

Comparing the right panel of Fig.~\ref{fig:dim_6_7_capped_main} to the allowed parameter regions from Fig.~\ref{fig:dim_6_capped_relic} (indicated by the grey dashed lines) does however reveal a number of differences. First of all, it is now possible to saturate the relic density bound for small $m_\chi$ (and small $\La$), thanks to the contribution of $\Q{3,q}{7}$ and $\Q{7,q}{7}$, which both give suppressed signals in direct and indirect detection experiments and are therefore largely unconstrained. Moreover, for $m_\chi > m_t$, we find that the predicted relic abundance can be substantially smaller than for the case with only dimension-6 operators, thanks to the contribution from the dimension-7 DM-quark operators $\Q{5\text{--}10,q}{7}$. The additional freedom in the annihilation cross-section also implies that the impact of imposing a strict relic density requirement is reduced compared to the case of dimension-6 operators only and will therefore not be discussed in further detail here.

We emphasize that global fits with 24 free parameters are computationally quite challenging, in particular when the best-fit region is not strongly constrained by data. As a result the contours in Fig. \ref{fig:dim_6_7_capped_main} are less smooth than for the case of dimension-6 operators only. This is particularly obvious in the right panel for DM masses around $150\,\mathrm{GeV}$. In this region many operators are strongly constrained by LHC data while annihilations into top quarks are kinematically forbidden. This makes it challenging to find parameter points that satisfy the relic density constraint, leading to comparably poor sampling. We have confirmed explicitly that this is not a physical effect, i.e.\ the allowed parameter region should be smooth and extend to $\Omega_\chi h^2 = 0.12$ everywhere.

\subsection{Full LHC likelihood}

\subsubsection*{Dimension-6 operators only (relic density upper bound)}

We now move onto the case where the full (rather than capped)
LHC likelihood is included in the scans. Fig.~\ref{fig:dim_6_full_main} shows the allowed parameter regions in terms of $m_\chi$ and $\La$ for the case where we introduce a hard cut-off in the missing energy spectrum for $\slashed{E}_T > \La$ (left panel), and the case where we introduce a smooth cut-off (right panel), as discussed in Sec.~\ref{sec:validity}. We see that in both cases, the results differ from Fig.~\ref{fig:dim_6_capped_main}, i.e.\ there is a preference for higher $\La$ values. This preference arises due to data excesses in a few high-$\slashed{E}_T$ bins in the ATLAS and CMS monojet searches.

\begin{figure*}[t]
	\centering
	\includegraphics[width=\columnwidth]{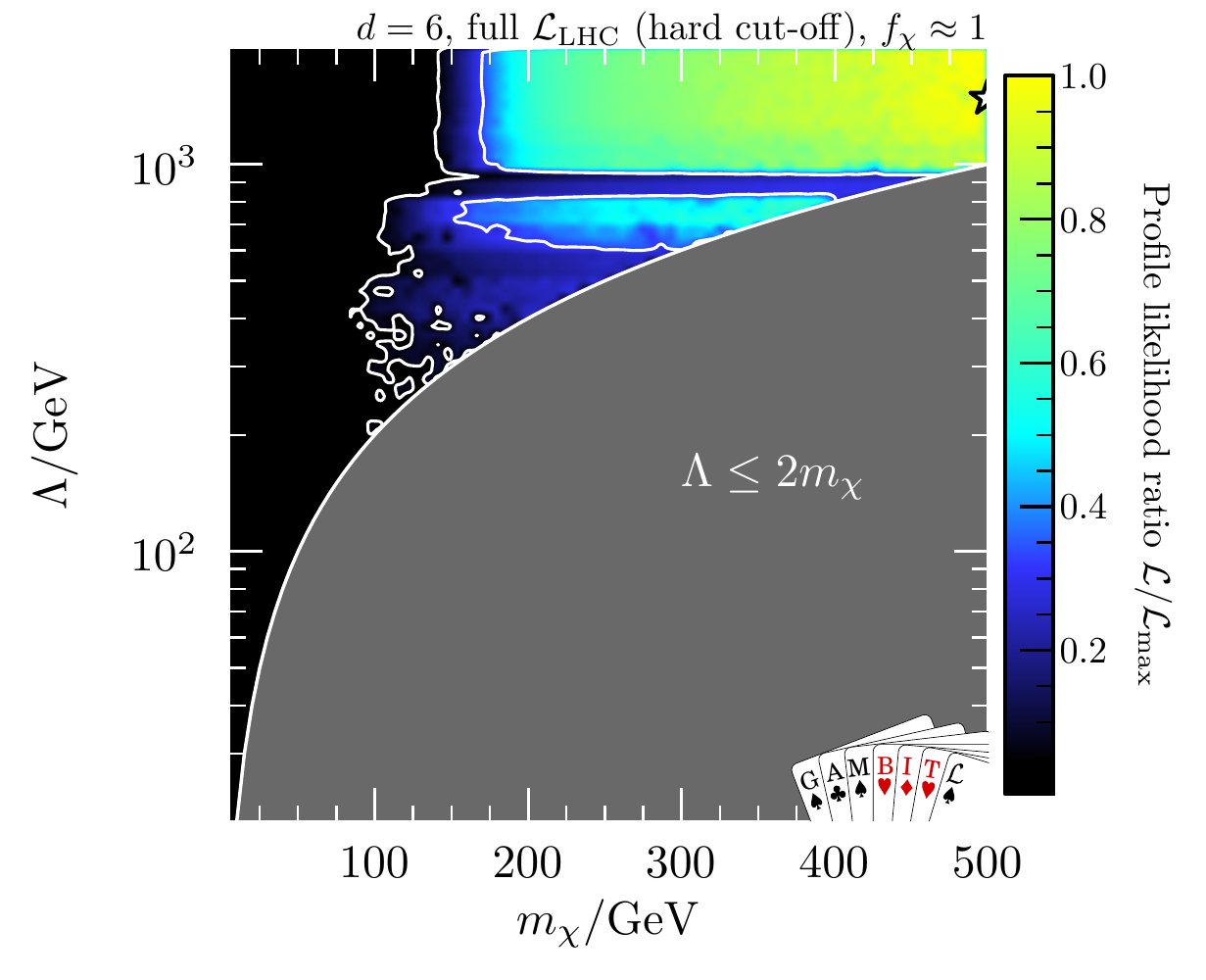} \hfill
	\includegraphics[width=\columnwidth]{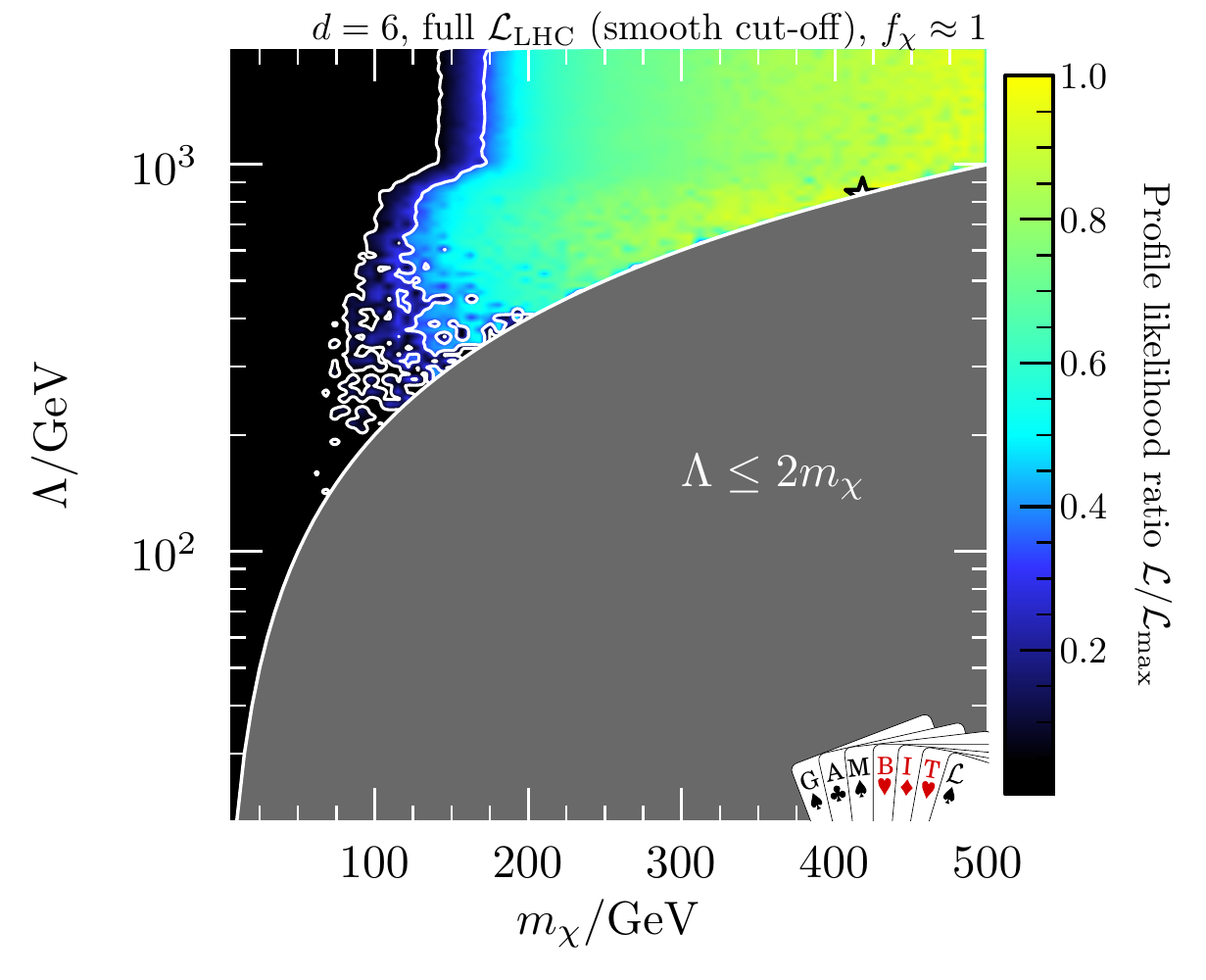}
	\caption{Same as Fig. \ref{fig:dim_6_full_main} but requiring the DM relic abundance to match the total observed DM abundance.}
	\label{fig:dim_6_full_RF_1}
\end{figure*}

The difference in the above two results can be understood as follows. For $\slashed{E}_T < \La$, the missing energy spectrum arising from DM is harder than the background, while for $\slashed{E}_T > \La$, we either set it to zero or assume that it drops rapidly. Thus, the ratio of signal-to-background is largest for $\slashed{E}_T \approx \La$, enabling our model to (partially) fit local excesses in the data. This is illustrated in Fig.~\ref{fig:MET_spectrum}, which shows the missing energy spectra for background and signal in CMS when applying different EFT validity prescriptions. As seen in the distribution of pulls in the bottom panel, the CMS search observes a couple of $1\sigma$--$2\sigma$ data excesses in bins around $\slashed{E} \approx 700\, \mathrm{GeV}$ (purple bars). By including a DM signal prediction on top of the SM background, these excesses can be reduced, thus reducing the pulls and improving the overall fit to the data (green bars). However, unless the signal spectrum dies off sufficiently fast above $\slashed{E} \approx 700\, \mathrm{GeV}$, the model will be penalized for causing larger pulls in the highest-$\slashed{E}_T$ bins, as seen for instance for the unmodified signal spectrum (lightest green bars, corresponding to $a=0$).

For the case where we impose a hard cut-off (left panel in Fig.~\ref{fig:dim_6_full_main}), we find (at the $1\sigma$ level) separate parameter regions preferred by the CMS analysis ($\La \approx 700 \, \mathrm{GeV}$) and the ATLAS analysis ($\La \gtrsim 1 \, \mathrm{TeV}$), with the overall best-fit point corresponding to the latter and being preferred relative to the background-only hypothesis by $2 \Delta \ln \mathcal{L} = 2.2$. When allowing for a smooth cut-off, on the other hand, the best-fit solution produces a partially improved fit to both excesses simultaneously, by suppressing the signal distribution approximately proportional to $(\slashed{E}_T / \La)^{-1}$. In this case, the best-fit point has $2 \Delta \ln \mathcal{L} = 2.6$.\footnote{We note that in both cases, the likelihood is very flat around the maximum and hence the precise location of the best-fit point is somewhat arbitrary.}~We refrain from translating these numbers into $p$-values, which would require extensive Monte Carlo simulations. For both choices of cut-off, the best-fit point predicts an annihilation cross-section that is slightly larger than the thermal cross-section, such that the DM particles in this case would only constitute a DM sub-component.

We emphasise that the preference for a non-zero signal contribution is to some degree an artefact of the way in which we have implemented the EFT validity requirement. Realistic UV completions typically do not introduce sharp features in the missing energy spectrum, making it harder to fit excesses observed in individual bins. Nevertheless, our findings emphasise the need to analyse missing energy searches at the LHC in terms of specific models in order to assess whether the signal preference found in the EFT approach can be recovered (at least partially) in a more complete setting.

\subsubsection*{Dimension-6 operators only (relic density saturated)}

We have also run scans with the full LHC likelihood and requiring the DM relic density to be saturated (see Fig.~\ref{fig:dim_6_full_RF_1}). We find the expected changes with respect to Fig.~\ref{fig:dim_6_full_main}, namely that small DM masses are disfavoured. For the case of a hard cut-off, the position of the best-fit point is unaffected, while for a smooth cut-off, it is pushed to slightly larger values of $m_\chi$ and $\La$. The respective preferences are reduced slightly to $2 \Delta \ln \mathcal{L} = 1.9$ and $2.0$. We also find that the best-fit point requires several Wilson coefficients to be non-zero. While the LHC signal can be fitted by either $\C{2}{6}$ or $\C{4}{6}$, the relic density can only be reproduced with a contribution from $\C{3}{6}$.
This is because $\Q{2,4}{6}$ lead to suppressed annihilation rates in the early universe, compared to $\Q{3}{6}$, while
$\Q{1}{6}$ is strongly constrained by direct detection (see also Table~\ref{tab:op_suppression}).

A summary of the various best-fit points from our scans with dimension-6 operators only is given in Table~\ref{tab:best-fit}. We note that essentially all of our scans require a non-zero contribution from $\C{3}{6}$ at the best-fit point in order to satisfy the relic density requirement. This is an interesting finding given that this operator is present only for Dirac fermion DM but not for Majorana fermion DM. In other words, we expect our results to change considerably for the case of Majorana fermion DM. Satisfying the relic density constraint with dimension-6 operators only while evading experimental constraints will be very challenging in this case.

%%%%%%%%%%%%%%%%%%%%%%%%%%%%%%%%%%%%%%%%%%%%%%%%%%%%%%%%%%%%%%%
\begin{table*}
  \centering
  \begin{tabular}{lccccc}
  \toprule
  \textbf{LHC likelihood} & \textbf{Relic density} & $\mathbf{2 \Delta \ln \mathcal{L}}$ & \textbf{Best-fit $m_\chi$} & \textbf{Best-fit $\La$} & \textbf{Best-fit constrained coupling} \\
  & \textbf{constraint} & &  \textbf{(GeV)} & \textbf{(GeV)} & \textbf{combination(s) $\mathbf{(TeV^{-2})}$}  \\ \midrule
  Capped & Upper bound & $0.3$ & $5.0$ & $< 200$ & $|\C{3}{6}| / \La^2 = 67$ \\[2mm]
  Capped & Saturated & $-0.5$ & $500$ & $> 1000$ &
  \begin{tabular}{cc}
  	$|\C{2}{6}| / \La^2 = 0.22$ \\[1.5mm]
  	$|\C{3}{6}| / \La^2 = 0.041$
  \end{tabular} \\[4mm]
  \midrule
  Full (hard cut-off) & Upper bound & $2.2$ & $500$ & $> 1250$ & $|\C{3}{6}|/\La^2 = 0.14$ \\[2mm]
  Full (smooth cut-off) & Upper bound & $2.6$ & $320$ & $640$ & $ |\C{3}{6}| / \La^2 = 0.18$ \\[1mm] \midrule
  Full (hard cut-off) & Saturated & $1.9$ & $500$ & $> 1250$ &
  \begin{tabular}{cc}
    $|\C{3}{6}| / \La^2 = 0.047$ \\[2mm]
    $\sqrt{(\C{2}{6})^2 + (\C{4}{6})^2}/\La^2 = 0.15$
  \end{tabular} \\[6mm]
  Full (smooth cut-off) & Saturated & 2.0 & 420 & 840 &
  \begin{tabular}{cc}
    $|\C{3}{6}|/\La^2 = 0.052$ \\[1.5mm]
    $\sqrt{(\C{2}{6})^2 + (\C{4}{6})^2}/\La^2 = 0.23$
  \end{tabular}
  \\[1mm] \bottomrule
\end{tabular}
 \caption{Best-fit points from our various scans involving dimension-6 operators with restricted parameter ranges ($5 \, \mathrm {GeV} \leq m_\chi \leq 500 \, \mathrm{GeV}$ and $20 \, \mathrm{GeV} \leq \La \leq 2 \, \mathrm{TeV}$). For most scans, there are degeneracies between different parameters around the best-fit point. In these cases, we only quote the combination that is well-constrained rather than each parameter individually. Parameters not stated explicitly are compatible with zero.}
 \label{tab:best-fit}
\end{table*}
%%%%%%%%%%%%%%%%%%%%%%%%%%%%%%%%%%%%%%%%%%%%%%%%%%%%%%%%%%%%%%%

\subsubsection*{Operators up to dimension 7}

\begin{figure*}[t]
	\centering
	\includegraphics[width=\columnwidth]{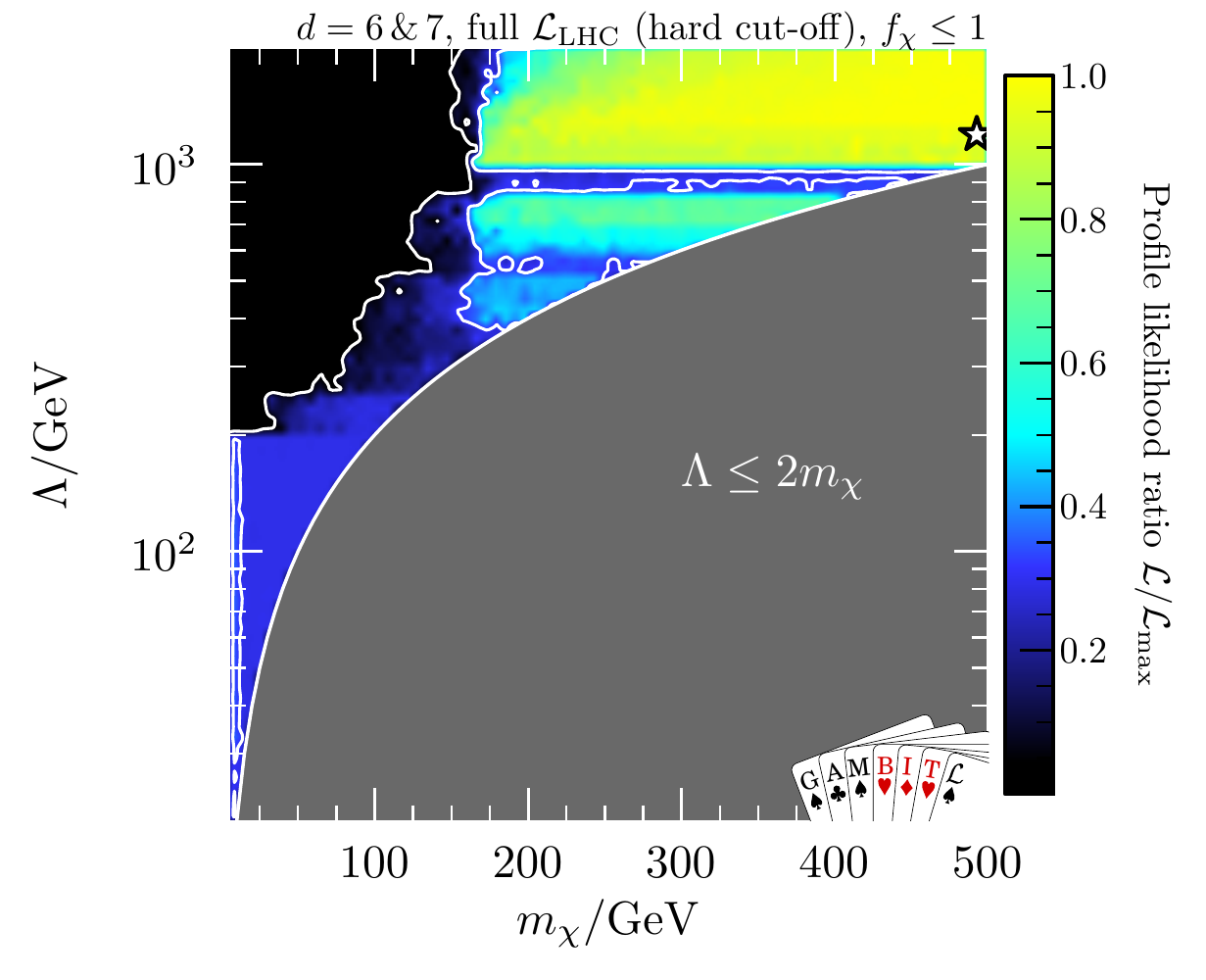} \hfill
	\includegraphics[width=\columnwidth]{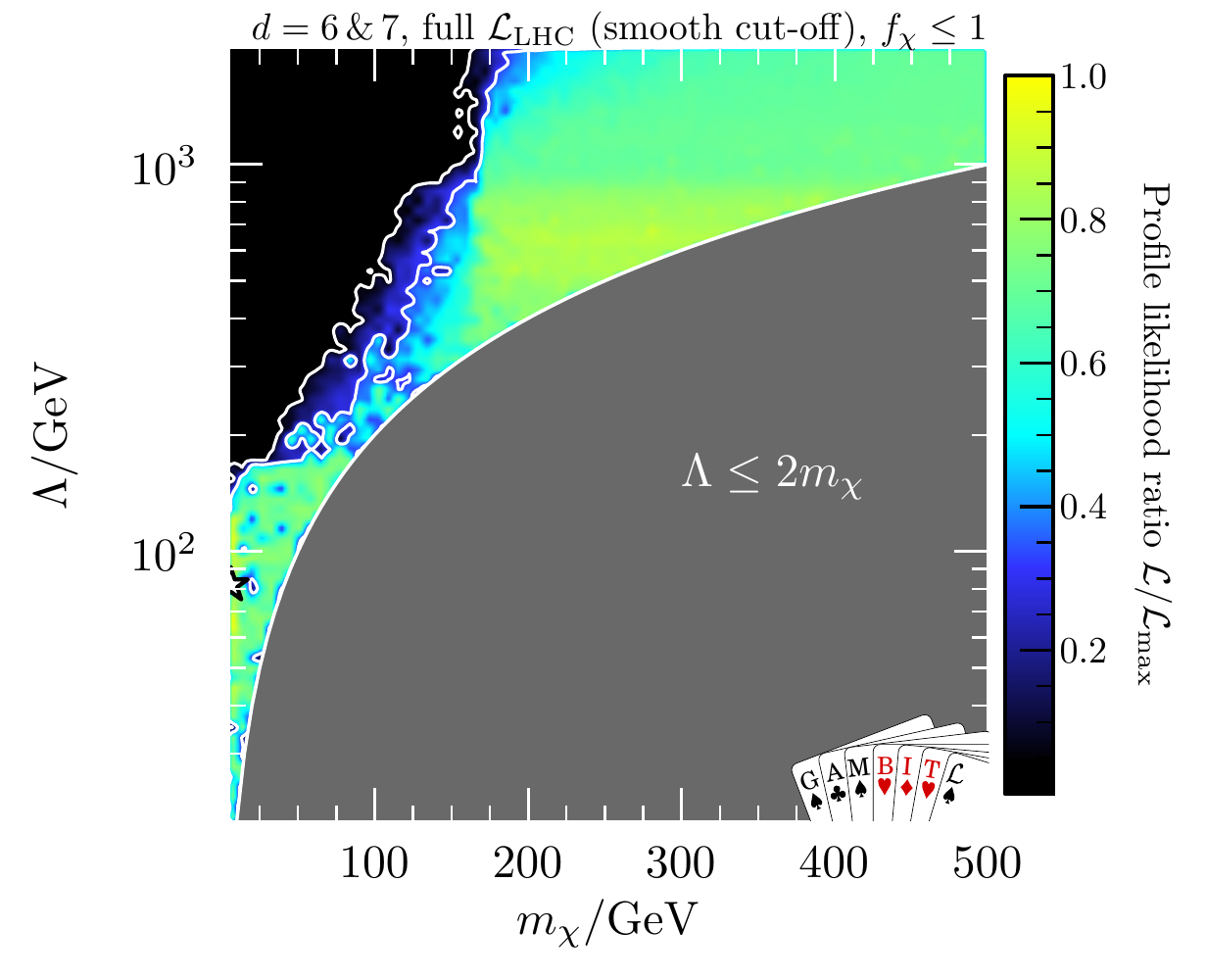}
	\caption{Profile likelihood in the $m_\chi$--$\La$ parameter plane when considering dimension-6 and dimension-7 operators and including the full LHC likelihood. In the left (right) panel, we impose a hard (smooth) cut-off in the predicted missing energy spectrum for $\slashed{E}_T > \La$ (see text for more details).}
	\label{fig:dim_6_7_full_main}
\end{figure*}

In Fig.~\ref{fig:dim_6_7_full_main}, we finally show the case where the full LHC likelihood is included when simultaneously considering all dimension-6 and dimension-7 operators, using either a hard cut-off (left) or profiling over possible smooth cut-offs (right). In the former case we find that the result looks very similar to the case of dimension-6 operators only (left panel of Fig.~\ref{fig:dim_6_full_main}) and also the likelihood at the best-fit point is very similar. In the latter case we find that it is now possible to simultaneously accommodate the upward fluctuations in the \emph{Fermi}-LAT data (as in Fig.~\ref{fig:dim_6_capped_main}) and in the LHC data (as in Fig.~\ref{fig:dim_6_full_main}). Doing so requires a small new-physics scale $\La \sim 80\,\mathrm{GeV}$ together with a rather soft cut-off $a \approx 1.7$ of the $\slashed{E}_T$ spectrum above $\La$. The resulting best-fit point has $2 \Delta \ln \mathcal{L} = 2.9$, which is the highest likelihood found in any of our scans.

A closer analysis reveals that the contribution of the dimension-6 operators is in fact not necessary to accommodate the small LHC excesses, because sufficiently large contributions can also be obtained from the gluon operators. For example, the operator $\Q{4}{7}$ is essentially unconstrained by direct detection and can induce sizeable LHC signals if $\C{4}{7}$ takes values close to the perturbativity bound. While it is challenging to satisfy the relic density requirement using only gluon operators, the allowed parameter space expands substantially when including a contribution from the dimension-7 DM-quark operators $\Q{5\text{--}8,q}{7}$. As a result, the allowed regions in $m_\chi$--$\La$ parameter space look very similar to the ones shown in Fig.~\ref{fig:dim_6_7_full_main} even when the Wilson coefficients of all dimension-6 operators are set to zero. For the same reason we expect no significant difference between Dirac and Majorana DM particles in this case. This complex interplay between different operators only becomes apparent in a global analysis and would be missed when studying individual operators separately.

%%%%%%%%%%%%%%%%%%%%%%%%%%%%%%%%%%%%%%%%%%%%%%%%%%%%%%%%%%%%%%%%%%%%%%%%%%%%%%%%%%%%%%%%%%%%%%%%%%%%%%%%%%%%%%%%%%%%%%%%
\section{Conclusions and outlook}\label{sec:summary}

%Technical developments:
In this work we have presented the first global analysis of the full set of effective operators up to dimension 7 involving a Dirac fermion DM particle and quarks or gluons. Key to enabling such an analysis were a number of technical developments:
\begin{itemize}
 \item We have fully automated the calculation of direct detection constraints, including mixing under RG evolution and matching onto non-relativistic effective operators at the hadronic scale, and indirect detection constraints, including cosmological constraints on energy injection;
 \item We have adopted a novel approach to address the issue of EFT validity at the LHC. Rather than performing a simple truncation procedure, we introduce a smooth cut-off for $\slashed{E}_T > \La$ and treat this parameter as a nuisance parameter to ensure that no artificially strong exclusions arise from the tails of the predicted distributions;
 \item We employ highly efficient likelihood calculations and sampling algorithms that make it possible to scan over up to 24 parameters (the DM mass $m_\chi$, the new physics scale $\La$, 14 Wilson coefficients and 8 nuisance parameters).
\end{itemize}
In combination, these developments enable us, for the first time, to include interference effects between different operators in all parts of the analysis.

Our main result is that it is typically possible to suppress the scattering and annihilation cross-sections in the non-relativistic limit, and thereby evade direct and indirect detection constraints while satisfying the relic density requirement. Doing so does not require finely tuned cancellations or interference effects but is a direct consequence of the spin structure of the operators that we consider. The LHC, however, plays a special role, because the production of relativistic DM particles is less sensitive to the specific spin structure of the operator. As a result, we find generally strong constraints on small DM masses and large $\La$, both for the case of dimension-6 operators only and also when including dimension-7 operators. Moreover, when allowing excesses in individual LHC bins to be fitted (rather than artificially capping the LHC likelihood), we find a slight preference for a DM signal with a relatively low new physics scale.
Given that the magnitude of this excess is sensitive to the precise EFT validity prescription that we adopt, we have not attempted to quantify its significance within the EFT.

We find that it is typically not necessary to have simultaneous contributions from many different operators  in order to find viable regions of parameter space. Indeed, large viable regions of parameter space are found both for the case when we consider only dimension-6 operators and only dimension-7 operators.
These sets of operators can easily be generated by integrating out a heavy mediator with spin 1 or spin 0, respectively. However, we typically do require sizeable contributions from operators that violate parity and/or CP, reflecting the pressure on the simplest WIMP models from the non-observation of a DM signal in direct and indirect detection experiments (see Ref.~\cite{HP} for a similar discussion in the context of Higgs portal models).

A particularly interesting observation is that it is generally not possible to have a large hierarchy between the DM mass and the new physics scale without violating the relic density requirement. In particular, for $m_\chi \lesssim 100 \, \mathrm{GeV}$, constraints from the LHC require $\La \lesssim 200 \, \mathrm{GeV}$, meaning that the EFT is no longer valid at LHC energies and additional new degrees of freedom should be kinematically accessible. Moreover, the well-known unitarity bound on the DM mass implies a robust upper bound on the scale of new physics of the order of $300 \, \mathrm{TeV}$.
We also note that for masses in the TeV range CTA will have a unique chance of probing part of the
currently inaccessible parameter space that is spanned between the EFT validity and the relic density constraints.

We emphasise that it is generally possible for the DM particle under consideration to constitute only a DM sub-component (in which case, constraints from direct and indirect detection experiments are correspondingly suppressed),
but large regions of viable parameter space also remain when requiring the relic density to be saturated. In future studies, it will be interesting to modify the way in which the relic density calculation is included. For example, one could consider an initial particle-antiparticle asymmetry in the dark sector, which would make it possible to saturate the relic density in parameter regions that would normally predict an underabundance, while at the same time suppressing constraints from indirect detection experiments. A more radical approach would be to not perform a relic density calculation at all and simply assume that the observed relic abundance (with $f_\chi =1$) is achieved through some unspecified modification of standard cosmology.
A detailed analysis of direct detection constraints on such a scenario is in preparation.

An exciting direction for future investigation is to embed the EFTs considered here into a more complete approach based on UV-complete (or simplified) models. Almost all of the machinery developed for the present work will also be directly applicable in this case. The main difference arises in the interpretation of the LHC signals. If the mediator of the DM interactions is kinematically accessible at LHC energies, it will be essential to not only consider the resulting changes in the missing energy spectra, but also additional signatures arising from visible decays of the mediator~\cite{Chala:2015ama,Fairbairn:2016iuf} (see Ref.~\cite{Bischer:2020sop} for a recent discussion of how to connect DM EFTs and UV-complete models).
Furthermore, close to the EFT validity boundary the presence of the mediator will also modify the results of the
relic density calculation, thus affecting the target couplings for these signals.
It will also be interesting to see to what extent the slight LHC excesses can be accommodated in such a set-up.

Another important extension of the present work will be to also consider operators coupling DM to leptons as well as electroweak gauge and Higgs bosons in order to embed our approach into a framework that respects the unbroken electroweak symmetry. Given that the relevant RG evolution is known (and already implemented in \textsf{DirectDM}) and that the relevant annihilation cross-sections and injection spectra can be calculated automatically, such an extension does not pose any conceptual difficulties regarding direct or indirect detection constraints and relic density calculations. Again, the most challenging part will be to include all relevant collider constraints (which in this case stem also from LEP). Given that these constraints are typically weaker than the corresponding ones for quarks, it will be interesting to see whether some of the conclusions found in the present work can be relaxed and additional viable parameter space opens up.

Finally, it will be very interesting to consider DM EFTs with non-trivial flavour structure, for example with couplings predominantly to the third generation. In such a set-up, one generally expects sizeable flavour-changing neutral currents and hence it will be essential to connect the EFTs used to study DM to the ones employed in flavour physics. Such a study would be particularly exciting given the recently observed anomalies in various flavour observables~(see e.g. Refs.~\cite{Barbieri:2021wrc,Graverini:2018riw,Aaij:2021vac}). Moreover, the effects of electroweak operator mixing on the direct detection bounds are expected to be much more pronounced in such scenarios.

Of course, the most important outstanding task is to collect more data that may shed light on the nature of DM. Upcoming LHC analyses will improve the sensitivity to missing energy signatures of DM, the next generation of direct detection experiments~\cite{Zhang:2018xdp,Akerib:2018lyp,Aprile:2020vtw} will be able to probe substantially smaller scattering cross-sections, and ongoing~\cite{Ahnen:2016qkx,Abdallah:2016ygi,Aguilar:2021tos} and
planned~\cite{Acharyya:2020sbj} indirect detection experiments will probe the freeze-out paradigm with unprecedented precision.
Our present work has shown that this effort is highly worthwhile given the wide regions of parameter space that cannot currently be excluded in a model-independent way. Reducing the vast number of viable possibilities to explain DM therefore remains a key challenge for years to come.

%%%%%%%%%%%%%%%%%%%%%%%%%%%%%%%%%%%%%%%%%%%%%%%%%%%%%%%%%%%%%%%%%%%%%%%%%%%%%%%%%%%%%%%%%%%%%%%%%%%%%%%%%%%%%%%%%%%%%%%%
\begin{acknowledgements}
    We thank all members of the \gambit community as well as Fady Bishara for discussions and checks. For computing, we thank PRACE for awarding us access to Marconi at CINECA and Joliot-Curie at CEA.  This project was also undertaken with the assistance of resources and services from the National Computational Infrastructure, which is supported by the Australian Government.  We thank Astronomy Australia Limited for financial support of computing resources, and the Astronomy Supercomputer Time Allocation Committee for its generous grant of computing time. We thank Juan Fuster, Adrián Irles, Davide Melini and Marcel Vos for clarifications regarding Ref.~\cite{Aad:2019mkw}. PA is supported by the Australian Research Council Future Fellowship grant FT160100274, and PA, CB, TEG and MW also acknowledge support from ARC Discovery Project DP180102209. NAK and ACV are supported by the Arthur B.~McDonald Canadian Astroparticle Physics Research Institute and NSERC, with equipment funded by the Canada Foundation for Innovation and the Province of Ontario, and supported by the Queen's Centre for Advanced Computing. Research at Perimeter Institute is supported by the Government of Canada through the Department of Innovation, Science, and Economic Development, and by the Province of Ontario. AB acknowledges support by F.N.R.S.\ through the F.6001.19 convention, JB and JZ by DOE grant DE-SC0011784, BF by the Horizon 2020 Marie Sk\l{}odowska-Curie actions (EU; H2020-MSCA-IF-2016-752162), WH by a Royal Society University Research Fellowship, FK, TEG and PSt from the DFG Emmy Noether Grant No.\ KA 4662/1-1 and Grant 396021762 -- TRR 257, JJR by Katherine Freese through a grant from the Swedish Research Council (Contract No. 638-2013-8993). MTP is supported by the Argelander Starter-Kit Grant of the University of Bonn and BMBF Grant No. 05H19PDKB1. AF is supported by an NSFC Research Fund for International Young Scientists grant 11950410509. PS acknowledges funding support from the Australian Research Council under Future Fellowship FT190100814. MW and AS are further supported by the Australian Research Council under Centre of Excellence CE200100008. This article made use of \textsf{pippi v2.1} \cite{pippi}.
\end{acknowledgements}

%%%%%%%%%%%%%%%%%%%%%%%%%%%%%%%%%%%%%%%%%%%%%%%%%%%%%%%%%%%%%%%%%%%%%%%%%%%%%%%%%%%%%%%%%%%%%%%%%%%%%%%%%%%%%%%%%%%%%%%%
\begin{appendices}

\renewcommand \thetocsection {\Alph{section}}

%%%%%%%%%%%%%%%%%%%%%%%%%%%%%%%%%%%%%%%%%%%%%%%%%%%%%%%%%%%%%%%%%%%%%%%%%%%%%%%%%%%%%%%%%%%%%%%%%%%%%%%%%%%%%%%%%%%%%%%%

\section{DirectDM interface} \label{app:directdm}

We briefly describe the \GB interface to the new backend \ddm, its interface to \ddcalc, and how to interface a new model to \ddm. For more background on the technical aspects of the \GB framework, please refer to the original \GB manual~\cite{gambit}, and the \gum paper~\cite{Gonzalo:2021cnq,GUM}.

\ddm matches Wilson coefficients of a relativistic EFT onto a non-relativistic EFT valid at the nuclear scale. The \GB implementation interfaces with the \python version of this package.

Relativistic Wilson coefficients can be defined at the 3-, 4- or 5-quark scale, with the capability \cpp{DD_rel_WCs_flavscheme}. For a given model, a new module function providing this capability should be written, returning the type \cpp{map_str_dbl} (\cpp{std::map<std::string,double>}). Once this capability has been fulfilled, \GB uses the module function \cpp{DD_nonrel_WCs_flavscheme} to call the \ddm backend via the convenience function \cpp{get_NR_WCs_flav}. This provides the capability \cpp{DD_nonrel_WCs} which can be connected to the \ddcalc backend.

This module function providing the capability \cpp{DD_nonrel_WCs} depends on the capability \cpp{WIMP_properties}, of native \GB type \cpp{WIMPprops}. \cpp{WIMP_properties} supplies the particle information about the WIMP candidate, such as its spin, mass, and whether or not it is self-conjugate, extracted from the particle database and either the spectrum or model parameters.

As an example, consider a simplified model where a vector mediator governs the interaction between $d$-type quarks and a fermionic DM candidate $\chi$, with the following interaction Lagrangian,
\begin{equation}
  \mathcal{L}_{\rm int} \supset g_\chi \overline{\chi} \gamma_\mu \chi V^\mu + g_b \sum_{q = d,s,b} \overline{q} \gamma_\mu q V^\mu \,.
  \label{eq:ddmlagr}
\end{equation}
The model implementation within \GB will contain four free parameters: the couplings $g_\chi$ and $g_b$, the DM mass $m_\chi$, and the mediator mass $m_V$. The model definition for the above simplified model looks like:

\begin{lstcpp}
#define MODEL NewModel
  START_MODEL

  // Standard model definition in Gambit
  DEFINEPARS(mchi, mV, gchi, gb)

#undef MODEL
\end{lstcpp}
The information about the WIMP properties should be added to the particle database, if it does not exist already, in the following format
\begin{lstyaml}
  - name: chi
    conjugate: chi~
    chargex3: 0
    color: 1
    spinx2: 1
    description: Dirac DM Singlet
    PDG_context: [62, 0]
\end{lstyaml}
and the \cpp{WIMP_properties} module function should be modified accordingly, adding the current model as allowed
\begin{lstcpp}
  #define CAPABILITY WIMP_properties
  START_CAPABILITY
    #define FUNCTION WIMP_properties
    START_FUNCTION(WIMPprops)
    ...
    ALLOW_MODELS(NewModel)
    #undef FUNCTION
  #undef CAPABILITY
\end{lstcpp}
and providing a source for the mass of the DM candidate, in this case from the model parameters, as
\begin{lstcpp}
   void WIMP_properties(WIMPprops &props)
   {
     ...
     if(ModelInUse("NewModel"))
       props.mass = *Param["mchi"];
   }
\end{lstcpp}
If we integrate out the mediator in Eq.~\eqref{eq:ddmlagr}, the interaction term becomes
\begin{equation}
  \mathcal{L}_{\rm int}^{\rm eff} \supset \frac{g_\chi g_b}{m_V^2} \, \left( \overline{\chi} \gamma_\mu \chi \sum_{q \,=\, d,s,b} \overline{q} \gamma^\mu q \right)\,.
\end{equation}
The operator in \ddm corresponding to this interaction is $\Q{1,q}{6} = (\overline\chi \gamma_\mu \chi)(\overline{q} \gamma^\mu q)$. We identify the relevant coefficient to pass to \ddm as $g_\chi g_b / m_V^2$. This is simply implemented in \darkbit by the following source code:

\begin{lstcpp}
/// Relativistic Wilson Coefficients for direct
/// detection. Map from the model parameters
/// to relativistic EFT for DirectDM
void DD_rel_WCs_flavscheme_NewModel(map_str_dbl& result)
{
  using namespace Pipes::DD_rel_WCs_flavscheme_NewModel;

  // The spectrum object associated with
  // the new model.
  const Spectrum& spec = *Dep::NewModel_spectrum;

  // Get some parameters from the Spectrum object
  double mV = spec.get(Par::Pole_Mass, "V");
  double gchi = spec.get(Par::dimensionless, "gchi");
  double gq = spec.get(Par::dimensionless, "gq");

  double prefactor = gchi * gq / pow(mphi, 2.);

  // Wilson coefficients in DirectDM are
  // *dimensionful*. Note that there is
  // a different operator for each quark.
  result["C61d"] = prefactor;
  result["C61s"] = prefactor;
  result["C61b"] = prefactor;
}
\end{lstcpp}
plus a new matching entry in \cpp{DarkBit_rollcall.hpp},

\begin{lstcpp}

// Relativistic Wilson coefficients defined
// in the 5 (or 4 or 3) flavscheme
#define CAPABILITY DD_rel_WCs_flavscheme
START_CAPABILITY

  #define FUNCTION DD_rel_WCs_flavscheme_NewModel
  START_FUNCTION(map_str_dbl)
  DEPENDENCY(NewModel_spectrum, Spectrum)
  ALLOW_MODEL(NewModel)
  #undef FUNCTION

#undef CAPABILITY
\end{lstcpp}
For a full definition of the operator basis used in \ddm, we refer the reader to Refs.~\cite{Bishara:2017nnn,Brod:2017bsw}.

When \ddm is used, the user must also scan over the model \textsf{nuclear\_params\_ChPT\_sigmapiN}, which contains (nuisance) parameters used in the matching and running routines in \ddm. These are defined in Table 1 of Ref.~\cite{Bishara:2017nnn}. We provide a \YAML file containing the default values used in \ddm (see the \term{nuclear_params_ChPT_sigmapiN_fixed.yaml} file in the \termalt{\$GAMBIT/yaml_files/include/} directory).

%%%%%%%%%%%%%%%%%%%%%%%%%%%%%%%%%%%%%%%%%%%%%%%%%%%%%%%%%%%%%%%%%%%%%%%%%%%%%%%%%%%%%%%%%%%%%%%%%%%%%%%%%%%%%%%%%%%%%%%%
\section{UFO to CalcHEP} \label{app:ufo_to_mdl}

\utc is a simple \Python tool distributed with \GB \textsf{v2.1} and above, and is integrated in the \gum framework. \utc is located at \termalt{\$GAMBIT/gum/src/ufo_to_mdl.py}. It can also be run as a standalone tool, using either \textsf{Python2} or \textsf{Python3}. Below we briefly describe the motivation for \utc and how to use it.

The purpose of \utc is to generate \CH input (\texttt{.mdl} files) from \ufo files. The motivation for this tool's creation is that \fr does not generate four-fermion \CH output, but it can create such output for \MG. In fact, at the time of writing, \textsf{LanHEP}~\cite{Semenov:2008jy} is the only package that supports automatic generation of four-fermion contact interactions for \CH files. \utc allows the user to study four-fermion interactions using \CH (and correspondingly, \micromegas), effectively creating a pathway from \fr to \CH for effective theories of this kind. In the context of \GB and the \gum pipeline, \utc allows the user to study EFTs of DM using the routines provided by \micromegas and \CH inside of the \GB framework, such as relic density calculations, direct detection rates, and indirect detection via the Process Catalogue (see the \darkbit manual~\cite{DarkBit} for details).

Usage of \utc is straightforward. There are two modes \utc can be operated in: comparison mode and conversion mode. The mode integrated into the \gum pipeline is the comparison mode, which compares two directories containing \texttt{.ufo} and \texttt{.mdl} files generated by \fr:
\begin{lstterm}
python ufo_to_mdl.py <UFODir> <MDLDir>
\end{lstterm}
This ensures that all vertices in the \MG files are present in the \CH files. \utc does not explicitly check that the vertex functions and Lorentz indices are in agreement; it solely checks the particle content of the vertices. If there are vertices missing from the \CH files,\footnote{If \MG and \CH output is generated from a fully functional \fr model implementation with trivial colour structures, the only missing vertices should be four-fermion vertices.} \utc generates these vertices and writes a set of corrected \CH files to a new directory \term{<MDLDir>_ufo2mdl}.

In the case of four-fermion operators, \utc adds an additional auxiliary field to the particle content, and creates two 3-field interactions by way of this new auxiliary mediator particle, following the prescription described in Chapter 8 of the \CH manual~\cite{Belyaev:2012qa}. An auxiliary field has no momentum dependence and serves only to split the vertex into a form in which \CH can use. The order of fields generated by \utc will be identical to those in the \MG files, i.e.\ a vertex
\begin{equation}
i \left( \overline{\chi} \Gamma_\chi \chi \right) \left(\overline{\psi} \Gamma_\psi \psi \right)
\end{equation}
would be broken up into two vertices,
\begin{align}
i \left(\overline{\chi} \Gamma_\chi \chi\right) \phi \,\,\, \mathrm{and} \,\,\, i \left(\overline{\psi} \Gamma_\psi \psi\right) \phi \,,
\end{align}
where $\Gamma_\chi$ is a generic Dirac structure contracted with the field $\chi$, and $\phi$ is the auxiliary field, with Lorentz indices corresponding to $\Gamma$ (either scalar, vector or tensor). As a result, operators in \fr files should be written pairwise.

As noted above, \utc can also be used as a standalone tool independent of the \gum pipeline. Running \utc in conversion mode, with only a directory containing \MG files as input,
\begin{lstterm}
python ufo_to_mdl.py <UFODir>
\end{lstterm}
will generate \texttt{.mdl} files from scratch and save them in a new directory with name \term{<UFODir>_ufo2mdl}. The version of \utc released with \GB \textsf{v2.1} does not support non-trivial colour structures and will throw an error if it is asked to generate a vertex with implicit colour structure.

%%%%%%%%%%%%%%%%%%%%%%%%%%%%%%%%%%%%%%%%%%%%%%%%%%%%%%%%%%%%%%%%%%%%%%%%%%%%%%%%%%%%%%%%%%%%%%%%%%%%%%%%%%%%%%%%%%%%%%%%
\section{CMB energy injection} \label{app:energy_injection}

In order to provide CMB constraints from energy injection through decays and annihilation of DM, the yields $dN/dE$ of photons, positrons and electrons produced in these processes need to be known. With \GB~\textsf{v2.1}, the existing capabilities for the calculation of photon yields (\cpp{GA_Yield}\footnote{Since \GB~\textsf{v2.0}, decaying DM is supported, such that the capability \cpp{GA_AnnYield} was generalised and renamed.}) were generalised and capabilities that calculate the yields of positrons (\cpp{positron_Yield}) and electrons (\cpp{electron_Yield}) were introduced. To support future analyses of charged cosmic rays, we also introduced the capabilities \cpp{antiproton_Yield} and \cpp{antideuteron_Yield} that calculate the yields of anti-protons and anti-deuterons, respectively. These capabilities are, however, not used for the CMB energy injection calculations.

Once the yields are known, they need to be passed to \textsf{DarkAges} via the capability \cpp{energy_injection_spectrum} to derive the effective efficiency function $ f_{\rm eff} (z) $. For maximal flexibility, we have implemented the function \cpp{energy_injection_spectrum_ProcessCatalog} that automatically provides the inputs for \textsf{DarkAges} based on the model-dependent \cpp{TH_ProcessCatalog}, and the yields for photons, electrons and positrons. Once these capabilities have been provided, no further input from the user is needed.

To enable CMB energy injection constraints, the user also needs to declare that the model in question can be mapped to one of the energy injection ``flag'' models (\textsf{AnnihilatingDM\_general} or \textsf{DecayingDM\_general}) and their parameters. This can be done via a friend relationship to the appropriate ``flag'' model.

Assuming that the model under consideration contains annihilating DM particles, the user has to define a relation to \textsf{AnnihilatingDM\_general}, and its two parameters \term{sigmav} and \term{mass}. It is important to note that the model \textsf{AnnihilatingDM\_general} implicitly assumes that the DM particle constitutes all of DM ($ f_\chi=1 $) and that it is self-conjugate. In case that the particle is not self-conjugate, the parameter \term{sigmav} needs to be rescaled by $ \kappa=1/2 $. Likewise, if the DM candidate does not constitute all of DM, \term{sigmav} needs to be rescaled by $ f_\chi^2 $.

To define the translation function, the user has to make sure that the definition of the model \textsf{AnnihilatingDM\_general} is known, i.e.\ the following header is included:
\begin{lstcpp}
#include "gambit/Models/models/CosmoEnergyInjection.hpp"
\end{lstcpp}
Furthermore, the translation function and its dependencies need to be defined by including the following lines to the definition of the model in question:
\begin{lstcpp}
INTERPRET_AS_X_FUNCTION(AnnihilatingDM_general, NewModel_to_AnnihilatingDM_general)
INTERPRET_AS_X_DEPENDENCY(AnnihilatingDM_general, WIMP_properties, WIMPprops)
INTERPRET_AS_X_DEPENDENCY(AnnihilatingDM_general, sigmav, double)
INTERPRET_AS_X_DEPENDENCY(AnnihilatingDM_general, RD_fraction, double)
\end{lstcpp}
Note that this definition makes use of the \cpp{WIMP_properties} capability, described in App.~\ref{app:directdm}, in order to get the mass of the DM candidate and the information whether the DM candidate is self-conjugate or not. In case that this capability is not defined for the model in question, this dependency has to be replaced by equivalent dependencies. For the translation function defined above, the source code looks like this:
\begin{lstcpp}
void NewModel_to_AnnihilatingDM_general(const ModelParameters& /*unused*/, ModelParameters &fpars)
{
  USE_MODEL_PIPE(AnnihilatingDM_general)

  const auto wimp_props = *Dep::WIMP_properties;
  const double k = (wimp_props.sc) ? 1. : 0.5;
  const double f = *Dep::RD_fraction;

  // Set the mass
  fpars.setValue("mass", wimp_props.mass);

  // In AnnihilatingDM_general, f^2 and k are implicitly included in sigmav
  fpars.setValue("sigmav", k*f*f*(*Dep::sigmav));
}
\end{lstcpp}
Note that this has to be placed in the correct namespace:
\begin{lstcpp}
namespace Gambit {
  namespace Models {
    namespace MODEL {
      ...
    }
  }
}
\end{lstcpp}

\end{appendices}

%%%%%%%%%%%%%%%%%%%%%%%%%%%%%%%%%%%%%%%%%%%%%%%%%%%%%%%%%%%%%%%%%%%%%%%%%%%%%%%%%%%%%%%%%%%%%%%%%%%%%%%%%%%%%%%%%%%%%%%%
\bibliography{R2}

\providecommand{\href}[2]{#2}\begingroup\raggedright\begin{thebibliography}{100}

\bibitem{Lee:1977ua}
B.~W. Lee and S.~Weinberg, {\it {Cosmological Lower Bound on Heavy Neutrino
  Masses}},  {\em Phys. Rev. Lett.} {\bf 39} (1977) 165--168.

\bibitem{Arcadi:2017kky}
G.~Arcadi, M.~Dutra, {\em et.~al.}, {\it {The waning of the WIMP? A review of
  models, searches, and constraints}},  {\em Eur. Phys. J.} {\bf C78} (2018)
  203, [\href{http://arxiv.org/abs/1703.07364}{{\tt arXiv:1703.07364}}].

\bibitem{Leane:2018kjk}
R.~K. Leane, T.~R. Slatyer, J.~F. Beacom, and K.~C.~Y. Ng, {\it {GeV-scale
  thermal WIMPs: Not even slightly ruled out}},  {\em Phys. Rev. D} {\bf 98}
  (2018) 023016, [\href{http://arxiv.org/abs/1805.10305}{{\tt
  arXiv:1805.10305}}].

\bibitem{Fan:2010gt}
J.~Fan, M.~Reece, and L.-T. Wang, {\it {Non-relativistic effective theory of
  dark matter direct detection}},  {\em \jcap} {\bf 1011} (2010) 042,
  [\href{http://arxiv.org/abs/1008.1591}{{\tt arXiv:1008.1591}}].

\bibitem{Agrawal:2010fh}
P.~Agrawal, Z.~Chacko, C.~Kilic, and R.~K. Mishra, {\it {A Classification of
  Dark Matter Candidates with Primarily Spin-Dependent Interactions with
  Matter}},  \href{http://arxiv.org/abs/1003.1912}{{\tt arXiv:1003.1912}}.

\bibitem{Fitzpatrick:2010br}
A.~Fitzpatrick and K.~M. Zurek, {\it {Dark Moments and the DAMA-CoGeNT
  Puzzle}},  {\em Phys. Rev. D} {\bf 82} (2010) 075004,
  [\href{http://arxiv.org/abs/1007.5325}{{\tt arXiv:1007.5325}}].

\bibitem{Crivellin:2014gpa}
A.~Crivellin and U.~Haisch, {\it {Dark matter direct detection constraints from
  gauge bosons loops}},  {\em Phys. Rev. D} {\bf 90} (2014) 115011,
  [\href{http://arxiv.org/abs/1408.5046}{{\tt arXiv:1408.5046}}].

\bibitem{DEramo:2016gos}
F.~D'Eramo, B.~J. Kavanagh, and P.~Panci, {\it {You can hide but you have to
  run: direct detection with vector mediators}},  {\em JHEP} {\bf 08} (2016)
  111, [\href{http://arxiv.org/abs/1605.04917}{{\tt arXiv:1605.04917}}].

\bibitem{Hoferichter:2016nvd}
M.~Hoferichter, P.~Klos, J.~Men\'endez, and A.~Schwenk, {\it {Analysis
  strategies for general spin-independent WIMP-nucleus scattering}},  {\em
  Phys. Rev. D} {\bf 94} (2016) 063505,
  [\href{http://arxiv.org/abs/1605.08043}{{\tt arXiv:1605.08043}}].

\bibitem{Kahlhoefer:2016eds}
F.~Kahlhoefer and S.~Wild, {\it {Studying generalised dark matter interactions
  with extended halo-independent methods}},  {\em JCAP} {\bf 10} (2016) 032,
  [\href{http://arxiv.org/abs/1607.04418}{{\tt arXiv:1607.04418}}].

\bibitem{Goodman:2010qn}
J.~Goodman, M.~Ibe, {\em et.~al.}, {\it {Gamma Ray Line Constraints on
  Effective Theories of Dark Matter}},  {\em Nucl. Phys. B} {\bf 844} (2011)
  55--68, [\href{http://arxiv.org/abs/1009.0008}{{\tt arXiv:1009.0008}}].

\bibitem{Beltran:2008xg}
M.~Beltran, D.~Hooper, E.~W. Kolb, and Z.~C. Krusberg, {\it {Deducing the
  nature of dark matter from direct and indirect detection experiments in the
  absence of collider signatures of new physics}},  {\em Phys. Rev. D} {\bf 80}
  (2009) 043509, [\href{http://arxiv.org/abs/0808.3384}{{\tt
  arXiv:0808.3384}}].

\bibitem{Cheung:2011nt}
K.~Cheung, P.-Y. Tseng, and T.-C. Yuan, {\it {Gamma-ray Constraints on
  Effective Interactions of the Dark Matter}},  {\em JCAP} {\bf 06} (2011) 023,
  [\href{http://arxiv.org/abs/1104.5329}{{\tt arXiv:1104.5329}}].

\bibitem{Harnik:2008uu}
R.~Harnik and G.~D. Kribs, {\it {An Effective Theory of Dirac Dark Matter}},
  {\em Phys. Rev. D} {\bf 79} (2009) 095007,
  [\href{http://arxiv.org/abs/0810.5557}{{\tt arXiv:0810.5557}}].

\bibitem{DeSimone:2013gj}
A.~De~Simone, A.~Monin, A.~Thamm, and A.~Urbano, {\it {On the effective
  operators for Dark Matter annihilations}},  {\em JCAP} {\bf 02} (2013) 039,
  [\href{http://arxiv.org/abs/1301.1486}{{\tt arXiv:1301.1486}}].

\bibitem{Karwin:2016tsw}
C.~Karwin, S.~Murgia, T.~M.~P. Tait, T.~A. Porter, and P.~Tanedo, {\it {Dark
  Matter Interpretation of the Fermi-LAT Observation Toward the Galactic
  Center}},  {\em Phys. Rev. D} {\bf 95} (2017) 103005,
  [\href{http://arxiv.org/abs/1612.05687}{{\tt arXiv:1612.05687}}].

\bibitem{Carpenter:2016thc}
L.~M. Carpenter, R.~Colburn, J.~Goodman, and T.~Linden, {\it {Indirect
  Detection Constraints on s and t Channel Simplified Models of Dark Matter}},
  {\em Phys. Rev. D} {\bf 94} (2016) 055027,
  [\href{http://arxiv.org/abs/1606.04138}{{\tt arXiv:1606.04138}}].

\bibitem{Abdallah:2015ter}
J.~Abdallah {\em et.~al.}, {\it {Simplified Models for Dark Matter Searches at
  the LHC}},  {\em Phys. Dark Univ.} {\bf 9-10} (2015) 8--23,
  [\href{http://arxiv.org/abs/1506.03116}{{\tt arXiv:1506.03116}}].

\bibitem{Kahlhoefer:2017dnp}
F.~Kahlhoefer, {\it {Review of LHC Dark Matter Searches}},  {\em Int. J. Mod.
  Phys. A} {\bf 32} (2017) 1730006,
  [\href{http://arxiv.org/abs/1702.02430}{{\tt arXiv:1702.02430}}].

\bibitem{Alanne:2017oqj}
T.~Alanne and F.~Goertz, {\it {Extended Dark Matter EFT}},  {\em Eur. Phys. J.
  C} {\bf 80} (2020) 446, [\href{http://arxiv.org/abs/1712.07626}{{\tt
  arXiv:1712.07626}}].

\bibitem{Alanne:2020xcb}
T.~Alanne, G.~Arcadi, F.~Goertz, V.~Tenorth, and S.~Vogl, {\it
  {Model-independent constraints with extended dark matter EFT}},  {\em JHEP}
  {\bf 10} (2020) 172, [\href{http://arxiv.org/abs/2006.07174}{{\tt
  arXiv:2006.07174}}].

\bibitem{Bai:2010hh}
Y.~Bai, P.~J. Fox, and R.~Harnik, {\it {The Tevatron at the Frontier of Dark
  Matter Direct Detection}},  {\em JHEP} {\bf 12} (2010) 048,
  [\href{http://arxiv.org/abs/1005.3797}{{\tt arXiv:1005.3797}}].

\bibitem{Dreiner:2013vla}
H.~Dreiner, D.~Schmeier, and J.~Tattersall, {\it {Contact Interactions Probe
  Effective Dark Matter Models at the LHC}},  {\em EPL} {\bf 102} (2013) 51001,
  [\href{http://arxiv.org/abs/1303.3348}{{\tt arXiv:1303.3348}}].

\bibitem{Zhou:2013fla}
N.~Zhou, D.~Berge, and D.~Whiteson, {\it {Mono-everything: combined limits on
  dark matter production at colliders from multiple final states}},  {\em Phys.
  Rev. D} {\bf 87} (2013) 095013, [\href{http://arxiv.org/abs/1302.3619}{{\tt
  arXiv:1302.3619}}].

\bibitem{Fox:2012ee}
P.~J. Fox, R.~Harnik, R.~Primulando, and C.-T. Yu, {\it {Taking a Razor to Dark
  Matter Parameter Space at the LHC}},  {\em Phys. Rev. D} {\bf 86} (2012)
  015010, [\href{http://arxiv.org/abs/1203.1662}{{\tt arXiv:1203.1662}}].

\bibitem{Rajaraman:2011wf}
A.~Rajaraman, W.~Shepherd, T.~M. Tait, and A.~M. Wijangco, {\it {LHC Bounds on
  Interactions of Dark Matter}},  {\em Phys. Rev. D} {\bf 84} (2011) 095013,
  [\href{http://arxiv.org/abs/1108.1196}{{\tt arXiv:1108.1196}}].

\bibitem{Goodman:2010ku}
J.~Goodman, M.~Ibe, {\em et.~al.}, {\it {Constraints on Dark Matter from
  Colliders}},  {\em Phys. Rev. D} {\bf 82} (2010) 116010,
  [\href{http://arxiv.org/abs/1008.1783}{{\tt arXiv:1008.1783}}].

\bibitem{Fox:2011pm}
P.~J. Fox, R.~Harnik, J.~Kopp, and Y.~Tsai, {\it {Missing Energy Signatures of
  Dark Matter at the LHC}},  {\em Phys. Rev. D} {\bf 85} (2012) 056011,
  [\href{http://arxiv.org/abs/1109.4398}{{\tt arXiv:1109.4398}}].

\bibitem{Beltran:2010ww}
M.~Beltran, D.~Hooper, E.~W. Kolb, Z.~A. Krusberg, and T.~M. Tait, {\it
  {Maverick dark matter at colliders}},  {\em JHEP} {\bf 09} (2010) 037,
  [\href{http://arxiv.org/abs/1002.4137}{{\tt arXiv:1002.4137}}].

\bibitem{Buchmueller:2013dya}
O.~Buchmueller, M.~J. Dolan, and C.~McCabe, {\it {Beyond Effective Field Theory
  for Dark Matter Searches at the LHC}},  {\em JHEP} {\bf 01} (2014) 025,
  [\href{http://arxiv.org/abs/1308.6799}{{\tt arXiv:1308.6799}}].

\bibitem{Belyaev:2016pxe}
A.~Belyaev, L.~Panizzi, A.~Pukhov, and M.~Thomas, {\it {Dark Matter
  characterization at the LHC in the Effective Field Theory approach}},  {\em
  JHEP} {\bf 04} (2017) 110, [\href{http://arxiv.org/abs/1610.07545}{{\tt
  arXiv:1610.07545}}].

\bibitem{Pobbe:2017wrj}
F.~Pobbe, A.~Wulzer, and M.~Zanetti, {\it {Setting limits on Effective Field
  Theories: the case of Dark Matter}},  {\em JHEP} {\bf 08} (2017) 074,
  [\href{http://arxiv.org/abs/1704.00736}{{\tt arXiv:1704.00736}}].

\bibitem{ATLAS:2012ky}
ATLAS: G.~Aad {\em et.~al.}, {\it {Search for dark matter candidates and large
  extra dimensions in events with a jet and missing transverse momentum with
  the ATLAS detector}},  {\em JHEP} {\bf 04} (2013) 075,
  [\href{http://arxiv.org/abs/1210.4491}{{\tt arXiv:1210.4491}}].

\bibitem{Chatrchyan:2012me}
CMS: S.~Chatrchyan {\em et.~al.}, {\it {Search for Dark Matter and Large Extra
  Dimensions in Monojet Events in $pp$ Collisions at $\sqrt{s}=7$ TeV}},  {\em
  JHEP} {\bf 09} (2012) 094, [\href{http://arxiv.org/abs/1206.5663}{{\tt
  arXiv:1206.5663}}].

\bibitem{Buckley:2011kk}
M.~R. Buckley, {\it {Asymmetric Dark Matter and Effective Operators}},  {\em
  Phys. Rev. D} {\bf 84} (2011) 043510,
  [\href{http://arxiv.org/abs/1104.1429}{{\tt arXiv:1104.1429}}].

\bibitem{Cheung:2012gi}
K.~Cheung, P.-Y. Tseng, Y.-L.~S. Tsai, and T.-C. Yuan, {\it {Global Constraints
  on Effective Dark Matter Interactions: Relic Density, Direct Detection,
  Indirect Detection, and Collider}},  {\em \jcap} {\bf 1205} (2012) 001,
  [\href{http://arxiv.org/abs/1201.3402}{{\tt arXiv:1201.3402}}].

\bibitem{MarchRussell:2012hi}
J.~March-Russell, J.~Unwin, and S.~M. West, {\it {Closing in on Asymmetric Dark
  Matter I: Model independent limits for interactions with quarks}},  {\em
  JHEP} {\bf 08} (2012) 029, [\href{http://arxiv.org/abs/1203.4854}{{\tt
  arXiv:1203.4854}}].

\bibitem{Zheng:2010js}
J.-M. Zheng, Z.-H. Yu, {\em et.~al.}, {\it {Constraining the interaction
  strength between dark matter and visible matter: I. fermionic dark matter}},
  {\em Nucl. Phys. B} {\bf 854} (2012) 350--374,
  [\href{http://arxiv.org/abs/1012.2022}{{\tt arXiv:1012.2022}}].

\bibitem{Belyaev:2018pqr}
A.~Belyaev, E.~Bertuzzo, {\em et.~al.}, {\it {Interplay of the LHC and non-LHC
  Dark Matter searches in the Effective Field Theory approach}},  {\em Phys.
  Rev. D} {\bf 99} (2019) 015006, [\href{http://arxiv.org/abs/1807.03817}{{\tt
  arXiv:1807.03817}}].

\bibitem{Bertuzzo:2017lwt}
E.~Bertuzzo, C.~J. Caniu~Barros, and G.~Grilli~di Cortona, {\it {MeV Dark
  Matter: Model Independent Bounds}},  {\em JHEP} {\bf 09} (2017) 116,
  [\href{http://arxiv.org/abs/1707.00725}{{\tt arXiv:1707.00725}}].

\bibitem{DelNobile:2013sia}
M.~Cirelli, E.~Del~Nobile, and P.~Panci, {\it {Tools for model-independent
  bounds in direct dark matter searches}},  {\em JCAP} {\bf 10} (2013) 019,
  [\href{http://arxiv.org/abs/1307.5955}{{\tt arXiv:1307.5955}}].

\bibitem{Kumar:2013iva}
J.~Kumar and D.~Marfatia, {\it {Matrix element analyses of dark matter
  scattering and annihilation}},  {\em \prd} {\bf 88} (2013) 014035,
  [\href{http://arxiv.org/abs/1305.1611}{{\tt arXiv:1305.1611}}].

\bibitem{Balazs:2014rsa}
C.~Bal\'azs, T.~Li, and J.~L. Newstead, {\it {Thermal dark matter implies new
  physics not far above the weak scale}},  {\em JHEP} {\bf 08} (2014) 061,
  [\href{http://arxiv.org/abs/1403.5829}{{\tt arXiv:1403.5829}}].

\bibitem{Liem:2016xpm}
S.~{Liem}, G.~{Bertone}, {\em et.~al.}, {\it {Effective field theory of dark
  matter: a global analysis}},  {\em \jhep} {\bf 9} (2016) 77,
  [\href{http://arxiv.org/abs/1603.05994}{{\tt arXiv:1603.05994}}].

\bibitem{Matsumoto:2014rxa}
S.~Matsumoto, S.~Mukhopadhyay, and Y.-L.~S. Tsai, {\it {Singlet Majorana
  fermion dark matter: a comprehensive analysis in effective field theory}},
  {\em \jhep} {\bf 10} (2014) 155, [\href{http://arxiv.org/abs/1407.1859}{{\tt
  arXiv:1407.1859}}].

\bibitem{Blennow:2015yca}
M.~Blennow, P.~Coloma, E.~Fernandez-Martinez, P.~A.~N. Machado, and
  B.~Zaldivar, {\it {Global constraints on vector-like WIMP effective
  interactions}},  {\em JCAP} {\bf 04} (2016) 015,
  [\href{http://arxiv.org/abs/1509.01587}{{\tt arXiv:1509.01587}}].

\bibitem{Matsumoto:2016hbs}
S.~Matsumoto, S.~Mukhopadhyay, and Y.-L.~S. Tsai, {\it {Effective Theory of
  WIMP Dark Matter supplemented by Simplified Models: Singlet-like Majorana
  fermion case}},  {\em \prd} {\bf 94} (2016) 065034,
  [\href{http://arxiv.org/abs/1604.02230}{{\tt arXiv:1604.02230}}].

\bibitem{gambit}
\GB Collaboration: P.~{Athron}, C.~{Bal{\'a}zs}, {\em et.~al.}, {\it {GAMBIT:
  The Global and Modular Beyond-the-Standard-Model Inference Tool}},  {\em
  \epjc} {\bf 77} (2017) 784, [\href{http://arxiv.org/abs/1705.07908}{{\tt
  arXiv:1705.07908}}]. Addendum in \cite{gambit_addendum}.

\bibitem{Duerr:2014wra}
M.~Duerr and P.~Fileviez~Perez, {\it {Theory for Baryon Number and Dark Matter
  at the LHC}},  {\em Phys. Rev. D} {\bf 91} (2015) 095001,
  [\href{http://arxiv.org/abs/1409.8165}{{\tt arXiv:1409.8165}}].

\bibitem{Dudas:2013sia}
E.~Dudas, L.~Heurtier, Y.~Mambrini, and B.~Zaldivar, {\it {Extra U(1),
  effective operators, anomalies and dark matter}},  {\em JHEP} {\bf 11} (2013)
  083, [\href{http://arxiv.org/abs/1307.0005}{{\tt arXiv:1307.0005}}].

\bibitem{Bauer:2018egk}
M.~Bauer, S.~Diefenbacher, T.~Plehn, M.~Russell, and D.~A. Camargo, {\it {Dark
  Matter in Anomaly-Free Gauge Extensions}},  {\em SciPost Phys.} {\bf 5}
  (2018) 036, [\href{http://arxiv.org/abs/1805.01904}{{\tt arXiv:1805.01904}}].

\bibitem{DarkBit}
\GB Dark Matter Workgroup: T.~{Bringmann}, J.~{Conrad}, {\em et.~al.}, {\it
  {DarkBit: A GAMBIT module for computing dark matter observables and
  likelihoods}},  {\em \epjc} {\bf 77} (2017) 831,
  [\href{http://arxiv.org/abs/1705.07920}{{\tt arXiv:1705.07920}}].

\bibitem{Fitzpatrick:2012ix}
A.~L. Fitzpatrick, W.~Haxton, E.~Katz, N.~Lubbers, and Y.~Xu, {\it {The
  Effective Field Theory of Dark Matter Direct Detection}},  {\em \jcap} {\bf
  1302} (2013) 004, [\href{http://arxiv.org/abs/1203.3542}{{\tt
  arXiv:1203.3542}}].

\bibitem{CosmoBit}
\GB Cosmology Workgroup: J.~J. {Renk}, P.~{St\"ocker}, {\em et.~al.}, {\it
  {CosmoBit: A GAMBIT module for computing cosmological observables and
  likelihoods}},  {\em \jcap} {\bf 02} (2021) 022,
  [\href{http://arxiv.org/abs/2009.03286}{{\tt arXiv:2009.03286}}].

\bibitem{Gonzalo:2021cnq}
T.~E. Gonzalo, {\it {GAMBIT: The Global and Modular BSM Inference Tool}},  in
  {\em {Tools for High Energy Physics and Cosmology}} (2021)
  [\href{http://arxiv.org/abs/2105.03165}{{\tt arXiv:2105.03165}}].

\bibitem{GUM}
S.~Bloor, T.~E. Gonzalo, {\em et.~al.}, {\it {The GAMBIT Universal Model
  Machine: from Lagrangians to Likelihoods}},
  \href{http://arxiv.org/abs/2107.00030}{{\tt arXiv:2107.00030}}.

\bibitem{Shoemaker:2011vi}
I.~M. Shoemaker and L.~Vecchi, {\it {Unitarity and Monojet Bounds on Models for
  DAMA, CoGeNT, and CRESST-II}},  {\em Phys. Rev. D} {\bf 86} (2012) 015023,
  [\href{http://arxiv.org/abs/1112.5457}{{\tt arXiv:1112.5457}}].

\bibitem{Busoni:2013lha}
G.~Busoni, A.~De~Simone, E.~Morgante, and A.~Riotto, {\it {On the Validity of
  the Effective Field Theory for Dark Matter Searches at the LHC}},  {\em Phys.
  Lett. B} {\bf 728} (2014) 412--421,
  [\href{http://arxiv.org/abs/1307.2253}{{\tt arXiv:1307.2253}}].

\bibitem{Busoni:2014sya}
G.~Busoni, A.~De~Simone, J.~Gramling, E.~Morgante, and A.~Riotto, {\it {On the
  Validity of the Effective Field Theory for Dark Matter Searches at the LHC,
  Part II: Complete Analysis for the $s$-channel}},  {\em JCAP} {\bf 06} (2014)
  060, [\href{http://arxiv.org/abs/1402.1275}{{\tt arXiv:1402.1275}}].

\bibitem{Busoni:2014haa}
G.~Busoni, A.~De~Simone, T.~Jacques, E.~Morgante, and A.~Riotto, {\it {On the
  Validity of the Effective Field Theory for Dark Matter Searches at the LHC
  Part III: Analysis for the $t$-channel}},  {\em JCAP} {\bf 09} (2014) 022,
  [\href{http://arxiv.org/abs/1405.3101}{{\tt arXiv:1405.3101}}].

\bibitem{Endo:2014mja}
M.~Endo and Y.~Yamamoto, {\it {Unitarity Bounds on Dark Matter Effective
  Interactions at LHC}},  {\em JHEP} {\bf 06} (2014) 126,
  [\href{http://arxiv.org/abs/1403.6610}{{\tt arXiv:1403.6610}}].

\bibitem{Bell:2016obu}
N.~Bell, G.~Busoni, A.~Kobakhidze, D.~M. Long, and M.~A. Schmidt, {\it
  {Unitarisation of EFT Amplitudes for Dark Matter Searches at the LHC}},  {\em
  \jhep} {\bf 08} (2016) 125, [\href{http://arxiv.org/abs/1606.02722}{{\tt
  arXiv:1606.02722}}].

\bibitem{Racco:2015dxa}
D.~Racco, A.~Wulzer, and F.~Zwirner, {\it {Robust collider limits on
  heavy-mediator Dark Matter}},  {\em \jhep} {\bf 05} (2015) 009,
  [\href{http://arxiv.org/abs/1502.04701}{{\tt arXiv:1502.04701}}].

\bibitem{Bruggisser:2016nzw}
S.~Bruggisser, F.~Riva, and A.~Urbano, {\it {The Last Gasp of Dark Matter
  Effective Theory}},  {\em \jhep} {\bf 11} (2016) 069,
  [\href{http://arxiv.org/abs/1607.02475}{{\tt arXiv:1607.02475}}].

\bibitem{Griest:1989wd}
K.~Griest and M.~Kamionkowski, {\it {Unitarity Limits on the Mass and Radius of
  Dark Matter Particles}},  {\em Phys. Rev. Lett.} {\bf 64} (1990) 615.

\bibitem{Zenodo_DMEFT}
\GB Collaboration, \textit{Supplementary Data: Thermal WIMPs and the Scale of
  New Physics: Global Fits of Dirac Dark Matter Effective Field Theories.},
  (2021),
  \href{https://zenodo.org/record/4836397}{\nolinkurl{https://zenodo.org/record/4836397}}.

\bibitem{Bishara:2017nnn}
F.~Bishara, J.~Brod, B.~Grinstein, and J.~Zupan, {\it {DirectDM: a tool for
  dark matter direct detection}},  \href{http://arxiv.org/abs/1708.02678}{{\tt
  arXiv:1708.02678}}.

\bibitem{Brod:2017bsw}
J.~Brod, A.~Gootjes-Dreesbach, M.~Tammaro, and J.~Zupan, {\it {Effective Field
  Theory for Dark Matter Direct Detection up to Dimension Seven}},  {\em \jhep}
  {\bf 10} (2018) 065, [\href{http://arxiv.org/abs/1710.10218}{{\tt
  arXiv:1710.10218}}].

\bibitem{Kopp:2009et}
J.~Kopp, V.~Niro, T.~Schwetz, and J.~Zupan, {\it {DAMA/LIBRA and leptonically
  interacting Dark Matter}},  {\em Phys. Rev. D} {\bf 80} (2009) 083502,
  [\href{http://arxiv.org/abs/0907.3159}{{\tt arXiv:0907.3159}}].

\bibitem{Fox:2011fx}
P.~J. Fox, R.~Harnik, J.~Kopp, and Y.~Tsai, {\it {LEP Shines Light on Dark
  Matter}},  {\em Phys. Rev. D} {\bf 84} (2011) 014028,
  [\href{http://arxiv.org/abs/1103.0240}{{\tt arXiv:1103.0240}}].

\bibitem{Weiner:2012gm}
N.~Weiner and I.~Yavin, {\it {UV completions of magnetic inelastic and Rayleigh
  dark matter for the Fermi Line(s)}},  {\em Phys. Rev. D} {\bf 87} (2013)
  023523, [\href{http://arxiv.org/abs/1209.1093}{{\tt arXiv:1209.1093}}].

\bibitem{Frandsen:2012db}
M.~T. Frandsen, U.~Haisch, F.~Kahlhoefer, P.~Mertsch, and K.~Schmidt-Hoberg,
  {\it {Loop-induced dark matter direct detection signals from gamma-ray
  lines}},  {\em JCAP} {\bf 10} (2012) 033,
  [\href{http://arxiv.org/abs/1207.3971}{{\tt arXiv:1207.3971}}].

\bibitem{Paz:2020pbc}
G.~Paz, A.~A. Petrov, M.~Tammaro, and J.~Zupan, {\it {Shining dark matter in
  Xenon1T}},  {\em \prd} {\bf 103} (2021) L051703,
  [\href{http://arxiv.org/abs/2006.12462}{{\tt arXiv:2006.12462}}].

\bibitem{Kavanagh:2018xeh}
B.~J. Kavanagh, P.~Panci, and R.~Ziegler, {\it {Faint Light from Dark Matter:
  Classifying and Constraining Dark Matter-Photon Effective Operators}},  {\em
  JHEP} {\bf 04} (2019) 089, [\href{http://arxiv.org/abs/1810.00033}{{\tt
  arXiv:1810.00033}}].

\bibitem{Arina:2020mxo}
C.~Arina, A.~Cheek, K.~Mimasu, and L.~Pagani, {\it {Light and Darkness:
  consistently coupling dark matter to photons via effective operators}},  {\em
  Eur. Phys. J. C} {\bf 81} (2021) 223,
  [\href{http://arxiv.org/abs/2005.12789}{{\tt arXiv:2005.12789}}].

\bibitem{Haisch:2016usn}
U.~Haisch, F.~Kahlhoefer, and T.~M.~P. Tait, {\it {On Mono-W Signatures in
  Spin-1 Simplified Models}},  {\em Phys. Lett.} {\bf B760} (2016) 207--213,
  [\href{http://arxiv.org/abs/1603.01267}{{\tt arXiv:1603.01267}}].

\bibitem{Hill:2014yxa}
R.~J. Hill and M.~P. Solon, {\it {Standard Model anatomy of WIMP dark matter
  direct detection II: QCD analysis and hadronic matrix elements}},  {\em \prd}
  {\bf 91} (2015) 043505, [\href{http://arxiv.org/abs/1409.8290}{{\tt
  arXiv:1409.8290}}].

\bibitem{Bishara:2018vix}
F.~Bishara, J.~Brod, B.~Grinstein, and J.~Zupan, {\it {Renormalization Group
  Effects in Dark Matter Interactions}},  {\em JHEP} {\bf 03} (2020) 089,
  [\href{http://arxiv.org/abs/1809.03506}{{\tt arXiv:1809.03506}}].

\bibitem{Brod:2018ust}
J.~Brod, B.~Grinstein, E.~Stamou, and J.~Zupan, {\it {Weak mixing below the
  weak scale in dark-matter direct detection}},  {\em JHEP} {\bf 02} (2018)
  174, [\href{http://arxiv.org/abs/1801.04240}{{\tt arXiv:1801.04240}}].

\bibitem{Haisch:2013uaa}
U.~Haisch and F.~Kahlhoefer, {\it {On the importance of loop-induced
  spin-independent interactions for dark matter direct detection}},  {\em
  \jcap} {\bf 1304} (2013) 050, [\href{http://arxiv.org/abs/1302.4454}{{\tt
  arXiv:1302.4454}}].

\bibitem{Crivellin:2014qxa}
A.~Crivellin, F.~D'Eramo, and M.~Procura, {\it {New Constraints on Dark Matter
  Effective Theories from Standard Model Loops}},  {\em Phys. Rev. Lett.} {\bf
  112} (2014) 191304, [\href{http://arxiv.org/abs/1402.1173}{{\tt
  arXiv:1402.1173}}].

\bibitem{Haisch:2012kf}
U.~Haisch, F.~Kahlhoefer, and J.~Unwin, {\it {The impact of heavy-quark loops
  on LHC dark matter searches}},  {\em \jhep} {\bf 07} (2013) 125,
  [\href{http://arxiv.org/abs/1208.4605}{{\tt arXiv:1208.4605}}].

\bibitem{Berlin:2014cfa}
A.~Berlin, T.~Lin, and L.-T. Wang, {\it {Mono-Higgs Detection of Dark Matter at
  the LHC}},  {\em JHEP} {\bf 06} (2014) 078,
  [\href{http://arxiv.org/abs/1402.7074}{{\tt arXiv:1402.7074}}].

\bibitem{Agnese:2015nto}
SuperCDMS: R.~Agnese {\em et.~al.}, {\it {New Results from the Search for
  Low-Mass Weakly Interacting Massive Particles with the CDMS Low Ionization
  Threshold Experiment}},  {\em \prl} {\bf 116} (2016) 071301,
  [\href{http://arxiv.org/abs/1509.02448}{{\tt arXiv:1509.02448}}].

\bibitem{Angloher:2015ewa}
CRESST: G.~Angloher {\em et.~al.}, {\it {Results on light dark matter particles
  with a low-threshold CRESST-II detector}},  {\em Eur. Phys. J.} {\bf C76}
  (2016) 25, [\href{http://arxiv.org/abs/1509.01515}{{\tt arXiv:1509.01515}}].

\bibitem{Abdelhameed:2019hmk}
CRESST: A.~H. Abdelhameed {\em et.~al.}, {\it {First results from the
  CRESST-III low-mass dark matter program}},  {\em \prd} {\bf 100} (2019)
  102002, [\href{http://arxiv.org/abs/1904.00498}{{\tt arXiv:1904.00498}}].

\bibitem{Agnes:2018fwg}
DarkSide: P.~Agnes {\em et.~al.}, {\it {DarkSide-50 532-day Dark Matter Search
  with Low-Radioactivity Argon}},  {\em \prd} {\bf 98} (2018) 102006,
  [\href{http://arxiv.org/abs/1802.07198}{{\tt arXiv:1802.07198}}].

\bibitem{LUXrun2}
LUX: D.~S. Akerib {\em et.~al.}, {\it {Results from a search for dark matter in
  the complete LUX exposure}},  {\em \prl} {\bf 118} (2017) 021303,
  [\href{http://arxiv.org/abs/1608.07648}{{\tt arXiv:1608.07648}}].

\bibitem{Amole:2017dex}
PICO: C.~Amole {\em et.~al.}, {\it {Dark Matter Search Results from the PICO-60
  C$_3$F$_8$ Bubble Chamber}},  {\em \prl} {\bf 118} (2017) 251301,
  [\href{http://arxiv.org/abs/1702.07666}{{\tt arXiv:1702.07666}}].

\bibitem{Amole:2019fdf}
PICO: C.~Amole {\em et.~al.}, {\it {Dark Matter Search Results from the
  Complete Exposure of the PICO-60 C$_3$F$_8$ Bubble Chamber}},  {\em \prd}
  {\bf 100} (2019) 022001, [\href{http://arxiv.org/abs/1902.04031}{{\tt
  arXiv:1902.04031}}].

\bibitem{Tan:2016zwf}
PandaX-II: A.~Tan {\em et.~al.}, {\it {Dark Matter Results from First 98.7 Days
  of Data from the PandaX-II Experiment}},  {\em \prl} {\bf 117} (2016) 121303,
  [\href{http://arxiv.org/abs/1607.07400}{{\tt arXiv:1607.07400}}].

\bibitem{Cui:2017nnn}
PandaX-II: X.~Cui {\em et.~al.}, {\it {Dark Matter Results From 54-Ton-Day
  Exposure of PandaX-II Experiment}},  {\em \prl} {\bf 119} (2017) 181302,
  [\href{http://arxiv.org/abs/1708.06917}{{\tt arXiv:1708.06917}}].

\bibitem{Aprile:2018dbl}
XENON: E.~Aprile {\em et.~al.}, {\it {Dark Matter Search Results from a One
  Ton-Year Exposure of XENON1T}},  {\em \prl} {\bf 121} (2018) 111302,
  [\href{http://arxiv.org/abs/1805.12562}{{\tt arXiv:1805.12562}}].

\bibitem{Aad:2021egl}
ATLAS: G.~Aad {\em et.~al.}, {\it {Search for new phenomena in events with an
  energetic jet and missing transverse momentum in $pp$ collisions at $\sqrt{s}
  = 13$ TeV with the ATLAS detector}},
  \href{http://arxiv.org/abs/2102.10874}{{\tt arXiv:2102.10874}}.

\bibitem{Sirunyan:2017jix}
CMS: A.~M. Sirunyan {\em et.~al.}, {\it {Search for new physics in final states
  with an energetic jet or a hadronically decaying $W$ or $Z$ boson and
  transverse momentum imbalance at $\sqrt{s}=13\text{ }\text{ }\mathrm{TeV}$}},
   {\em Phys. Rev. D} {\bf 97} (2018) 092005,
  [\href{http://arxiv.org/abs/1712.02345}{{\tt arXiv:1712.02345}}].

\bibitem{LATdwarfP8}
Fermi-LAT: M.~Ackermann {\em et.~al.}, {\it {Searching for Dark Matter
  Annihilation from Milky Way Dwarf Spheroidal Galaxies with Six Years of Fermi
  Large Area Telescope Data}},  {\em \prl} {\bf 115} (2015) 231301,
  [\href{http://arxiv.org/abs/1503.02641}{{\tt arXiv:1503.02641}}].

\bibitem{IC79_SUSY}
IceCube Collaboration: M.~G. {Aartsen} {\em et.~al.}, {\it {Improved limits on
  dark matter annihilation in the Sun with the 79-string IceCube detector and
  implications for supersymmetry}},  {\em \jcap} {\bf 04} (2016) 022,
  [\href{http://arxiv.org/abs/1601.00653}{{\tt arXiv:1601.00653}}].

\bibitem{Aghanim:2018eyx}
Planck: N.~Aghanim {\em et.~al.}, {\it {Planck 2018 results. VI. Cosmological
  parameters}},  {\em Astron. Astrophys.} {\bf 641} (2020) A6,
  [\href{http://arxiv.org/abs/1807.06209}{{\tt arXiv:1807.06209}}].

\bibitem{Anand:2013yka}
N.~Anand, A.~L. Fitzpatrick, and W.~C. Haxton, {\it {Weakly interacting massive
  particle-nucleus elastic scattering response}},  {\em \prc} {\bf 89} (2014)
  065501, [\href{http://arxiv.org/abs/1308.6288}{{\tt arXiv:1308.6288}}].

\bibitem{Bishara:2016hek}
F.~Bishara, J.~Brod, B.~Grinstein, and J.~Zupan, {\it {Chiral Effective Theory
  of Dark Matter Direct Detection}},  {\em \jcap} {\bf 1702} (2017) 009,
  [\href{http://arxiv.org/abs/1611.00368}{{\tt arXiv:1611.00368}}].

\bibitem{Bishara:2017pfq}
F.~Bishara, J.~Brod, B.~Grinstein, and J.~Zupan, {\it {From quarks to nucleons
  in dark matter direct detection}},  {\em \jhep} {\bf 11} (2017) 059,
  [\href{http://arxiv.org/abs/1707.06998}{{\tt arXiv:1707.06998}}].

\bibitem{PDG20}
Particle Data Group: P.~A. Zyla {\em et.~al.}, {\it {Review of Particle
  Physics}},  {\em Prog. Theor. Exp. Phys.} {\bf 083} (2020) C01.

\bibitem{Crivellin:2013ipa}
A.~Crivellin, M.~Hoferichter, and M.~Procura, {\it {Accurate evaluation of
  hadronic uncertainties in spin-independent WIMP-nucleon scattering:
  Disentangling two- and three-flavor effects}},  {\em Phys. Rev. D} {\bf 89}
  (2014) 054021, [\href{http://arxiv.org/abs/1312.4951}{{\tt
  arXiv:1312.4951}}].

\bibitem{Djukanovic:2019jtp}
D.~Djukanovic, K.~Ottnad, J.~Wilhelm, and H.~Wittig, {\it {Strange
  electromagnetic form factors of the nucleon with $N_f = 2 + 1$
  $\mathcal{O}(a)$-improved Wilson fermions}},  {\em Phys. Rev. Lett.} {\bf
  123} (2019) 212001, [\href{http://arxiv.org/abs/1903.12566}{{\tt
  arXiv:1903.12566}}].

\bibitem{Sufian:2016pex}
R.~S. Sufian, Y.-B. Yang, {\em et.~al.}, {\it {Strange Quark Magnetic Moment of
  the Nucleon at the Physical Point}},  {\em Phys. Rev. Lett.} {\bf 118} (2017)
  042001, [\href{http://arxiv.org/abs/1606.07075}{{\tt arXiv:1606.07075}}].

\bibitem{Gupta:2018lvp}
R.~Gupta, B.~Yoon, {\em et.~al.}, {\it {Flavor diagonal tensor charges of the
  nucleon from (2+1+1)-flavor lattice QCD}},  {\em Phys. Rev. D} {\bf 98}
  (2018) 091501, [\href{http://arxiv.org/abs/1808.07597}{{\tt
  arXiv:1808.07597}}].

\bibitem{Aoki:2019cca}
Flavour Lattice Averaging Group: S.~Aoki {\em et.~al.}, {\it {FLAG Review 2019:
  Flavour Lattice Averaging Group (FLAG)}},  {\em Eur. Phys. J. C} {\bf 80}
  (2020) 113, [\href{http://arxiv.org/abs/1902.08191}{{\tt arXiv:1902.08191}}].

\bibitem{Liang:2018pis}
J.~Liang, Y.-B. Yang, T.~Draper, M.~Gong, and K.-F. Liu, {\it {Quark spins and
  Anomalous Ward Identity}},  {\em Phys. Rev. D} {\bf 98} (2018) 074505,
  [\href{http://arxiv.org/abs/1806.08366}{{\tt arXiv:1806.08366}}].

\bibitem{Pasquini:2005dk}
B.~Pasquini, M.~Pincetti, and S.~Boffi, {\it {Chiral-odd generalized parton
  distributions in constituent quark models}},  {\em Phys. Rev.} {\bf D72}
  (2005) 094029, [\href{http://arxiv.org/abs/hep-ph/0510376}{{\tt
  hep-ph/0510376}}].

\bibitem{HP}
GAMBIT Collaboration: P.~Athron {\em et.~al.}, {\it {Global analyses of Higgs
  portal singlet dark matter models using GAMBIT}},  {\em \epjc} {\bf 79}
  (2019) 38, [\href{http://arxiv.org/abs/1808.10465}{{\tt arXiv:1808.10465}}].

\bibitem{Horsley:2011wr}
QCDSF-UKQCD: R.~Horsley, Y.~Nakamura, {\em et.~al.}, {\it {Hyperon sigma terms
  for 2+1 quark flavours}},  {\em Phys. Rev. D} {\bf 85} (2012) 034506,
  [\href{http://arxiv.org/abs/1110.4971}{{\tt arXiv:1110.4971}}].

\bibitem{Durr:2015dna}
S.~Durr {\em et.~al.}, {\it {Lattice computation of the nucleon scalar quark
  contents at the physical point}},  {\em Phys. Rev. Lett.} {\bf 116} (2016)
  172001, [\href{http://arxiv.org/abs/1510.08013}{{\tt arXiv:1510.08013}}].

\bibitem{Yang:2015uis}
xQCD: Y.-B. Yang, A.~Alexandru, T.~Draper, J.~Liang, and K.-F. Liu, {\it
  {$\pi$N and strangeness sigma terms at the physical point with chiral
  fermions}},  {\em Phys. Rev. D} {\bf 94} (2016) 054503,
  [\href{http://arxiv.org/abs/1511.09089}{{\tt arXiv:1511.09089}}].

\bibitem{Abdel-Rehim:2016won}
ETM: A.~Abdel-Rehim, C.~Alexandrou, {\em et.~al.}, {\it {Direct Evaluation of
  the Quark Content of Nucleons from Lattice QCD at the Physical Point}},  {\em
  \prl} {\bf 116} (2016) 252001, [\href{http://arxiv.org/abs/1601.01624}{{\tt
  arXiv:1601.01624}}].

\bibitem{Bali:2016lvx}
RQCD: G.~S. Bali, S.~Collins, {\em et.~al.}, {\it {Direct determinations of the
  nucleon and pion $\sigma$ terms at nearly physical quark masses}},  {\em
  \prd} {\bf 93} (2016) 094504, [\href{http://arxiv.org/abs/1603.00827}{{\tt
  arXiv:1603.00827}}].

\bibitem{Alexandrou:2019brg}
C.~Alexandrou, S.~Bacchio, {\em et.~al.}, {\it {Nucleon axial, tensor, and
  scalar charges and $\sigma$-terms in lattice QCD}},  {\em Phys. Rev. D} {\bf
  102} (2020) 054517, [\href{http://arxiv.org/abs/1909.00485}{{\tt
  arXiv:1909.00485}}].

\bibitem{Yamanaka:2018uud}
JLQCD: N.~Yamanaka, S.~Hashimoto, T.~Kaneko, and H.~Ohki, {\it {Nucleon charges
  with dynamical overlap fermions}},  {\em Phys. Rev. D} {\bf 98} (2018)
  054516, [\href{http://arxiv.org/abs/1805.10507}{{\tt arXiv:1805.10507}}].

\bibitem{Borsanyi:2020bpd}
S.~Borsanyi, Z.~Fodor, {\em et.~al.}, {\it {Ab-initio calculation of the proton
  and the neutron's scalar couplings for new physics searches}},
  \href{http://arxiv.org/abs/2007.03319}{{\tt arXiv:2007.03319}}.

\bibitem{Alarcon:2011zs}
J.~M. Alarcon, J.~Martin~Camalich, and J.~A. Oller, {\it {The chiral
  representation of the $\pi N$ scattering amplitude and the pion-nucleon sigma
  term}},  {\em \prd} {\bf 85} (2012) 051503,
  [\href{http://arxiv.org/abs/1110.3797}{{\tt arXiv:1110.3797}}].

\bibitem{Hoferichter:2015dsa}
M.~Hoferichter, J.~Ruiz~de Elvira, B.~Kubis, and U.-G. Meissner, {\it
  {High-Precision Determination of the Pion-Nucleon \ensuremath{\sigma} Term
  from Roy-Steiner Equations}},  {\em Phys. Rev. Lett.} {\bf 115} (2015)
  092301, [\href{http://arxiv.org/abs/1506.04142}{{\tt arXiv:1506.04142}}].

\bibitem{Dmitrasinovic:2016hup}
V.~Dmitra\v{s}inovi\'c, H.-X. Chen, and A.~Hosaka, {\it {Baryon fields with
  $U_L(3)\texttimes U_R(3)$ chiral symmetry. V. Pion-nucleon and kaon-nucleon
  \ensuremath{\Sigma} terms}},  {\em Phys. Rev. C} {\bf 93} (2016) 065208,
  [\href{http://arxiv.org/abs/1812.03414}{{\tt arXiv:1812.03414}}].

\bibitem{RuizdeElvira:2017stg}
J.~Ruiz~de Elvira, M.~Hoferichter, B.~Kubis, and U.-G. Meissner, {\it
  {Extracting the $\sigma$-term from low-energy pion-nucleon scattering}},
  {\em J. Phys. G} {\bf 45} (2018) 024001,
  [\href{http://arxiv.org/abs/1706.01465}{{\tt arXiv:1706.01465}}].

\bibitem{Friedman:2019zhc}
E.~Friedman and A.~Gal, {\it {The pion-nucleon \ensuremath{\sigma} term from
  pionic atoms}},  {\em Phys. Lett. B} {\bf 792} (2019) 340--344,
  [\href{http://arxiv.org/abs/1901.03130}{{\tt arXiv:1901.03130}}].

\bibitem{Gondolo:1990dk}
P.~Gondolo and G.~Gelmini, {\it {Cosmic abundances of stable particles:
  Improved analysis}},  {\em \nphysa} {\bf 360} (1991) 145--179.

\bibitem{Binder:2017rgn}
T.~Binder, T.~Bringmann, M.~Gustafsson, and A.~Hryczuk, {\it {Early kinetic
  decoupling of dark matter: when the standard way of calculating the thermal
  relic density fails}},  {\em \prd} {\bf 96} (2017) 115010,
  [\href{http://arxiv.org/abs/1706.07433}{{\tt arXiv:1706.07433}}].

\bibitem{Kaplan:2009ag}
D.~E. Kaplan, M.~A. Luty, and K.~M. Zurek, {\it {Asymmetric Dark Matter}},
  {\em \prd} {\bf 79} (2009) 115016,
  [\href{http://arxiv.org/abs/0901.4117}{{\tt arXiv:0901.4117}}].

\bibitem{Pukhov:2004ca}
A.~Pukhov, {\it {CalcHEP 2.3: MSSM, structure functions, event generation,
  batchs, and generation of matrix elements for other packages}},
  \href{http://arxiv.org/abs/hep-ph/0412191}{{\tt hep-ph/0412191}}.

\bibitem{Belyaev:2012qa}
A.~Belyaev, N.~D. Christensen, and A.~Pukhov, {\it {CalcHEP 3.4 for collider
  physics within and beyond the Standard Model}},  {\em \cpc} {\bf 184} (2013)
  1729--1769, [\href{http://arxiv.org/abs/1207.6082}{{\tt arXiv:1207.6082}}].

\bibitem{Arbey:2021gdg}
A.~Arbey and F.~Mahmoudi, {\it {Dark matter and the early Universe: A review}},
   {\em Prog. Part. Nucl. Phys.} {\bf 119} (2021) 103865,
  [\href{http://arxiv.org/abs/2104.11488}{{\tt arXiv:2104.11488}}].

\bibitem{darksusy}
T.~Bringmann, J.~Edsjö, P.~Gondolo, P.~Ullio, and L.~Bergström, {\it
  {DarkSUSY 6 : An Advanced Tool to Compute Dark Matter Properties
  Numerically}},  {\em \jcap} {\bf 1807} (2018) 033,
  [\href{http://arxiv.org/abs/1802.03399}{{\tt arXiv:1802.03399}}].

\bibitem{darksusy4}
P.~Gondolo, J.~Edsjo, {\em et.~al.}, {\it {DarkSUSY: Computing supersymmetric
  dark matter properties numerically}},  {\em \jcap} {\bf 0407} (2004) 008,
  [\href{http://arxiv.org/abs/astro-ph/0406204}{{\tt astro-ph/0406204}}].

\bibitem{Bell:2013wua}
N.~F. Bell, Y.~Cai, and A.~D. Medina, {\it {Co-annihilating Dark Matter:
  Effective Operator Analysis and Collider Phenomenology}},  {\em Phys. Rev. D}
  {\bf 89} (2014) 115001, [\href{http://arxiv.org/abs/1311.6169}{{\tt
  arXiv:1311.6169}}].

\bibitem{Baker:2015qna}
M.~J. Baker {\em et.~al.}, {\it {The Coannihilation Codex}},  {\em JHEP} {\bf
  12} (2015) 120, [\href{http://arxiv.org/abs/1510.03434}{{\tt
  arXiv:1510.03434}}].

\bibitem{Bringmann:2012ez}
T.~Bringmann and C.~Weniger, {\it {Gamma Ray Signals from Dark Matter:
  Concepts, Status and Prospects}},  {\em Phys. Dark Univ.} {\bf 1} (2012)
  194--217, [\href{http://arxiv.org/abs/1208.5481}{{\tt arXiv:1208.5481}}].

\bibitem{TheFermi-LAT:2017vmf}
Fermi-LAT: M.~Ackermann {\em et.~al.}, {\it {The Fermi Galactic Center GeV
  Excess and Implications for Dark Matter}},  {\em Astrophys. J.} {\bf 840}
  (2017) 43, [\href{http://arxiv.org/abs/1704.03910}{{\tt arXiv:1704.03910}}].

\bibitem{Acharyya:2020sbj}
CTA: A.~Acharyya {\em et.~al.}, {\it {Sensitivity of the Cherenkov Telescope
  Array to a dark matter signal from the Galactic centre}},  {\em JCAP} {\bf
  01} (2021) 057, [\href{http://arxiv.org/abs/2007.16129}{{\tt
  arXiv:2007.16129}}].

\bibitem{Choi:2015ara}
Super-Kamiokande: K.~Choi {\em et.~al.}, {\it {Search for neutrinos from
  annihilation of captured low-mass dark matter particles in the Sun by
  Super-Kamiokande}},  {\em Phys. Rev. Lett.} {\bf 114} (2015) 141301,
  [\href{http://arxiv.org/abs/1503.04858}{{\tt arXiv:1503.04858}}].

\bibitem{Aartsen:2016zhm}
IceCube: M.~G. Aartsen {\em et.~al.}, {\it {Search for annihilating dark matter
  in the Sun with 3 years of IceCube data}},  {\em Eur. Phys. J. C} {\bf 77}
  (2017) 146, [\href{http://arxiv.org/abs/1612.05949}{{\tt arXiv:1612.05949}}].
  [Erratum: Eur.Phys.J.C 79, 214 (2019)].

\bibitem{Kozar:2021iur}
N.~Avis~Kozar, A.~Caddell, L.~Fraser-Leach, P.~Scott, and A.~C. Vincent, {\it
  {Capt'n General: A generalized stellar dark matter capture and heat transport
  code}},  (2021) [\href{http://arxiv.org/abs/2105.06810}{{\tt
  arXiv:2105.06810}}].

\bibitem{Catena:2015uha}
R.~Catena and B.~Schwabe, {\it {Form factors for dark matter capture by the Sun
  in effective theories}},  {\em JCAP} {\bf 04} (2015) 042,
  [\href{http://arxiv.org/abs/1501.03729}{{\tt arXiv:1501.03729}}].

\bibitem{Vinyoles:2016djt}
N.~Vinyoles, A.~M. Serenelli, {\em et.~al.}, {\it {A new Generation of Standard
  Solar Models}},  {\em Astrophys. J.} {\bf 835} (2017) 202,
  [\href{http://arxiv.org/abs/1611.09867}{{\tt arXiv:1611.09867}}].

\bibitem{AGSS}
M.~{Asplund}, N.~{Grevesse}, A.~J. {Sauval}, and P.~{Scott}, {\it {The chemical
  composition of the Sun}},  {\em \araa} {\bf 47} (2009) 481--522,
  [\href{http://arxiv.org/abs/0909.0948}{{\tt arXiv:0909.0948}}].

\bibitem{IC79}
IceCube Collaboration: M.~G. {Aartsen}, R.~{Abbasi}, {\em et.~al.}, {\it
  {Search for Dark Matter Annihilations in the Sun with the 79-String IceCube
  Detector}},  {\em \prl} {\bf 110} (2013) 131302,
  [\href{http://arxiv.org/abs/1212.4097}{{\tt arXiv:1212.4097}}].

\bibitem{IC22Methods}
P.~{Scott}, C.~{Savage}, J.~{Edsj{\"o}}, and {the IceCube Collaboration:
  R.~Abbasi et al.}, {\it {Use of event-level neutrino telescope data in global
  fits for theories of new physics}},  {\em \jcap} {\bf 11} (2012) 57,
  [\href{http://arxiv.org/abs/1207.0810}{{\tt arXiv:1207.0810}}].

\bibitem{Slatyer15a}
T.~R. {Slatyer}, {\it {Indirect dark matter signatures in the cosmic dark ages.
  I. Generalizing the bound on s -wave dark matter annihilation from Planck
  results}},  {\em \prd} {\bf 93} (2016) 023527,
  [\href{http://arxiv.org/abs/1506.03811}{{\tt arXiv:1506.03811}}].

\bibitem{Slatyer:2015kla}
T.~R. Slatyer, {\it {Indirect Dark Matter Signatures in the Cosmic Dark Ages
  II. Ionization, Heating and Photon Production from Arbitrary Energy
  Injections}},  {\em \prd} {\bf 93} (2016) 023521,
  [\href{http://arxiv.org/abs/1506.03812}{{\tt arXiv:1506.03812}}].

\bibitem{Stocker:2018avm}
P.~St{\"o}cker, M.~Kr{\"a}mer, J.~Lesgourgues, and V.~Poulin, {\it {Exotic
  energy injection with ExoCLASS: Application to the Higgs portal model and
  evaporating black holes}},  {\em \jcap} {\bf 1803} (2018) 018,
  [\href{http://arxiv.org/abs/1801.01871}{{\tt arXiv:1801.01871}}].

\bibitem{Aghanim:2019ame}
Planck: N.~Aghanim {\em et.~al.}, {\it {Planck 2018 results. V. CMB power
  spectra and likelihoods}},  {\em Astron. Astrophys.} {\bf 641} (2020) A5,
  [\href{http://arxiv.org/abs/1907.12875}{{\tt arXiv:1907.12875}}].

\bibitem{2011MNRAS.416.3017B}
F.~{Beutler}, C.~{Blake}, {\em et.~al.}, {\it {The 6dF Galaxy Survey: baryon
  acoustic oscillations and the local Hubble constant}},  {\em \mnras} {\bf
  416} (2011) 3017--3032, [\href{http://arxiv.org/abs/1106.3366}{{\tt
  arXiv:1106.3366}}].

\bibitem{2015MNRAS.449..835R}
A.~J. {Ross}, L.~{Samushia}, {\em et.~al.}, {\it {The clustering of the SDSS
  DR7 main Galaxy sample - I. A 4 per cent distance measure at z = 0.15}},
  {\em \mnras} {\bf 449} (2015) 835--847,
  [\href{http://arxiv.org/abs/1409.3242}{{\tt arXiv:1409.3242}}].

\bibitem{Alam:2016hwk}
BOSS: S.~Alam {\em et.~al.}, {\it {The clustering of galaxies in the completed
  SDSS-III Baryon Oscillation Spectroscopic Survey: cosmological analysis of
  the DR12 galaxy sample}},  {\em \mnras} {\bf 470} (2017) 2617--2652,
  [\href{http://arxiv.org/abs/1607.03155}{{\tt arXiv:1607.03155}}].

\bibitem{Kopp:2013eka}
J.~Kopp, {\it {Constraints on dark matter annihilation from AMS-02 results}},
  {\em Phys. Rev. D} {\bf 88} (2013) 076013,
  [\href{http://arxiv.org/abs/1304.1184}{{\tt arXiv:1304.1184}}].

\bibitem{Bergstrom:2013jra}
L.~Bergstr{\"o}m, T.~Bringmann, I.~Cholis, D.~Hooper, and C.~Weniger, {\it {New
  limits on dark matter annihilation from AMS cosmic ray positron data}},  {\em
  \prl} {\bf 111} (2013) 171101, [\href{http://arxiv.org/abs/1306.3983}{{\tt
  arXiv:1306.3983}}].

\bibitem{Ibarra:2013zia}
A.~Ibarra, A.~S. Lamperstorfer, and J.~Silk, {\it {Dark matter annihilations
  and decays after the AMS-02 positron measurements}},  {\em Phys. Rev. D} {\bf
  89} (2014) 063539, [\href{http://arxiv.org/abs/1309.2570}{{\tt
  arXiv:1309.2570}}].

\bibitem{Bergstrom:1999jc}
L.~Bergstrom, J.~Edsjo, and P.~Ullio, {\it {Cosmic anti-protons as a probe for
  supersymmetric dark matter?}},  {\em Astrophys. J.} {\bf 526} (1999)
  215--235, [\href{http://arxiv.org/abs/astro-ph/9902012}{{\tt
  astro-ph/9902012}}].

\bibitem{Bringmann:2006im}
T.~Bringmann and P.~Salati, {\it {The galactic antiproton spectrum at high
  energies: Background expectation vs. exotic contributions}},  {\em Phys. Rev.
  D} {\bf 75} (2007) 083006, [\href{http://arxiv.org/abs/astro-ph/0612514}{{\tt
  astro-ph/0612514}}].

\bibitem{Cuoco:2016eej}
A.~Cuoco, M.~Kr\"amer, and M.~Korsmeier, {\it {Novel Dark Matter Constraints
  from Antiprotons in Light of AMS-02}},  {\em Phys. Rev. Lett.} {\bf 118}
  (2017) 191102, [\href{http://arxiv.org/abs/1610.03071}{{\tt
  arXiv:1610.03071}}].

\bibitem{Heisig:2020nse}
J.~Heisig, M.~Korsmeier, and M.~W. Winkler, {\it {Dark matter or correlated
  errors: Systematics of the AMS-02 antiproton excess}},  {\em Phys. Rev. Res.}
  {\bf 2} (2020) 043017, [\href{http://arxiv.org/abs/2005.04237}{{\tt
  arXiv:2005.04237}}].

\bibitem{Boudaud:2019efq}
M.~Boudaud, Y.~G\'enolini, {\em et.~al.}, {\it {AMS-02 antiprotons' consistency
  with a secondary astrophysical origin}},  {\em Phys. Rev. Res.} {\bf 2}
  (2020) 023022, [\href{http://arxiv.org/abs/1906.07119}{{\tt
  arXiv:1906.07119}}].

\bibitem{Johannesson:2016rlh}
G.~J\'ohannesson {\em et.~al.}, {\it {Bayesian analysis of cosmic-ray
  propagation: evidence against homogeneous diffusion}},  {\em Astrophys. J.}
  {\bf 824} (2016) 16, [\href{http://arxiv.org/abs/1602.02243}{{\tt
  arXiv:1602.02243}}].

\bibitem{Bauer:2017fsw}
M.~Bauer, M.~Klassen, and V.~Tenorth, {\it {Universal properties of
  pseudoscalar mediators in dark matter extensions of 2HDMs}},  {\em JHEP} {\bf
  07} (2018) 107, [\href{http://arxiv.org/abs/1712.06597}{{\tt
  arXiv:1712.06597}}].

\bibitem{Brennan:2016xjh}
A.~J. Brennan, M.~F. McDonald, J.~Gramling, and T.~D. Jacques, {\it {Collide
  and Conquer: Constraints on Simplified Dark Matter Models using Mono-X
  Collider Searches}},  {\em JHEP} {\bf 05} (2016) 112,
  [\href{http://arxiv.org/abs/1603.01366}{{\tt arXiv:1603.01366}}].

\bibitem{Alloul:2013bka}
A.~Alloul, N.~D. Christensen, C.~Degrande, C.~Duhr, and B.~Fuks, {\it
  {FeynRules 2.0 - A complete toolbox for tree-level phenomenology}},  {\em
  Comput. Phys. Commun.} {\bf 185} (2014) 2250--2300,
  [\href{http://arxiv.org/abs/1310.1921}{{\tt arXiv:1310.1921}}].

\bibitem{Alwall:2011uj}
J.~Alwall, M.~Herquet, F.~Maltoni, O.~Mattelaer, and T.~Stelzer, {\it {MadGraph
  5 : Going Beyond}},  {\em \jhep} {\bf 06} (2011) 128,
  [\href{http://arxiv.org/abs/1106.0522}{{\tt arXiv:1106.0522}}].

\bibitem{Sjostrand:2007gs}
T.~Sjostrand, S.~Mrenna, and P.~Z. Skands, {\it {A Brief Introduction to PYTHIA
  8.1}},  {\em Comput. Phys. Commun.} {\bf 178} (2008) 852--867,
  [\href{http://arxiv.org/abs/0710.3820}{{\tt arXiv:0710.3820}}].

\bibitem{DELPHES3}
DELPHES 3: J.~de~Favereau, C.~Delaere, {\em et.~al.}, {\it {DELPHES 3, A
  modular framework for fast simulation of a generic collider experiment}},
  {\em \jhep} {\bf 02} (2014) 057, [\href{http://arxiv.org/abs/1307.6346}{{\tt
  arXiv:1307.6346}}].

\bibitem{Collaboration:2242860}
CMS Collaboration, {\it {Simplified likelihood for the re-interpretation of
  public CMS results}},   CMS-NOTE-2017-001, 2017.

\bibitem{ColliderBit}
\GB Collider Workgroup: C.~{Bal{\'a}zs}, A.~{Buckley}, {\em et.~al.}, {\it
  {ColliderBit: a GAMBIT module for the calculation of high-energy collider
  observables and likelihoods}},  {\em \epjc} {\bf 77} (2017) 795,
  [\href{http://arxiv.org/abs/1705.07919}{{\tt arXiv:1705.07919}}].

\bibitem{EWMSSM}
GAMBIT Collaboration: P.~Athron {\em et.~al.}, {\it {Combined collider
  constraints on neutralinos and charginos}},  {\em \epjc} {\bf 79} (2019) 395,
  [\href{http://arxiv.org/abs/1809.02097}{{\tt arXiv:1809.02097}}].

\bibitem{Reid:2014boa}
M.~J. Reid {\em et.~al.}, {\it {Trigonometric Parallaxes of High Mass Star
  Forming Regions: the Structure and Kinematics of the Milky Way}},  {\em
  Astrophys. J.} {\bf 783} (2014) 130,
  [\href{http://arxiv.org/abs/1401.5377}{{\tt arXiv:1401.5377}}].

\bibitem{Deason:2019kgj}
A.~J. Deason, A.~Fattahi, {\em et.~al.}, {\it The local high-velocity tail and
  the galactic escape speed},  {\em \mnras} {\bf 485} (2019) 3514–3526,
  [\href{http://arxiv.org/abs/1901.02016}{{\tt arXiv:1901.02016}}].

\bibitem{Aad:2019mkw}
ATLAS: G.~Aad {\em et.~al.}, {\it {Measurement of the top-quark mass in
  $t\bar{t}+1$-jet events collected with the ATLAS detector in $pp$ collisions
  at $\sqrt{s}=8$ TeV}},  {\em \jhep} {\bf 11} (2019) 150,
  [\href{http://arxiv.org/abs/1905.02302}{{\tt arXiv:1905.02302}}].

\bibitem{ScannerBit}
\GB Scanner Workgroup: G.~D. {Martinez}, J.~{McKay}, {\em et.~al.}, {\it
  {Comparison of statistical sampling methods with ScannerBit, the GAMBIT
  scanning module}},  {\em \epjc} {\bf 77} (2017) 761,
  [\href{http://arxiv.org/abs/1705.07959}{{\tt arXiv:1705.07959}}].

\bibitem{Akerib:2018lyp}
LUX-ZEPLIN: D.~S. Akerib {\em et.~al.}, {\it {Projected WIMP sensitivity of the
  LUX-ZEPLIN dark matter experiment}},  {\em Phys. Rev. D} {\bf 101} (2020)
  052002, [\href{http://arxiv.org/abs/1802.06039}{{\tt arXiv:1802.06039}}].

\bibitem{Aalbers:2016jon}
DARWIN: J.~Aalbers {\em et.~al.}, {\it {DARWIN: towards the ultimate dark
  matter detector}},  {\em JCAP} {\bf 11} (2016) 017,
  [\href{http://arxiv.org/abs/1606.07001}{{\tt arXiv:1606.07001}}].

\bibitem{Aalseth:2017fik}
C.~E. Aalseth {\em et.~al.}, {\it {DarkSide-20k: A 20 tonne two-phase LAr TPC
  for direct dark matter detection at LNGS}},  {\em \epjp} {\bf 133} (2018)
  131, [\href{http://arxiv.org/abs/1707.08145}{{\tt arXiv:1707.08145}}].

\bibitem{Chala:2015ama}
M.~Chala, F.~Kahlhoefer, M.~McCullough, G.~Nardini, and K.~Schmidt-Hoberg, {\it
  {Constraining Dark Sectors with Monojets and Dijets}},  {\em JHEP} {\bf 07}
  (2015) 089, [\href{http://arxiv.org/abs/1503.05916}{{\tt arXiv:1503.05916}}].

\bibitem{Fairbairn:2016iuf}
M.~Fairbairn, J.~Heal, F.~Kahlhoefer, and P.~Tunney, {\it {Constraints on Z'
  models from LHC dijet searches and implications for dark matter}},  {\em
  JHEP} {\bf 09} (2016) 018, [\href{http://arxiv.org/abs/1605.07940}{{\tt
  arXiv:1605.07940}}].

\bibitem{Bischer:2020sop}
I.~Bischer, T.~Plehn, and W.~Rodejohann, {\it {Dark Matter EFT, the Third --
  Neutrino WIMPs}},  {\em SciPost Phys.} {\bf 10} (2021) 039,
  [\href{http://arxiv.org/abs/2008.04718}{{\tt arXiv:2008.04718}}].

\bibitem{Barbieri:2021wrc}
R.~Barbieri, {\it {A View of Flavour Physics in 2021}},  {\em Acta Phys. Polon.
  B} {\bf 52} (2021) 789, [\href{http://arxiv.org/abs/2103.15635}{{\tt
  arXiv:2103.15635}}].

\bibitem{Graverini:2018riw}
ATLAS, CMS, LHCb: E.~Graverini, {\it {Flavour anomalies: a review}},  {\em J.
  Phys. Conf. Ser.} {\bf 1137} (2019) 012025,
  [\href{http://arxiv.org/abs/1807.11373}{{\tt arXiv:1807.11373}}].

\bibitem{Aaij:2021vac}
LHCb: R.~Aaij {\em et.~al.}, {\it {Test of lepton universality in beauty-quark
  decays}},  \href{http://arxiv.org/abs/2103.11769}{{\tt arXiv:2103.11769}}.

\bibitem{Zhang:2018xdp}
PandaX: H.~Zhang {\em et.~al.}, {\it {Dark matter direct search sensitivity of
  the PandaX-4T experiment}},  {\em Sci. China Phys. Mech. Astron.} {\bf 62}
  (2019) 31011, [\href{http://arxiv.org/abs/1806.02229}{{\tt
  arXiv:1806.02229}}].

\bibitem{Aprile:2020vtw}
XENON: E.~Aprile {\em et.~al.}, {\it {Projected WIMP sensitivity of the XENONnT
  dark matter experiment}},  {\em JCAP} {\bf 11} (2020) 031,
  [\href{http://arxiv.org/abs/2007.08796}{{\tt arXiv:2007.08796}}].

\bibitem{Ahnen:2016qkx}
MAGIC, Fermi-LAT: M.~L. Ahnen {\em et.~al.}, {\it {Limits to Dark Matter
  Annihilation Cross-Section from a Combined Analysis of MAGIC and Fermi-LAT
  Observations of Dwarf Satellite Galaxies}},  {\em JCAP} {\bf 02} (2016) 039,
  [\href{http://arxiv.org/abs/1601.06590}{{\tt arXiv:1601.06590}}].

\bibitem{Abdallah:2016ygi}
H.E.S.S.: H.~Abdallah {\em et.~al.}, {\it {Search for dark matter annihilations
  towards the inner Galactic halo from 10 years of observations with H.E.S.S}},
   {\em Phys. Rev. Lett.} {\bf 117} (2016) 111301,
  [\href{http://arxiv.org/abs/1607.08142}{{\tt arXiv:1607.08142}}].

\bibitem{Aguilar:2021tos}
AMS: M.~Aguilar {\em et.~al.}, {\it {The Alpha Magnetic Spectrometer (AMS) on
  the international space station: Part II \textemdash{} Results from the first
  seven years}},  {\em Phys. Rept.} {\bf 894} (2021) 1--116.

\bibitem{pippi}
P.~{Scott}, {\it {Pippi -- painless parsing, post-processing and plotting of
  posterior and likelihood samples}},  {\em \epjp} {\bf 127} (2012) 138,
  [\href{http://arxiv.org/abs/1206.2245}{{\tt arXiv:1206.2245}}].

\bibitem{Semenov:2008jy}
A.~Semenov, {\it {LanHEP: A Package for the automatic generation of Feynman
  rules in field theory. Version 3.0}},  {\em Comput. Phys. Commun.} {\bf 180}
  (2009) 431--454, [\href{http://arxiv.org/abs/0805.0555}{{\tt
  arXiv:0805.0555}}].

\bibitem{gambit_addendum}
\GB Collaboration: P.~{Athron}, C.~{Bal{\'a}zs}, {\em et.~al.}, {\it {GAMBIT:
  The Global and Modular Beyond-the-Standard-Model Inference Tool. Addendum for
  GAMBIT 1.1: Mathematica backends, SUSYHD interface and updated likelihoods}},
   {\em \epjc} {\bf 78} (2018) 98, [\href{http://arxiv.org/abs/1705.07908}{{\tt
  arXiv:1705.07908}}]. Addendum to \cite{gambit}.

\end{thebibliography}\endgroup

\end{document}